\documentclass[12pt, a4paper]{article}
\usepackage[utf8]{inputenc}
\usepackage{geometry}
\usepackage{graphicx}
\usepackage{amsmath, amssymb}
\usepackage{authblk}

\usepackage{hyperref}
\usepackage{cite}
\usepackage{marvosym}
\usepackage{siunitx}
\usepackage{multirow}
\usepackage{subcaption}
\usepackage{todonotes}
\usepackage{mathrsfs}
\usepackage[title]{appendix}
\usepackage{textcomp}
\usepackage{manyfoot}
\usepackage{booktabs}
\usepackage{algorithm}
\usepackage{algorithmicx}
\usepackage{algpseudocode}
\usepackage{listings}
\usepackage{xr}
\usepackage{longtable}
\usepackage{wasysym}
\usepackage{makecell}
\usepackage{shellesc}
\usepackage{placeins}
\usepackage{etoc}
\usepackage{capt-of}
\usepackage{tabularx}
\newcolumntype{Y}{>{\raggedright\arraybackslash}X}
\usepackage{array}
\usepackage{ragged2e}
\newcolumntype{L}[1]{>{\RaggedRight\arraybackslash\hspace{0pt}}m{#1}}
\usepackage[font=footnotesize,labelfont=bf]{caption}
\hypersetup{
    colorlinks=true,  % 
    linkcolor=black,   %
    citecolor=black,   % 
    urlcolor=black      % 
}
\title{Perceived risk evolution in automated driving inferred from large-scale discrete ratings}

\author[a,b,c,*,\Letter]{Xiaolin He}
\author[c,d,*]{Zirui Li}
\author[e]{Xinwei Wang}
\author[a]{Riender Happee}
\author[c]{Meng Wang}

\affil[a]{Department of Cognitive Robotics, Faculty of Mechanical Engineering, Delft University of Technology, Delft, 2628 CD, The Netherlands}
\affil[b]{Department of Transport and Planning, Faculty of Civil Engineering and Geosciences, Delft University of Technology, Delft, 2628 CN, Delft, The Netherlands}
\affil[c]{Chair of Traffic Process Automation, ``Friedrich List'' Faculty of Transport and Traffic Sciences, Technische Universit\"at Dresden, Dresden, 01069, Germany}
\affil[d]{School of Mechanical Engineering, Beijing Institute of Technology, Beijing, 100081, China}
\affil[e]{School of Engineering and Materials Science, Queen Mary University of London, London, E1 4NS, The United Kingdom}
\date{}

\begin{document}
\addtocontents{toc}{\protect\setcounter{tocdepth}{-5}}
\maketitle
\begingroup
\renewcommand{\thefootnote}{}
\footnotetext{$^*$ These authors contributed equally to this work.}
\footnotetext{\textsuperscript{\Letter} Corresponding author. Email: x.he-2@tudelft.nl}
\endgroup
\vspace{-3em}
\begin{abstract}
Perceived risk in automated driving is often measured as discrete scores that summarise riding experience but this obscures volatile peaks from sustained elevation. Here we treat discrete clipwise ratings as constraints on an unobserved inferred evolution and apply a kernel constrained inverse model to infer the temporal evolution of perceived risk. Across 2,164 participants and 141,628 discrete clipwise ratings spanning 236 hours of scripted motorway interactions, we infer evolutions under kernel constraints whose shapes follow priors from independent handset-based ratings and whose timing is fixed by scripted manoeuvre markers. The inferred perceived risk evolutions differentiate accumulated perceived risk from within clip concentration, revealing scenario differences that are not identifiable from peak judgements alone. We then map these inferred evolutions from observable vehicle and relative motion cues under strict event level holdout using a deep neural network, enabling interpretable attribution analyses. Attribution shows distinct patterns between risk rising and falling segments, with a shift toward conflict cues in the rising phase, and a rebound toward stability cues in the falling phase. Attribution concentration increases only modestly at high perceived risk levels. These results move beyond treating perceived risk as a single severity score by characterising within episode dynamics and phase dependent cue associations in scripted motorway interactions.
\end{abstract}
% \clearpage
%%%%%%%%%%%%%%%%%%%%%%%%%%%%%%%%%%%%%%%%%%%%%%%%%%%%%%%%%%%%%%%%%%%%%%%%%%%%%%%%%%%%%%%%%%%%%%%%%%%%%%%%%%%%%%%%%%%%%%%%%%%%%%%

\section{Introduction}
\begin{figure}[t!]
    \centering
    \includegraphics[width=1\textwidth]{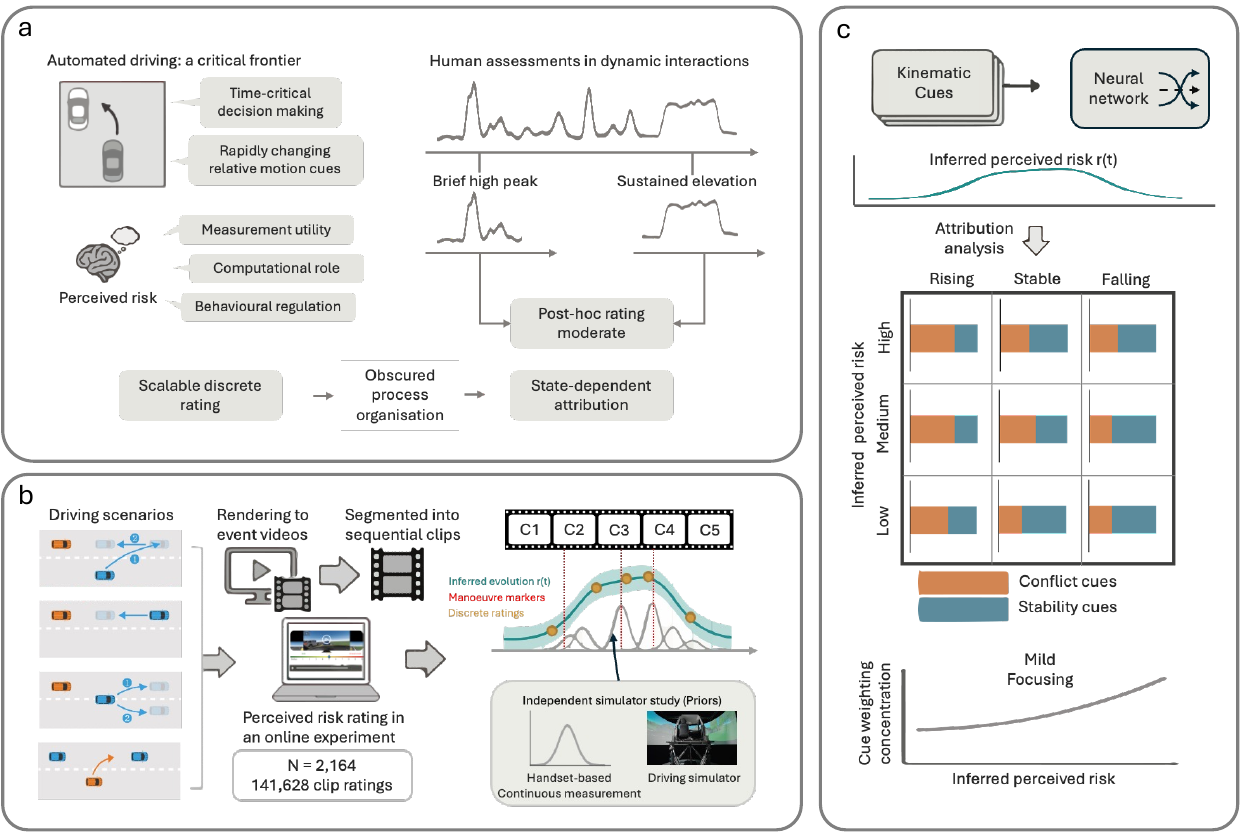}
    \caption{Schematic overview of the study motivation, the full workflow from data collection to inference, and the decoding and attribution analyses. \textbf{a}, Automated driving as a safety critical model system for human assessments in dynamic interactions. Safety relevant judgements must be formed under stringent control demands and rapidly changing relative motion cues, yet retrospective summaries can assign similar overall judgements to episodes that differ in within episode dynamics, such as a brief high peak versus sustained elevation. Perceived risk is used as an interpretable user side appraisal with measurement utility, computational relevance, and links to behaviour regulation. \textbf{b}, End to end workflow from scripted motorway scenarios to inferred evolution. Scripted events are rendered as videos, segmented into sequential clips, and rated online with instructions to reflect the most dangerous moment within each clip. The resulting clip ratings are treated as constraints on an unobserved event level inferred evolution $r(t)$, expressed as a superposition of response kernels whose locations are anchored to pre specified manoeuvre timing markers and whose shapes are grounded by an independent driving simulator study with handset based continuous ratings collected throughout events. \textbf{c}, Decoding the inferred evolution from kinematic cues and testing cue utilisation through attribution analysis. A neural network predicts $\hat r(t)$ from kinematic cues under strict event level holdout, and attribution to the validated predictor is summarised by inferred intensity and by local direction of change, rising, stable, and falling. This analysis tests whether cue utilisation differs across periods and whether attribution concentration shows mild focussing at higher inferred risk. Colours indicate cue families, orange denotes conflict related cues and green denotes stability related cues.}
    \label{fig:fig1 framework}
\end{figure}

Evaluating safety during dynamic interactions is a fundamental challenge for automated driving, yet human perception of risk evolves as vehicles move  \cite{Straub2022_PNAS,Brunton2013_Science,markkula2023explaining}. Classic work on retrospective evaluation shows that post-hoc ratings can be disproportionately shaped by salient moments while being less sensitive to duration \cite{FredricksonKahneman1993DurationNeglect,RedelmeierKahneman1996PatientsMemories}, an effect that has been confirmed across a broad set of domains \cite{Alaybek2022PeakEndMeta}. For automated systems in an interactive environment, this temporal compression matters because identical overall evaluation can arise from different within episode phases, such as a momentary high intensity deviation versus a longer period of moderate elevation \cite{fuller2011driver,lee2004trust}(Fig.~\ref{fig:fig1 framework}a).
%, which can imply different subsequent choices and different demands on behaviour regulation 
Scripted motorway automated driving interactions provide a useful model system for studying these questions because safety-relevant assessments must be formed under changing relative motion cues and evolving manoeuvre phases, making within-episode dynamics consequential rather than incidental \cite{markkula2023explaining}.

As the primary proxy for these assessments, self-reporting perceived risk represents an attractive target because it provides a low-effort measure that can be elicited with brief instructions, while remaining interpretable as a user side assessment rather than a direct physical quantity \cite{slovic2016perception,ledoux2017higher}. In automated driving contexts, perceived risk has been modelled as a function of observable interaction cues \cite{ping2018modeling,deWinter2023predicting,Song2024_TITS}, and it has also been formalised as an internal variable in computational driver models and planning objectives \cite{Fulvio2013_PNAS}. When behaviour is regulated by such an internal risk variable rather than objective risk measures, models can reproduce aspects of human-like adaptation across scenarios \cite{kolekar2020human,Xia2024_GDRF,Yuan2024_IET,He2024_PCAD}. This dual role makes perceived risk attractive for studying how subjective assessment relates to observable cues and computational accounts of behaviour in safety critical interactions (Fig.~\ref{fig:fig1 framework}a).

A key scientific question is therefore not only how strongly perceived risk is expressed, but how it develops within an interaction and how its cue associations fluctuate between periods. Even within the same interaction type, risk perception may depend on distinct combinations of implicit motion cues and contextual cues \cite{Felbel2025LaneChangeCues}, and cognitive process models of driving decisions have highlighted the inherently dynamic nature of  cue integration during an unfolding interaction \cite{Mobbs2010_PNAS,Zgonnikov2024LeftTurnDDM}. These observations motivate tests of whether fixed cue weighting can capture perceived risk throughout an interaction process, or whether the mapping implied by the data 

\noindent differs between periods with rising, stable, or falling risk levels (Fig.~\ref{fig:fig1 framework}c). They also motivate the quest for whether high intensity periods are associated with a more concentrated attribution profile characterised by greater reliance on a smaller set of cues (Fig.~\ref{fig:fig1 framework}c) \cite{mather2011arousal}. Such a shift would suggest that mapping from physical cues to subjective risk assessment in these interactions does not rely on a fixed accumulation of evidence, but involves a state-dependent weighting where the relevance of information is dynamically gated by the current context.

However, perceived risk is commonly measured using static summaries such as post event ratings or clip level judgements (Fig.~\ref{fig:fig1 framework}b) \cite{Xu2018,Nordhoff2021,Stapel2022}. These formats assign an aggregate scalar to a process with internal dynamics and therefore cannot discern risk peaks from sustained elevation, nor resolve the dynamic progression of risk escalation and recovery (Fig.~\ref{fig:fig1 framework}a)\cite{zhu2023real,trende2024trust}. Richer traces can be obtained using continuous response methods or physiological recordings, yet these approaches are harder to scale and introduce interpretive and logistical constraints, especially when the construct of interest is latent and context dependent \cite{cleij2017continuous,wintersberger2018let,du2020psychophysiological,petit2021risk,perello2022physiological,Petit2023SharedSpaceRiskEDA}. This leaves a gap between the scientific quest to unravel within-process dynamics in subjective risk assessments and the methodological deficiency to resolve the within-episode dynamics of subjective appraisal in complex environments \cite{atcheson2017gaussian,nicolaou2011continuous}.
Here we address this gap by combining crowdsourced perceived risk ratings with a kernel-constrained inference procedure to estimate the evolution of perceived risk from discrete constraints. In a large-scale video-based online experiment, $N = 2{,}164$ participants viewed scripted motorway interaction events presented as sequences of short video clips and provided discrete ratings intended to reflect the most dangerous moment within each clip. Across $105$ scripted events, quality controlled data yielded $141{,}628$ clip-level ratings for analysis (Fig.~\ref{fig:fig1 framework}b). We treat the collection of clip ratings as constraints on an unobserved process and estimate its  risk evolution with response shape priors derived from handset-based ratings recorded throughout events in an independent simulator study \cite{He2022}. Kernel locations are anchored to pre-specified manoeuvre timing markers defined by the scripted events, and the inferred evolution is expressed as a superposition of response kernels with constrained rise and fall profiles, yielding an inference anchored by independent priors rather than an unconstrained interpolation.

This inferred evolution then enables two complementary analyses that jointly unravel the characteristics of perceived risk. First, it supports summary metrics that separate cumulative perceived risk from temporal concentration (Fig.~\ref{fig:fig1 framework}c), allowing interaction types to be compared in terms of how perceived risk is distributed within events even when clip ratings are similar. Second, it provides a target for decoding latent dynamics from observable kinematics under strict event level holdout, enabling interpretable attribution analyses that test whether cue utilisation varies between rising, stable, and falling periods (Fig.~\ref{fig:fig1 framework}c). In the studied motorway events, we find that perceived risk judgements vary not only in overall intensity but also within episodes, and that the kinematic mapping implied by the data is clustered by phases, with systematic reallocation between conflict related cues and stability related cues across rising and falling periods. We also observe a modest increase in attribution concentration at higher inferred risk intensity, consistent with mild focusing in high intensity periods. Together, these results move beyond treating perceived risk as a single severity score by characterising how it unfolds within episodes and how its cue associations shift across periods. By providing empirical targets for models of human-compatible behaviour in automated driving interactions, this work serves as a potential starting point for examining the temporal structure of subjective judgments in domains where human-system interaction is integral \cite{Alaybek2022PeakEndMeta}.

\section{Results}\label{chap:results}
We explored perceived risk from the perspective of being a user of an automated vehicle. We evaluated four common motorway traffic scenario types, namely the subject automated vehicle (AV) reacting to hard braking (HB), merging with hard braking (MB), a merging vehicle with lateral control (LC), and the subject AV merging onto the main road (SVM), and systematically varied manoeuvre intensity to cover a wide range of interaction kinematics \cite{Zhao2017Trafficnet} (\textit{\hyperlink{chap:methods}{Materials and Methods}} and \textit{SI Appendix}, \textit{Section~1A--1C}). To bridge the gap between large-scale subjective sampling efficiency and the need to resolve evolution, we developed a kernel-constrained inference approach. By incorporating temporal constraints informed by independent, high-fidelity driving simulator measurements (\textit{SI Appendix}, \textit{Section~2A--2C}), our approach infers about 236 hours of the temporal evolution of perceived risk from 141,628 clip-level ratings (N = 2,164, 200–300 participants per event). 

To demonstrate the necessity of this temporal inference, we derived complementary metrics for cumulative perceived risk ($A$), time-averaged perceived risk ($E$) and temporal concentration ($F$). Analysis of these metrics reveals that distinct interaction scenarios exhibit unique temporal signatures, ranging from brief, concentrated spikes to sustained risk periods that are mathematically not observable in both discrete ratings ($y_k$) and peak-based summaries ($P$) (Fig. \ref{fig:fig3}). We then confirmed the predictive validity of the inferred perceived risk evolution by evaluating a unified Deep Neural Network (DNN) against two physics-based baseline models using rigorous event-based four-fold cross-validation. The DNN's robust predictive performance demonstrates that the inferred perceived risk evolution is reliably grounded in interaction kinematics (Fig. \ref{fig:fig4}). Finally, we employed this DNN to decode the underlying perceptual strategy. By stratifying the risk evolution into nine distinct states (based on intensity and rate of change), we applied SHapley Additive exPlanations (SHAP)-based feature attribution to characterise the state-dependent cue reweighting. This attribution analysis shows that the kinematic predictor DNN relies on different kinematic cues depending on whether the inferred evolution is rising, stable, or falling, revealing a state structured mapping from kinematics to inferred perceived risk (Fig. \ref{fig:fig5}).

\subsection{Kernel-constrained inference yields inferred evolution of perceived risk}
We inferred the temporal evolution of perceived risk from discrete clip-level perceived risk ratings using a kernel-constrained inverse model. Kernel locations were anchored to pre-specified manoeuvre timing markers defined by the scripted events. Because $\{\tau_{e,i}\}$ are fixed by the scripted event specification, kernel placement does not depend on the observed ratings or on any data-driven peak finding in the kinematic traces (\textit{\hyperlink{chap:methods}{Materials and Methods}}). The perceived risk evolution for each event was modelled as a weighted combination of response kernels. We employ Gamma-distributed kernels, a standard choice in cognitive neuroscience and psychophysiology for modelling arousal and physiological responses to discrete stimuli, as they naturally capture the temporal asymmetry of a rapid onset followed by a slower recovery tail \cite{bach2010time, boynton1996linear}. To determine the shape of these kernels, we imposed constraints derived from continuous handset-based perceived risk ratings recorded in an independent high-fidelity driving simulator study \cite{He2022}. This ensures that the inferred evolution captures the characteristic rise-and-return dynamics of human perceived risk, consistent with real-time subjective responses. (\textit{\hyperlink{chap:methods}{Materials and Methods}}; \textit{SI Appendix}, \textit{Sections~2A--2B}). 
\begin{figure}[t!]
    \includegraphics[width=\textwidth]{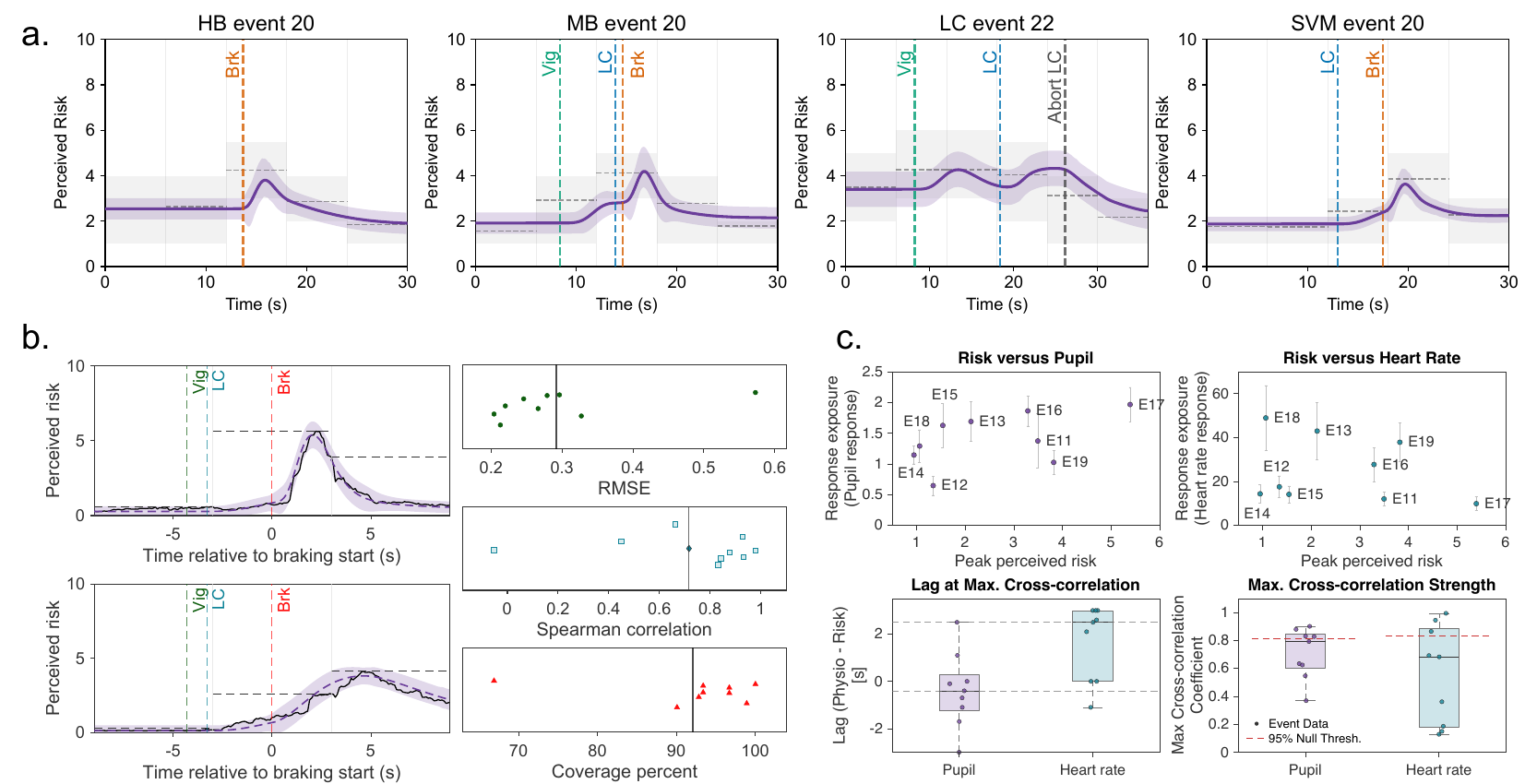}
    \caption{Kernel-constrained inference of perceived risk evolution with validation against time-continuous perceived risk ratings and reference psychological measurements in an independent simulator experiment.
\textbf{a}, Representative online study events from each scenario show inferred evolution of perceived risk (solid black) with uncertainty intervals (grey shading). Events were selected algorithmically to maximise typicality (minimising deviation from the scenario group mean) and inference error (minimising error against discrete ratings), ensuring the plots illustrate characteristic dynamics (see \textit{SI Appendix}, \textit{Section 2B.6} for selection criteria). Horizontal dashed segments indicate the clipwise discrete ratings, and vertical dashed lines mark pre-specified manoeuvre timing markers from the scripted events. Vig marks the scripted interaction onset (neighbouring vehicle appearance or approach), LC marks lane change onset, and Brk marks braking onset of the neighbouring vehicle.
\textbf{b}, Validation against independent driving simulator data for the merging with hard braking scenario. For held-out simulator events (nine events E11-E19), discrete constraints were constructed to match the online protocol by taking the within clip maxima of the 10Hz time-continuous ratings in three six-second clips, together with boundary constraints. The perceived risk evolution was then inferred under the same kernel constraints and simulator informed priors. Example panels show the collected time-continuous ratings alongside the inferred evolutions and uncertainty intervals. The scatter plots summarise event wise agreement metrics across the nine held-out events, including RMSE, rank correlation, and uncertainty coverage, with mean RMSE 0.29, mean Spearman correlation 0.72, and mean coverage 92.1\%. Results for all nine simulator events are in \textit{SI Appendix}, \textit{Section 2C}.
\textbf{c}, Correspondence with reference measurements in the simulator study. For the same held-out events, peak inferred perceived risk shows a clearer association with peak pupil response than with peak heart rate, which is also more variable across events, despite using the same fixed time window aligned to brake onset. Event-level temporal correspondence is quantified using cross-correlations between each physiological signal and the inferred evolution across time lags. We report, for each signal, the lag that maximises correlation (negative indicates physiology leading the inferred risk) and the corresponding maximal correlation coefficient. Across events, pupil dynamics tend to precede the inferred risk (median lag \(=-0.40\,\mathrm{s}\)), whereas heart rate tends to follow (median lag \(=+2.50\,\mathrm{s}\)).
}
\label{fig:fig2}
\end{figure}
Across the online study dataset, the inferred temporal evolution of perceived risk exhibited coherent event-level perceived risk dynamics. In most events, the inferred perceived risk rose around the scripted manoeuvre phase and fell more gradually, while preserving clear differences across events in peak magnitude and return profile. Fig. \ref{fig:fig2}a shows algorithmically selected representative events from each scenario (\textit{SI Appendix}, \textit{Section~2B}), illustrating the correspondence between the inferred perceived risk evolution, the observed clipwise ratings, and the scripted manoeuvre timing markers. Uncertainty bands reflect a simulator calibrated inference error model whose magnitude varies with the inferred perceived risk level and local temporal variability, and are shown as a time dependent interval around the inferred evolution (\textit{\hyperlink{chap:methods}{Materials and Methods}}; \textit{SI Appendix}, \textit{Section~2B}). 

We next evaluated inference accuracy using an independent driving simulator study ($N=25$) with collected time-continuous perceived risk ratings. The kernel shape constraints and the uncertainty calibration were estimated using the remaining simulator events, and the nine events analysed below were excluded from this estimation and used only for evaluation. The simulator study covered only the merging with hard braking (MB) scenario, which was identical to the corresponding scenario in the online study. For each of the 9 MB events included in the inference validation (out of 18 MB events in the simulator dataset), we constructed discrete constraints that matched the online rating protocol (\textit{SI Appendix}, \textit{Section~2B}). The time-continuous perceived risk rating (10 Hz, 18 s duration) was segmented into three 6 s segments, and we extracted the maximum within each segment, reflecting the instruction in the online study to rate the most dangerous moment within each clip. We additionally fixed the first and last samples of the time-continuous perceived risk rating as boundary constraints, and then inferred latent perceived risk evolution again under the same kernel constraints and simulator informed priors (\textit{\hyperlink{chap:methods}{Materials and Methods}}). Across 9 events, the re-inferred temporal perceived risk evolution closely tracked the collected time-continuous perceived risk ratings, with a mean RMSE of 0.29 on a scale of 0-10 and a mean Spearman correlation of 0.72 (Fig. \ref{fig:fig2}b). The inferred temporal perceived risk evolutions were statistically significantly correlated with the collected time-continuous ratings in 8 out of 9 validation events ($p < 0.001$). The single non-significant case Event 10 corresponded to a low-amplitude event where the variance of the collected ratings was insufficient to drive correlation metrics, yet the inference maintained high fidelity in absolute magnitude (RMSE = 0.27). The mean uncertainty coverage was 92.1\%, where coverage denotes the proportion of time points at which the collected time-continuous perceived risk ratings within the estimated uncertainty interval (\textit{\hyperlink{chap:methods}{Materials and Methods}}; \textit{SI Appendix}, \textit{Sections~2B.5}). Fig. \ref{fig:fig2}b shows representative recoveries and the corresponding quantitative agreement metrics. 

Finally, we examined convergent validity using physiological signals recorded in the simulator study (both pupil and heart rate: $N = 25$; \textit{\hyperlink{chap:methods}{Materials and Methods}}; \textit{SI Appendix}, \textit{Section~2C}). For the 9 held-out events, we analysed the physiological responses relative to the inferred perceived risk evolution. While peak pupil dilation showed a positive but non-significant trend with peak inferred risk ($\rho = 0.47, p = 0.21$), peak heart rate associations were inconsistent and negative ($\rho = -0.35, p = 0.36$). This divergence likely reflects the known variability in autonomic magnitude responses and habituation effects ($N=9$). However, despite these magnitude-level discrepancies, the temporal dynamics exhibited strong convergence. Event-level cross correlation analyses further indicated closer temporal correspondence between pupil dilation and the inferred perceived risk evolution than between heart rate and risk. The median peak cross-correlation coefficient was 0.79 for pupil diameter. Permutation tests confirmed that this temporal alignment was statistically significant ($p<0.001$) in 4 out of 9 validation events, indicating that the shape of the inferred perceived risk evolution closely matches the physiological response. Crucially, the temporal structure is aligned with the expected physiological cascade: pupil dilation preceded the inferred risk with a median lead of 0.40 s (median lag $= -0.40\,\mathrm{s}$), consistent with the rapid onset of subcortical arousal relative to conscious motor reporting \cite{bauer_pupillometry_2022}. In contrast, heart rate responses followed the inferred risk (median lag $=+2.50\,\mathrm{s}$), reflecting the slower dynamics of autonomic regulation \cite{mokrane_dynamics_1998}. These results demonstrate that the inferred risk evolution is temporally situated between rapid physiological alerting and slower autonomic response. Fig. \ref{fig:fig2}c reports these associations and lag estimates (\textit{SI Appendix}, \textit{Section 2C}).
\begin{figure}[t!]
\centering
    \includegraphics[width=0.6\textwidth]{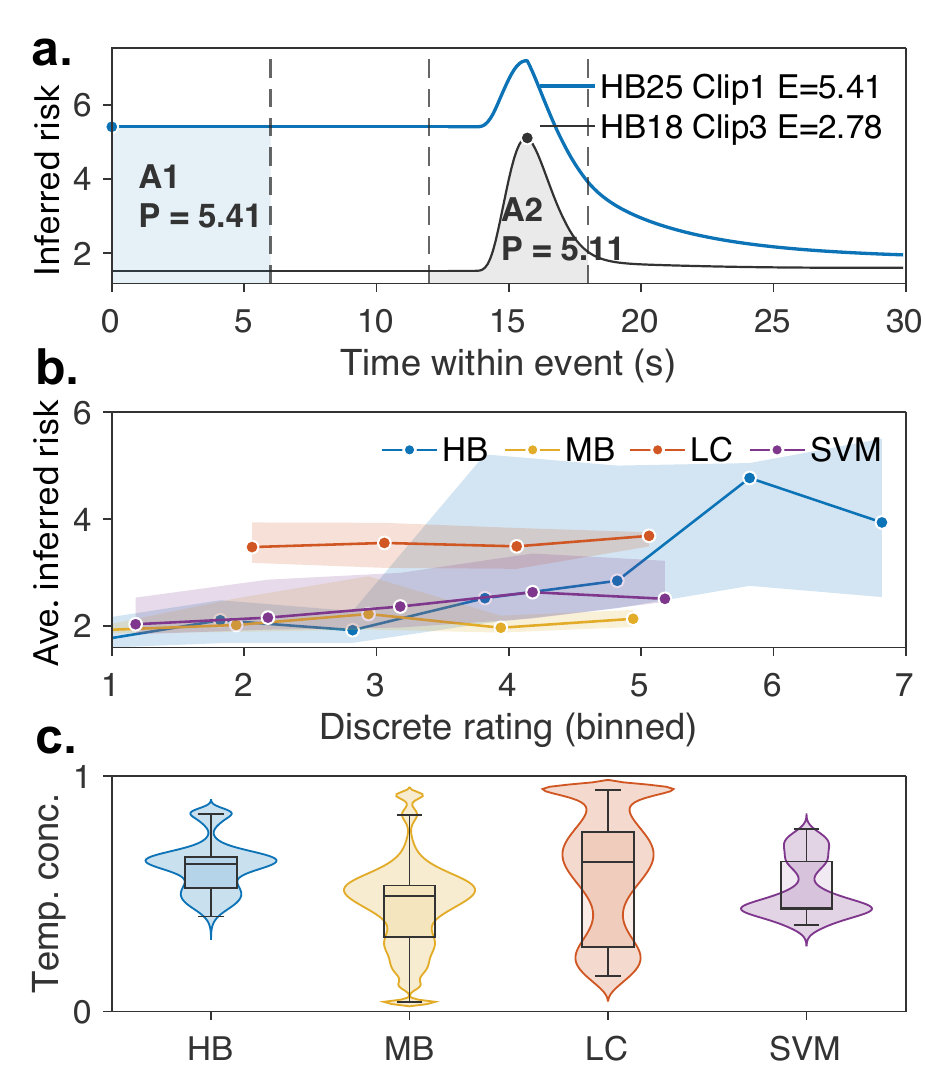}
  \caption{Why peak judgements are insufficient and why inferred evolution is needed.
    \textbf{a}, Two clips can exhibit similar peak perceived risk \(P\) while cumulative perceived risk \(A\) and time-averaged perceived risk \(E\) differ, showing that peak summaries can miss sustained elevation within the same clip duration.
    \textbf{b}, The mapping from the discrete clip rating \(y_k\) to time-averaged perceived risk \(E_k\) depends on interaction type. Clips were binned by \(y_k\) and \(E_k\) was summarised within each bin for each scenario. Points denote bin medians and vertical bars denote interquartile ranges.
    \textbf{c}, Temporal concentration \(F_k\) differs across interaction types. Violin plots summarise \(F_k\) for clips with \(F_k<1\), with horizontal bars indicating medians. The fraction of clips at the boundary value \(F_k=1\) is reported separately for each scenario. Global and pairwise statistics are reported in the main text, with full post hoc results reported in \textit{SI Appendix}, \textit{Table~S5}.
    Together, \textbf{a--c} show that discrete ratings and peak summaries do not capture differences in time-averaged perceived risk and temporal concentration, motivating inference of a latent evolution of perceived risk from which complementary clip level summaries of \(P\), \(A\), \(E\), and \(F\) can be derived.}
    \label{fig:fig3}
\end{figure}

\subsection{Cumulative, mean, and concentration metrics compare risk profiles masked by discrete ratings}

Discrete perceived risk ratings collapse the dynamic experience of risk into a single scalar. Although participants were instructed to report their maximum risk for each clip, this scalar reporting format inherently masks the within-clip temporal evolution. Consequently, it fails to differentiate between acute transient deviations (high-amplitude, short-duration) and sustained risk accumulation (moderate-amplitude, long-duration). To resolve this ambiguity, we derived three metrics from the inferred temporal evolution of perceived risk $r(t)$: Cumulative Perceived Risk ($A$), the time-integral of the evolution representing the total psychological burden; Time-averaged perceived risk ($E$), which normalises this burden $A$ by duration to reflect mean intensity; and the Temporal Density Index ($F$), a dimensionless metric quantifying the evolution's degree of temporal concentration (\textit{\hyperlink{chap:methods}{Materials and Methods}}; \textit{SI Appendix}, \textit{Section~3B -- 3C}).

To ensure that perceived risk magnitudes were comparable across distinct driving scenarios, we calibrated the inferred perceived risk evolution using independent relative severity ratings that are provided by participants to judge each scenario's severity relative to the others ($N=2,164$). In this calibration task, participants explicitly rated the general danger level of each scenario type, allowing us to normalise subjective baselines across contexts (\textit{\hyperlink{chap:methods}{Materials and Methods}}; \textit{SI Appendix}, \textit{Section~3A}). Even after this alignment, we find that discrete ratings fail to distinguish between different dynamic profiles. Crucially, this limitation exists even within the same scenario. As illustrated in Fig. \ref{fig:fig3}a, two clips from HB scenario elicit nearly identical peak risk judgements ($P \approx 5.41$ vs. $5.11$), which would appear indistinguishable if risk were judged solely on extrema. However, their cumulative risk profiles diverge significantly. The first, Event HB25 Clip 1 (close car following), maintains a consistently high risk level, resulting in a high time-averaged perceived risk ($E = 5.41$). In contrast, Event HB18 Clip3 (lead vehicle braking) is characterised by a sudden rise and fall, yielding nearly half the time-averaged perceived risk ($E = 2.78$).

When clips are aligned by the clip-level discrete rating $y_k$, the time-averaged $E$ separates interaction types in a way that the discrete ratings cannot reveal. By binning all clips by their discrete rating $y_k$, we analysed the distribution of time-averaged perceived risk ($E$) across scenarios (Fig. \ref{fig:fig3}b)(\textit{\hyperlink{chap:methods}{Materials and Methods}}). The mapping from subjective rating to time-averaged perceived risk magnitude is scenario dependent. This discrepancy is particularly pronounced between LC and MB scenarios. For a moderate rating bin $y_k \in (3.5, 4.5]$, the median $E$ is 3.30 in LC ($n=35$) but 1.97 in MB ($n=22$), a gap of 1.33 units. This pattern is consistent with duration neglect in discrete ratings, in that a high amplitude, short duration transient deviation and a moderate amplitude, long duration sustained accumulation can receive comparable peak judgements while implying different cumulative burdens.

Temporal concentration differed across interaction types (Fig.~\ref{fig:fig3}c). Many clips showed little within-clip change in the inferred evolution, so $F_k$ takes its boundary value of 1 by definition, reflecting negligible temporal variation rather than low perceived risk. Because the pile-up of values at $F_k=1$  primarily reflects the prevalence of approximately constant clips rather than graded temporal concentration within clips that exhibit a structured rise and fall, we analysed the continuous component of the distribution by restricting to $F_k<1$, and found that $F_k$ varied across scenarios (Kruskal--Wallis $H=56.61$, $p<0.001$). Median $F_k$ was 0.63 in HB and 0.64 in LC, compared with 0.49 in MB and 0.44 in SVM. Post-hoc Dunn comparisons with Holm correction showed that HB and LC exceeded MB and SVM (all $p <0.01$), whereas HB and LC did not differ ($p=0.32$) and MB and SVM were not distinguishable ($p=0.14$). Separately, the prevalence of clips with $F_k=1$ differed across scenarios ($\chi^2=43.39$, $p<0.001$), occurring more often in HB and SVM (40\%) than in LC (16.7\%) and MB (13.3\%). The association corresponded to Cram\'er's \(V=0.28\) (\textit{SI Appendix}, \textit{Table~S5}). Together, these results show that interaction type modulates whether inferred risk is concentrated into brief episodes or distributed more evenly over the clip, a distinction that is not captured by peak judgements alone.

\subsection{Kinematics to inferred evolution mapping holds under event holdout}\label{sec:kinematic_predictability}

\begin{figure}[t!]
    \centering
    \includegraphics[width=\textwidth]{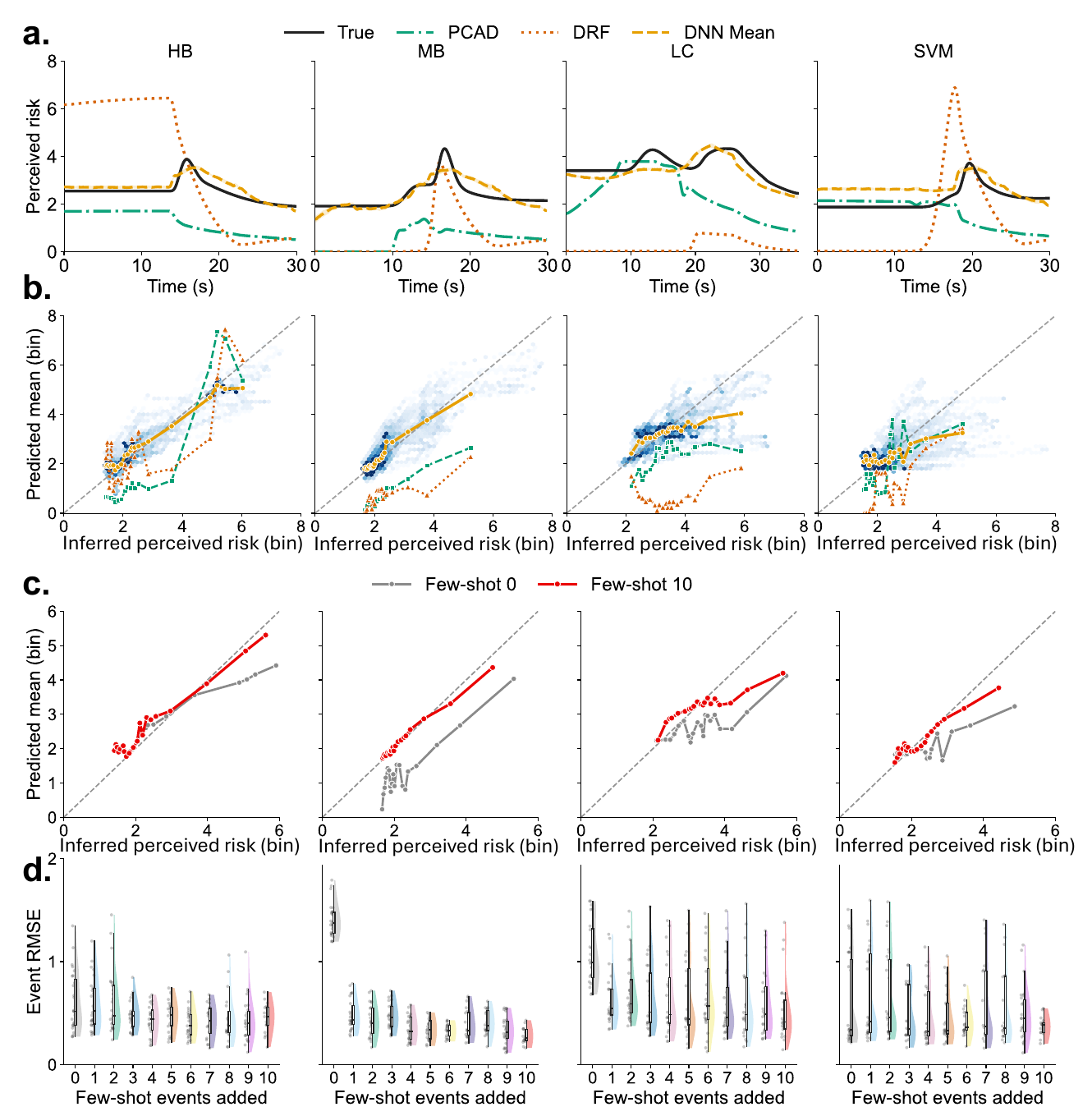}
    \caption{Kinematics to inferred evolution mapping holds under event holdout. 
\textbf{a}, Unified scenario decoding under strict event level four fold cross validation. Top row shows representative held out events from HB, MB, LC, and SVM comparing the inferred perceived risk evolution (black) with predictions from the unified DNN (orange dashed, mean across independent initialisations) and physics based baselines PCAD (green dash dot) and DRF (red dotted). The DNN follows both the onset and recovery profile, whereas the baselines frequently show abrupt spikes or misaligned decays.
\textbf{b}, summarises calibration within each scenario using binned conditional means, where the horizontal axis is the inferred risk mean within a bin and the vertical axis is the predicted mean within the same bin. Shaded points denote the density of time samples, the dashed diagonal indicates the identity relation, and overlaid curves show bin means for each model. Marginal histograms show the distribution of binned inferred and predicted means.
\textbf{c}, Cross-scenario stress test. The row shows quantile binned conditional means for the held out scenario after adding one event or ten events from the target scenario to the training set, with the dashed diagonal indicating the identity relation. 
\textbf{d}, Few-shot adaptation. The row shows raincloud plots of event RMSE as the number of few shot events added increases from one to ten, combining a half violin density, jittered event points, and a box plot summary. Box plots indicate the median and interquartile range, and whiskers extend to 1.5 times the interquartile range.}
\label{fig:fig4}
\end{figure}

We examined whether the inferred evolution is systematically predictable from kinematic features under strict event holdout, as a test of kinematic grounding. We trained a unified deep neural network (DNN) to predict the inferred evolution $r(t)$ from vehicle kinematics and evaluated it with strict event level four-fold cross-validation across all 105 events, such that no time samples from a held-out event were visible during training, model selection, or calibration (\textit{\hyperlink{chap:methods}{Materials and Methods}} and \textit{SI Appendix}, \textit{Section 4D}). Across held-out events, the DNN achieved a mean event RMSE of 0.51 with a median of 0.45, and a mean within event Pearson correlation of 0.81 with a median of 0.88 between predicted and inferred evolution, demonstrating that the inferred evolution carries a stable kinematic signature that generalises across events. Representative held out events illustrate that the DNN follows both the rise and the return toward baseline in the inferred evolution (Fig.~\ref{fig:fig4}a). We repeated training with five independent random initialisations and report the mean prediction across runs to reduce dependence on any single random seed (\textit{SI Appendix}, \textit{Section~4H}).
Physics-based models provide an interpretable reference for what is captured by simple kinematic mappings. We therefore report two such physics-based references, PCAD \cite{He2024_PCAD} and DRF \cite{kolekar2020human} to contextualise the decoding task and to provide a physically anchored point of comparison for the learned kinematic mapping. Importantly, the purpose of these references is not to identify a uniquely correct functional form, but to show what can be obtained by limited flexibility mappings derived from established risk surrogates. To ensure comparability across methods, within each cross-validation fold, the DNN was trained on the training events, whereas PCAD and DRF were calibrated by fitting their model parameters on the same training events only, and all three models were evaluated on held out events only (\textit{SI Appendix}, \textit{Section 4D}). On held out events, mean event RMSE was 0.51 for the DNN, compared with 1.62 for PCAD and 2.12 for DRF (\textit{SI Appendix}, \textit{Section~4J}; \textit{Table~S10}). Agreement on held out time samples was further summarised using quantile binned conditional means within each scenario (Fig.~\ref{fig:fig4}b). Averaged across scenarios, the mean absolute deviation of these binned conditional means from the identity relation was 0.28 for the DNN, compared with 1.17 for PCAD and 1.66 for DRF (\textit{SI Appendix}, \textit{Section~4M}).

As an upper bound on within type predictability, we also trained scenario specific DNNs and evaluated them with strict leave-one-event-out testing within each scenario, which yields modestly tighter error distributions and confirms that the unified model is not artificially advantaged by mixing interaction types (\textit{SI Appendix, Section 4N}). Across held out events pooled over scenarios, these scenario specific predictors achieved a median event RMSE of 0.28 and a median within event Pearson correlation of 0.93, providing an upper bound relative to the unified model evaluated across interaction types (\textit{SI Appendix}, \textit{Table S11}). We then tested robustness under interaction type shift by holding out one scenario type, training the unified DNN predictor on the remaining scenarios, and evaluating it on the unseen scenario. Zero shot transfer widened the error distributions, consistent with scenario specific differences in kinematic composition and, in particular, uncertainty in the baseline for an unseen interaction type, whereas few shot adaptation rapidly tightened these distributions (Fig.~\ref{fig:fig4}c-d). Increasing the number of added events from zero to ten reduced the median event RMSE by 11.0\% for HB from 0.52 to 0.47, 40.6\% for MB from 0.43 to 0.26, 24.5\% for LC from 0.55 to 0.41, and 7.7\% for SVM from 0.42 to 0.38. In parallel, agreement in the quantile binned conditional means improved from zero shot to few shot 10, with the mean absolute deviation from the identity relation across 20 bins decreasing from 0.45 to 0.33 in HB, from 0.97 to 0.09 in MB, from 0.68 to 0.35 in LC, and from 0.45 to 0.19 in SVM (\textit{SI Appendix}, Table S12).

Together, these predictability and stress test results suggest that the inferred evolution carries a reproducible kinematic signature under strict event level evaluation. We next apply SHapley Additive exPlanations (SHAP) to the unified predictor, and compute attribution summaries under the same event level separation used for Fig.~\ref{fig:fig4}, to quantify which kinematic cues dominate its predictions over time, and how these attributions differ between time points where the inferred evolution is increasing and time points where it is decreasing.

\subsection{State dependent cue reweighting differs across rising, stable, and falling segments}
\begin{figure*}[htbp]
\centering
    \includegraphics[width=0.95\textwidth]{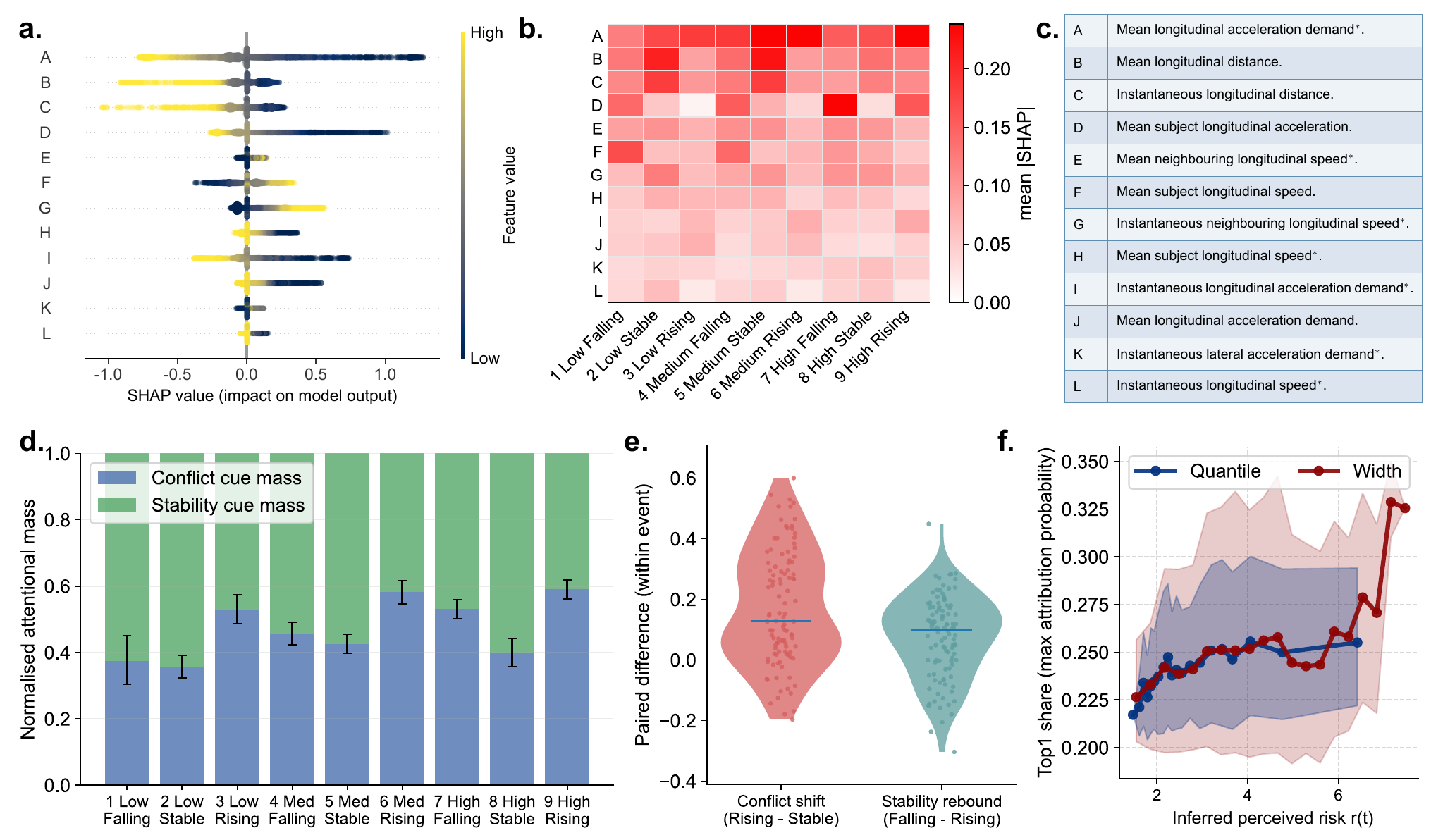}
    \caption{State dependent cue reweighting with a physical anchoring check.
    \textbf{a}, Model physical anchoring and state dependence for the same ordered feature set. The beeswarm shows signed Shapley values for the top features, with points coloured by feature value. \textbf{b}, The state heat map reports mean absolute Shapley values within each of the nine risk states using the identical feature ordering. \textbf{c}, The table provides descriptions for the same ordered features. All signed kinematic variables are expressed in a vehicle fixed frame, with longitudinal positive forward and lateral positive left. Variables carrying \(\mathrm{mean}\) denote a local average computed within a 50 time step window (5 s), whereas variables without \(\mathrm{mean}\) denote the instantaneous value at the same time point. An asterisk (*) marks the PCAD manoeuvre uncertainty term, which provides a conservative anticipatory surrogate beyond instantaneous kinematics (Definitions of all features are detailed in \textit{SI Appendix}, \textit{Section 4D}).
    \textbf{d}, Cue family allocation across nine risk states. For each event, absolute Shapley values were normalised to sum to one, then summed within each cue family to yield a cue family share, and finally aggregated across events. Risk states are defined by inferred risk intensity (low, medium, high) combined with local phase (falling, stable, rising). Bars show the mean across events and error bars indicate 95\% bootstrap intervals with event resampling.
    \textbf{e}, Paired within event contrasts. The left distribution shows the within event difference in conflict cue share between rising and stable segments within the same intensity tier, and the right distribution shows the corresponding difference in stability cue share between falling and rising segments. Points denote events and violin envelopes show empirical distributions; horizontal bars indicate medians.
    \textbf{f}, Modest increase in attribution concentration with inferred risk intensity. At each time point, absolute Shapley values were normalised to sum to one across features, and the top one share was defined as the largest resulting feature share. The mean top one share was summarised as a function of inferred risk using two binning schemes, quantile bins and equal width bins (legend). Within each bin and for each scheme, we used matched event count resampling to aggregate the same number of events per bin and to obtain uncertainty intervals. Lines show bin means and shaded regions indicate 95\% intervals.
}
    \label{fig:fig5}
\end{figure*}

To interpret feature attributions as cue utilisation rather than dataset specific associations, we first checked that the learned kinematic mapping exhibits directionality consistent with collision avoidance intuition by inspecting signed Shapley patterns in the beeswarm summaries (\ref{fig:fig5}a, \textit{SI Appendix}, \textit{Fig. S13}). In the held out events, larger braking demand as captured by DRAC (an action demand, see \textit{SI Appendix}, \textit{Section 4C}) tends to contribute positively to the predicted perceived risk, reduced distance to a neighbouring vehicle contributes positively, and higher speeds tend to contribute positively for a given margin. We stratified the inferred evolution into nine risk states defined by intensity and local rate of change, and examined Shapley attributions both globally and within states (Fig.~\ref{fig:fig5}a-c). Using the identical ordered feature set, the state map in Fig.~\ref{fig:fig5}b shows that rising states exhibit a more concentrated attribution profile, whereas stable and falling states exhibit a more distributed profile across features. We quantify these statewise shifts with the cue share and concentration analyses below.

On this basis, we used the model as a probe for adaptive processing. We grouped the kinematic features into two cue families. Beyond instantaneous kinematics, the ordered feature set also includes conservative anticipatory surrogates, captured by collision avoidance demand indices and by an explicit manoeuvre uncertainty term (marked in Fig.~\ref{fig:fig5}c; definitions in \textit{SI Appendix}, \textit{Section 5A}). Conflict cues capture immediate control demand and closing tendency, including longitudinal and lateral components of DRAC terms, acceleration, and relative speed. Stability cues capture safety margin and vehicle state, including spacing margin and speed terms. For each time point, we normalised absolute Shapley values across features so that they sum to one and summed within each family to obtain a cue family share, then aggregated within events and within risk states (\textit{SI Appendix}, \textit{Section 6F}).
Fig.~\ref{fig:fig5}d shows a systematic shift toward conflict cues in the rising states of perceived risk. Within the same risk level, rising states assign a larger share of Shapley values to conflict cues than stable states. For example, the conflict cue share increases from 0.425 in the medium stable state (state 5 in Fig.~\ref{fig:fig5}d) to 0.582 in the medium rising state (state 6 in Fig.~\ref{fig:fig5}d), and from 0.400 in the high stable state to 0.591 in the high rising state. This difference is not an artefact of comparing different events. Fig.~\ref{fig:fig5}e (left) quantifies the within event change by comparing, within each event, the rising state against the stable state at the same risk level, and then averaging the resulting differences across levels within that event. Across events, the median increase in conflict cue share is 0.128, with a 95\% bootstrap interval of [0.083, 0.240]; a paired one sided Wilcoxon signed rank test confirms that the distribution is greater than zero ($n=105, p<0.001$).

A complementary pattern appears when comparing falling states against rising states at the same perceived risk level. Within events, the stability cue share is higher in falling than in rising. In Fig.~\ref{fig:fig5}e (right), we compute this effect within each event by taking, for each risk level, the difference in stability cue share between falling and rising, and then average these levelwise differences to obtain one value per event. Across events, the median increase in stability cue share is 0.101, with a 95\% bootstrap interval of [0.049, 0.120]; a paired one sided Wilcoxon signed rank test confirms that the distribution is greater than zero ($n=105, p<0.001$). Together, these two within event contrasts show that the cue shares depend on state, not only on risk level.

Finally, we tested whether attribution becomes more concentrated at higher inferred risk levels. Fig.~\ref{fig:fig5}f reports the top one share, defined as the largest feature share after normalising absolute Shapley values to sum to one across features at each time point. To summarise how this quantity varies with inferred risk, we used both quantile binning and equal width binning as a robustness check, and in each case used matched event count resampling to obtain uncertainty intervals for the mean curve. Both binnings show a modest upward trend with inferred perceived risk. To quantify this trend without relying on any particular binning choice, we additionally estimated, within each event, the slope of top one share as a function of inferred perceived risk using all time points. Eventwise slope tests computed on the unbinned time samples confirm that this upward trend is greater than zero across events (median slope $=0.011$, $n=105$; one sided Wilcoxon signed rank test $p<0.001$). Complementary concentration summaries, including attribution entropy and effective cue count, are reported in the \textit{SI Appendix}, \textit{Section 6D}.

\section{Discussion}\label{chap:discussion}
Perceived risk of automated driving in motorway interactions cannot be reduced to a single severity score. Across the four interaction types, clipwise judgements that appear similar at the summary level can correspond to meaningfully different within event phases, separating sustained exposure from volatile concentrated peaks. When the inferred evolutions are decoded from kinematics under strict event holdout, the resulting cue associations are phase structured, with systematic reallocation between conflict related cues and stability related cues across rising, stable, and falling segments within the same events. We also observe only a modest increase in attribution concentration at higher inferred intensity, suggesting mild focusing rather than near exclusive reliance on a single cue. 

A central contribution is therefore not simply that we interpolate a curve through discrete ratings, but that we pose an inverse inference problem with explicit, externally grounded structure. Clipwise maxima constraints alone do not identify a unique within event profile, and unconstrained interpolation would entangle the data with arbitrary smoothness choices. We avoid that failure mode by restricting the solution class to superpositions of response kernels. The rise and fall profiles of these kernels are informed by independent handset-based continuous ratings. Furthermore, their placement is anchored to scripted manoeuvre markers rather than being inferred from kinematic time series. Under this model, the data primarily inform kernel weights and baseline terms, while the temporal development is constrained by priors that can be checked in held out reference measurements. The result is an inference grounded in response shape priors estimated from independent handset based ratings and in kernel timing fixed by scripted manoeuvre markers, and it should be interpreted as an inferred evolution consistent with the measurement protocol rather than as a direct observation of internal state at each instant. It supports comparisons of within episode dynamics and recovery shape, but it should not be used to draw claims about fine scale within clip peak timing or other dynamics that are not identifiable from discrete maxima constraints.

The time-averaged and temporal concentration analyses clarify why this inference matters beyond presentation. Even under peak instructed reporting, clips with similar discrete ratings can carry different time-averaged perceived risk $E$ and different temporal concentration $F$, indicating that scalar summaries do not uniquely determine accumulated perceived risk or within-clip risk dynamics. The scenario dependent mapping from $y_k$ to $E_k$ in Fig.~\ref{fig:fig3}b is consistent with duration neglect and shows that interaction type can shift how a given rating relates to integrated experience. Temporal concentration $F$ further distinguishes clips dominated by short peak episodes from clips with more distributed elevation. Together, these results imply that peak oriented summaries such as $P$, and the clip rating $y_k$, can miss distinctions that are explicit in the inferred evolution and relevant to how an episode is experienced and remembered. In many dynamic tasks, global ratings can be efficient but can also conflate experiences that differ within events. 

The credibility of the inferred evolution is supported by three independent checks, and we keep interpretation bounded to what these checks directly support. First, in the simulator recovery validation, the kernel-constrained inference tracks collected time continuous handset ratings when only discrete constraints matching the online protocol are provided, with statistically significant correspondence in most held out events. The one non-significant event occurred in a low amplitude case where correlation metrics were limited by variance, which highlights a boundary condition for inference evaluation rather than a contradiction of the reconstruction. Second, physiological analyses are based on a small number of held-out events, and the mixed peak magnitude associations underscore that autonomic responses are noisy and context dependent. Consequently, convergent validity with physiology is mixed at the level of peak magnitude, which cautions against interpreting inferred peak risk as a proxy for autonomic magnitude. At the same time, the dynamic correspondence was stronger and the ordering of lags aligns with a plausible cascade in which pupil dynamics precede conscious reporting, followed by heart rate response \cite{bauer_pupillometry_2022,mokrane_dynamics_1998}. Third, kinematic decoding under strict event holdout shows that the inferred evolution carries a reproducible kinematic signature and is predicted more accurately by the unified DNN than by physics-based baseline models of perceived risk (PCAD and DRF), supporting validity in the minimal sense that the inference is systematically related to closing and recovery patterns in the kinematics rather than being arbitrary. The cross scenario stress test complements this point by showing that interaction type shifts baselines and kinematic composition, while few shot adaptation can align a predictor to a new interaction type without changing the event based evaluation protocol.

Within this validated setting, the attribution analyses quantify state dependent cue utilisation in the learned kinematic mapping. At matched inferred perceived risk level, rising segments allocate a larger share of attribution to conflict cues, whereas falling segments allocate relatively more to stability cues within the same events. The paired within-event contrasts reduce the chance that these shifts are driven only by between event differences in manoeuvre severity. These attributions characterise how the predictor uses the available kinematic inputs to reproduce the inferred evolution, and they should not be interpreted as direct evidence about neural computation. Nevertheless, the phase structured shifts are consistent with the idea that different information is emphasised when perceived risk is increasing versus decreasing \cite{wickens2005attentional,lee1976theory,ernst2002humans,fetsch2009dynamic}. We also observe only a modest increase in attribution concentration with higher inferred intensity, suggesting mild focusing rather than near exclusive reliance on a single cue.

These results motivate phase-aware design questions for motorway automated driving \cite{Lan2017_LKAS}. If perceived risk varies within events and cue utilisation differs between rising, stable, and falling segments, then control objectives that only penalise peaks can miss sustained perceived risk and overlook how return towards baseline is shaped. The inferred evolution supports complementary targets, such as reducing time averaged exposure while avoiding overly concentrated peaks and supporting credible return profiles. We use the term bandwidth alignment to denote a phase-aware matching between the control objective emphasis and the cue utilisation in that phase, while recognising that interaction type can shift both baselines and cue utilisation.

Several constraints bound the scope of the present evidence. The online study used screen based video ratings, so participants had no control authority and did not experience vestibular cues or steering effort; accordingly, we emphasise within-event dynamics and relative comparisons rather than absolute baseline levels that may shift with context, presentation format, or user role. Automated vehicles can place occupants in a rider role with limited control over the vehicle and limited access to why the automation acts as it does. In that role, perceived risk may partly reflect perceived controllability and system transparency, including dread related concern about rare severe outcomes, alongside the kinematic cues shown in the scene. Because our video protocol did not manipulate or measure these components separately, we interpret the present results as evidence about the dynamics of perceived risk and its cue associations within the scripted motorway episodes, rather than as estimates of absolute perceived risk levels in deployed automated services. We also do not attempt to calibrate perceived risk against objective risk in this study, so statements about multiplicative gaps between perceived and actual risk are outside the scope of the present data. Although simulator anchoring data were available for one motorway interaction type, that interaction engaged the same core kernels used throughout this work, including vigilance related onset, lane change onset, and braking onset, spanning longitudinal and lateral manoeuvre components common across the scripted motorway events studied here; extension to driving scenes with substantially different perceptual structure will require additional anchoring and evaluation. The present analyses are at the group level and the physiological convergence check is based on a limited holdout set, so stable individual differences and autonomic magnitude relations were not a focus. Finally, viewpoint, role instruction, and display geometry may shift both baselines and cue utilisation, motivating targeted tests under alternative presentation conditions and more naturalistic settings.

Despite these constraints, the study indicates that, in scripted motorway automated driving interactions, perceived risk should be evaluated with attention to within-event development and phase dependence rather than peak intensity alone. More generally, it motivates a testable question in other dynamic human--system interactions, namely whether discrete summary judgements obscure within-process variation that can be examined using domain-specific anchoring and independent validation.
\section{Materials and Methods}\label{chap:methods}
\subsection{Design of driving scenarios}\label{chap:Design_of_driving_scenarios}
We created four predefined motorway scenario types in which the subject automated vehicle (AV) interacted with one or more neighbouring vehicles --- hard braking (HB), merging with hard braking (MB), reacting to a merging vehicle with lateral control (LC), and subject AV merging onto the main road (SVM) (Fig.~\ref{fig:fig1 framework}\textbf{a}, left). In HB, a lead vehicle braked hard and then returned to cruising. In MB, a vehicle merged in front of the subject AV from an on-ramp and then braked hard. In LC, a vehicle approached from an adjacent lane and executed a lane change toward the subject AV with systematically varied lateral behaviour. In SVM, the subject AV merged into dense traffic and adjusted speed to maintain spacing. Within each interaction type, scenario parameters such as initial spacing, merging distance, cruising speed, and braking intensity were varied at discrete levels to span a wide range of criticality, yielding 105 unique events in total (event specifications are provided in \textit{SI Appendix}, \textit{Table~S1}).

All events were implemented in IPG CarMaker to generate vehicle kinematics, which were recorded at 10 Hz. Each event video was segmented into sequential six second clips without overlap for online rating.

\subsection{Procedure}\label{chap_methods_procedure}
The online study was approved by the Human Research Ethics Committee of Delft University of Technology under application number 1245. Participants were recruited via Prolific \cite{Prolific2023} and provided digital consent before participation. After a brief training module, participants rated perceived risk after each video clip using a slider scale from 0 to 10, recorded as integer responses. Each participant rated 16 events in total, four events per scenario type, with event selection and presentation order randomised across participants. Videos were embedded in an online questionnaire (see \textit{SI Appendix}, \textit{Section 1B} for questionnaire materials).

\subsection{Data validity}\label{chap_EoCPoDS}
We applied quality control to exclude incomplete sessions and implausibly fast completion consistent with not viewing the clips, and we additionally removed clip sequences that were inconsistent with the group level within event pattern, yielding a final sample of $N=2{,}164$ participants and 141,628 clip level ratings. To confirm that the scripted parameter variations produced meaningful perceptual variation, we tested whether controlled scenario parameters systematically changed ratings using a mixed effects regression model with participant level random intercepts. Full exclusion criteria, thresholds, sensitivity checks, and regression outputs are reported in the \textit{SI Appendix}, \textit{Section 1C}.
\subsection{Kernel-constrained inverse inference of perceived risk evolution}\label{chap:methods_inference_perceived_risk}
The online study provides one perceived risk rating per clip, which we treat as discrete constraints on an unobserved, inferred evolution of perceived risk $r_{p,e}(t)$ for each participant $p$ within each event $e$. We represent $r_{p,e}(t)$ on a fixed temporal grid (e.g., $10\,\mathrm{Hz}$) for numerical convenience, noting that this discretisation is an oversampling relative to the response bandwidth implied by the kernels and does not imply sub-second psychological dynamics.

We model $r_{p,e}(t)$ as a superposition of response kernels anchored to prespecified scripted manoeuvre timing markers (for example braking onset and lane change onset), together with a baseline and return component:
\begin{equation}
r_{p,e}(t) = b_{p,e}(t;\theta_b) + \mathcal{F}\!\Big(\{w_{p,e,i}\,k_i(t-\tau_{e,i};\theta_i)\}_{i=1}^{M_e}\Big),
\end{equation}
where $\{\tau_{e,i}\}$ are scripted timing markers defined by the event specification $e$ and held fixed for all participants. These markers are not estimated from the kinematic time series and no kinematic feature values are used to place kernels or to assign within-clip peak times. Kernel locations $\{\tau_{e,i}\}$ are read directly from the scenario script used to render each event and were fixed before any rating data were analysed, so the clipwise ratings only inform kernel weights and baseline terms. $w_{p,e,i}\ge 0$ are inferred participant- and event-specific weights, and $\mathcal{F}(\cdot)$ is a scenario-specific fusion rule (e.g., SoftMax aggregation). Kernel shapes $\theta_i$ are fixed by response-shape priors estimated from an independent simulator study with time-continuous handset-based ratings \cite{He2022}, rather than from physiological signals. The kernel family, the prior estimation procedure, and the scenario-specific fusion definitions are provided in the \textit{SI Appendix}, \textit{Section 2A}.

The observation operator follows the online instruction to rate the most dangerous moment within each clip. For clip interval $[t_k,t_{k+1}]$ with observed rating $y_{p,e,k}$, we model
\begin{equation}
y_{p,e,k} \approx \max_{t\in[t_k,t_{k+1}]} r_{p,e}(t),
\end{equation}
and infer $\{w_{p,e,i}\}$ and parameters of $b_{p,e}$ by minimising a hinge squared reconstruction loss with tolerance bands, together with weak regularisation to enforce plausible rise and return behaviour. Optimisation is performed at the participant level to obtain inferred evolutions. Full objective definitions, tolerance settings, and implementation details are reported in the \textit{SI Appendix}, \textit{Section 2B}. Note that physiological signals are not used to constrain the inference, to place kernel anchors, or to train the kinematic decoder, and are analysed only as independent reference measurements for a convergent check.
\subsection{Cross-scenario perceived risk alignment}
Perceived risk scales can vary across driving contexts. To improve comparability across scenario types, we collected independent relative severity ratings in which participants rated the general danger of the four scenarios (HB, MB, LC, SVM). For each scenario $s$, we computed a global robust rescaling factor $\bar{\alpha}_s$ as the median of participant-specific scaling factors. These individual factors were derived from ratios relative to a participant-specific anchor, defined as the second largest rating provided by that participant across the four scenarios. We applied a linear rescaling around the event median:
\begin{equation}
r_{\mathrm{cs}}(t) = c + \bar{\alpha}_s \cdot (r(t) - c),
\end{equation}
where $c$ is the temporal median of the inferred evolution $r(t)$ and $r_{\mathrm{cs}}(t)$ denotes the cross-scenario aligned risk. Full task materials, the definition of individual factors $\alpha_{i,s}$, and sensitivity checks are reported in the \textit{SI Appendix}, \textit{Section 3A}.
\subsection{Definition of perceived risk metrics}
To characterise the temporal topology of perceived risk, we computed three derived metrics at the clip level, with $T$ denoting the clip duration and $r_{\mathrm{cs}}(t)$ denoting the cross-scenario rescaled inferred evolution within that clip. We defined Cumulative Perceived Risk ($A$) as the definite integral of the risk intensity, $A = \int_{0}^{T} r_{\mathrm{cs}}(t) dt$, representing the total psychological burden accumulated throughout the event. This metric was normalised by the event duration to obtain the Mean Risk Intensity ($E = A/T$), which reflects the time-averaged magnitude of sustained pressure.

Furthermore, to distinguish between acute transient deviations and sustained accumulation, we derived the dimensionless Temporal Density Index ($F$). We computed $F$ on a within-clip baseline-removed signal $\tilde r(t)=r_{\mathrm{cs}}(t)-\min_{t\in[0,T]} r_{\mathrm{cs}}(t)$, so that temporal concentration is quantified independently of an arbitrary offset. Let $\tilde E = \frac{1}{T}\int_{0}^{T}\tilde r(t)\,dt$ and $\widetilde{M}_2=\frac{1}{T}\int_{0}^{T}\tilde r(t)^2\,dt$. We then defined
\begin{equation}
F=\frac{\tilde E^2}{\widetilde{M}_2}.
\end{equation}
When \(\tilde r(t)\) exhibits negligible temporal variation within a clip, \(F\) collapses to the boundary case \(F=1\) by construction, reflecting approximate constancy rather than low intensity. As demonstrated in the \textit{SI Appendix}, Section~3C, $F$ is mathematically related to the coefficient of variation; physically, values approaching unity indicate a uniform, sustained risk experience, while values approaching zero indicate risk concentrated in brief, high-intensity spikes.
\subsection{Statistical analysis of time-averaged perceived risk and temporal concentration}
For Fig.~\ref{fig:fig3}b, clips were binned by the observed discrete rating $y_k$ and scenario-specific distributions of $E_k$ were summarised by bin medians and interquartile ranges. We used unit width bins with half unit edges, for example $y_k \in (3.5,4.5]$. For Fig.~\ref{fig:fig3}c, scenario differences in $F_k$ were tested on the continuous component $F_k<1$ using a Kruskal--Wallis test, followed by Dunn pairwise comparisons with Holm correction. The prevalence of the boundary case $F_k=1$ across scenarios was tested using a chi-square test of independence, and association strength was summarised by Cram\'er's $V$.

\subsection{Kinematic predictability test}
We tested kinematic grounding by training a unified deep neural network to predict the inferred perceived risk evolution $r_{\mathrm{cs}}(t)$ from vehicle kinematics under strict event level holdout. The target $r_{\mathrm{cs}}(t)$ was obtained from kernel constrained inference applied to clipwise ratings and was defined on a fixed time grid within each event. Kinematic features were computed on the same grid and standardised using statistics estimated on the training events only. The full kinematic input feature set and feature availability for PCAD, DRF, and DNN are listed in \textit{SI Appendix}, \textit{Section 4D}.

The predictor was a fully connected feedforward network that maps a 36-dimensional per-time-sample kinematic feature vector to an output $\hat r(t)$, where $r_{\mathrm{cs}}(t)$ denotes the cross-scenario rescaled inferred perceived risk evolution used throughout the paper. The network comprised two hidden layers of width 1,000 with a two-dimensional output head. The predicted perceived risk evolution $\hat r(t)$ was taken as the first output channel, while the second output channel parametrised the predictive variance used by the training objective (\textit{SI Appendix}, \textit{Section 4E}). The model was trained to minimise the Gaussian negative log likelihood loss, computed over time samples from training events only. The stochastic Gradient Descent optimisation algorithm with an initial learning rate of 0.001 is applied in the training process (\textit{SI Appendix}, \textit{Section 4R}). The weight decay dropout was considered for regularisation. All reported predictions, metrics, and figures use the predictive $\hat r(t)$ only, and the variance output is not analysed further.

Hyperparameters were fixed in a separate grid search, in which we varied dropout rate, hidden-layer width, learning rate, and window size used to construct the window summary features. We selected a conservative configuration using event level RMSE on held out development events drawn from the training portion only, prioritising stability over absolute error minimisation (\textit{SI Appendix}, \textit{Section~4G}). Final reported performance is always computed on the held out test events under the event level four fold cross validation protocol used throughout this subsection (\textit{SI Appendix}, \textit{Section 4G}). 

To prevent leakage arising from temporally correlated samples within an event, the event was treated as the smallest unit for splitting and evaluation. We used event-level four-fold cross validation over all 105 events, stratified by scenario. In each fold, all time samples from held-out events were excluded from training and all preprocessing steps were fitted on the training events only. Held-out predictions from the four folds were pooled to obtain performance over the full dataset. Predictive performance was summarised using event-level RMSE and within-event Pearson correlation, where each event contributed one value regardless of its number of time samples. Training was repeated with five independent random initialisations using the same folds; for reporting, we averaged predictions across runs at each time sample and computed all event-level metrics from the averaged predictions. Leave-one-scenario-out robustness and few-shot adaptation (Fig.~\ref{fig:fig4}c -- d) are described in the \textit{SI Appendix}, \textit{Section 4O}.
\subsection{State stratification and SHAP based cue utilisation summaries}\label{sec:methods_shap_state}

We interpreted DNN attributions as cue utilisation by computing Shapley values for held out events in each cross validation fold. At each $10\,\mathrm{Hz}$ time sample, we obtained Shapley values $\phi_i(t)$ for all relevant input features (excluding bookkeeping columns) and analysed attribution magnitude using absolute values. Signed Shapley values were inspected only to assess qualitative directionality in beeswarm plots, whereas all reported cue shares and concentration metrics were computed from absolute Shapley magnitudes. We summarised cue utilisation using feature shares obtained by normalising absolute Shapley values to sum to one across the $K$ active features at each time sample,
\begin{equation}
p_i(t)=\frac{|\phi_i(t)|}{\sum_{j=1}^{K}|\phi_j(t)|+\epsilon},
\end{equation}
where $\epsilon=10^{-12}$ ensures numerical stability.

To test dependence on local dynamics rather than intensity alone, we stratified time samples into nine states defined by three intensity strata of the inferred evolution $r(t)$ and three categories of its local rate of change $\dot r(t)$. State labels were constructed on the $10\,\mathrm{Hz}$ grid using a smoothed $r(t)$ and a centred finite difference estimate of $\dot r(t)$; intensity thresholds and the derivative dead band were computed separately within each scenario from pooled time samples. Features were grouped into conflict cues and stability cues (\textit{SI Appendix, Table~S6}), and cue family shares were computed by summing $p_i(t)$ within each family and averaging within event and within state. State dependence was quantified by paired within event contrasts at matched intensity, using paired one sided Wilcoxon signed rank tests and bootstrap intervals.

Attribution concentration was summarised by the top one share $Top1(t)=\max_i p_i(t)$ and complementary entropy-based metrics reported in the \textit{SI Appendix, Section 6D}. To characterise trends with respect to inferred risk intensity while controlling for varying event coverage across risk levels, we employed a matched event count resampling procedure. This method aggregates metrics into risk bins (using both quantile and equal-width boundaries) by repeatedly sampling a fixed number of contributing events per bin, ensuring that aggregate trends are not biased by events with longer dwell times. Full details on the resampling protocol and sensitivity checks are provided in the \textit{SI Appendix, Section 6E}.

\section*{Acknowledgments}
This research is supported by the SHAPE-IT project funded by the European Union’s Horizon 2020 research and innovation programme under the Marie Skłodowska-Curie grant agreement 860410. 

\section*{Data availability}
The data supporting the findings of this study are available on 4TU.ResearchData at \url{https://doi.org/10.4121/242d9474-e522-4518-8917-8f284fc8a7a8}.

\section*{Code availability}
The code used in this study is available on 4TU.ResearchData at \url{https://doi.org/10.4121/242d9474-e522-4518-8917-8f284fc8a7a8}.

\section*{Additional information}
Supplementary Information is provided with this paper.
% \bibliography{sn-bibliography}
% \bibliographystyle{unsrt}

% ==========================================
% Devider: Supplementary Information
% ==========================================
\clearpage
\vspace*{2em}
\begin{center}
    \textbf{\Large Supporting Information}
\end{center}
\vspace{2em}

\setcounter{section}{0}
\setcounter{subsection}{0}
\setcounter{subsubsection}{0}
\setcounter{figure}{0}
\setcounter{table}{0}
\setcounter{equation}{0}

\renewcommand{\thesection}{\arabic{section}} 
\renewcommand{\thesubsection}{\Alph{subsection}} 
\renewcommand{\thesubsubsection}{\thesubsection.\arabic{subsubsection}}

\renewcommand{\thefigure}{S\arabic{figure}} 
\renewcommand{\thetable}{S\arabic{table}}  
\renewcommand{\theequation}{S\arabic{equation}} 

\addtocontents{toc}{\protect\setcounter{tocdepth}{3}}
\renewcommand{\contentsname}{\large Table of Contents}
\tableofcontents
\clearpage
% ==========================================
% \SItext
\section{Scenario and online experiment design}
%%%%%%%%%%%%%%% Table: Scenario parameters
\subsection{Controlled parameters for scenario design}
\begin{table}[h]
    \centering
    \small
    \renewcommand{\arraystretch}{1.4}
    \begin{tabularx}{\textwidth}{>{\raggedright\arraybackslash}m{3.5cm} X >{\centering\arraybackslash}m{2cm}}
        \toprule
        \textbf{Scenarios} & \textbf{Varied parameters} & \textbf{Events} \\
        \midrule
        \multirow{3}{=}{MB: Merging with hard braking}
        & Initial merging distance (m): 5, 15, 25 & \multirow{3}{*}{27} \\
        & Desired cruising speed (km/h): 80, 100, 120 & \\
        & Braking intensity (m/s$^2$): $-2, -5, -8$ & \\
        \hline
        
        \multirow{3}{=}{HB: Hard braking}
        & Initial distance (m): 5, 15, 25 & \multirow{3}{*}{27} \\
        & Desired cruising speed (km/h): 80, 100, 120 & \\
        & Braking intensity (m/s$^2$): $-2, -5, -8$ & \\
        \hline

        \multirow{3}{=}{LC: Reacting to merging vehicles}
        & Initial merging distance (m): 5, 15 & \multirow{3}{*}{24} \\
        & Lateral: 1 or 3 m/s, fragmented, aborted & \\
        & ACC categories: cautious, mild, aggressive & \\
        \hline

        \multirow{3}{=}{SVM: Subject AV merging}
        & Initial distance (m): 5, 15, 25 & \multirow{3}{*}{27} \\
        & Desired cruising speed (km/h): 80, 100, 120 & \\
        & Braking intensity (\SI{}{m/s^2}): $-2, -5, -8$ & \\
        \bottomrule
    \end{tabularx}
    \caption{Designed scenarios for the perceived risk experiments.}
    \label{tab:designed_scenarios}
\end{table}
\subsection{Online questionnaire materials and instructions}
The online questionnaire can be accessed at \\\href{https://tudelft.fra1.qualtrics.com/jfe/form/SV_bQ2JkNfOCMp2FLM}{\text{https://tudelft.fra1.qualtrics.com/jfe/form/SV\_bQ2JkNfOCMp2FLM}}.
\subsection{Sampling, exclusions, and response quality control}\label{si_1c_sampling}
Self reported perceived risk ratings were collected in a large scale online study conducted from 11 July 2023 to 19 September 2023 using the Prolific platform \cite{Prolific2023}. Participation was restricted to countries with right hand traffic to align with the visual conventions of the motorway stimuli. A total of 2,341 participants started the study. Each participant rated 16 events in total, with four events per interaction type, presented as sequential six second clips in their original order. For MB, HB, and SVM events, each event comprised five clips, whereas LC events comprised six clips. Ratings were provided after each clip on an integer slider from 0 to 10.

We applied predefined quality control to exclude incomplete sessions and implausibly fast completion consistent with not viewing the clips. Because each participant was required to view the full set of assigned clips, the minimum feasible viewing time was 504 seconds, excluding any response time. We therefore excluded participants who did not complete the study or whose completion time was below 504 seconds. This removed 177 participants and yielded a retained sample of $N=2{,}164$.

We next applied event specific response pattern screening to identify inattentive sequences at the event level. Within each event, a participant provided a short rating sequence across clips, five ratings for MB, HB, and SVM and six ratings for LC. For each event, we computed the Pearson correlation between an individual participant sequence and the event mean sequence across participants. If this correlation was below $r<0.3$, we removed that participant sequence for that event only, while retaining the same participant data for other events when their sequences passed screening. This procedure targets sequences that are weakly aligned with the within event temporal pattern and is not conditioned on rating magnitude. After screening, 141,628 clip level ratings remained for analysis.

\begin{figure}[t!]
    \centering
    \captionsetup[subfigure]{labelformat=simple, labelsep=period, font=bf}
     \begin{subfigure}{0.4\textwidth}
     \centering
    \includegraphics[width=\textwidth]{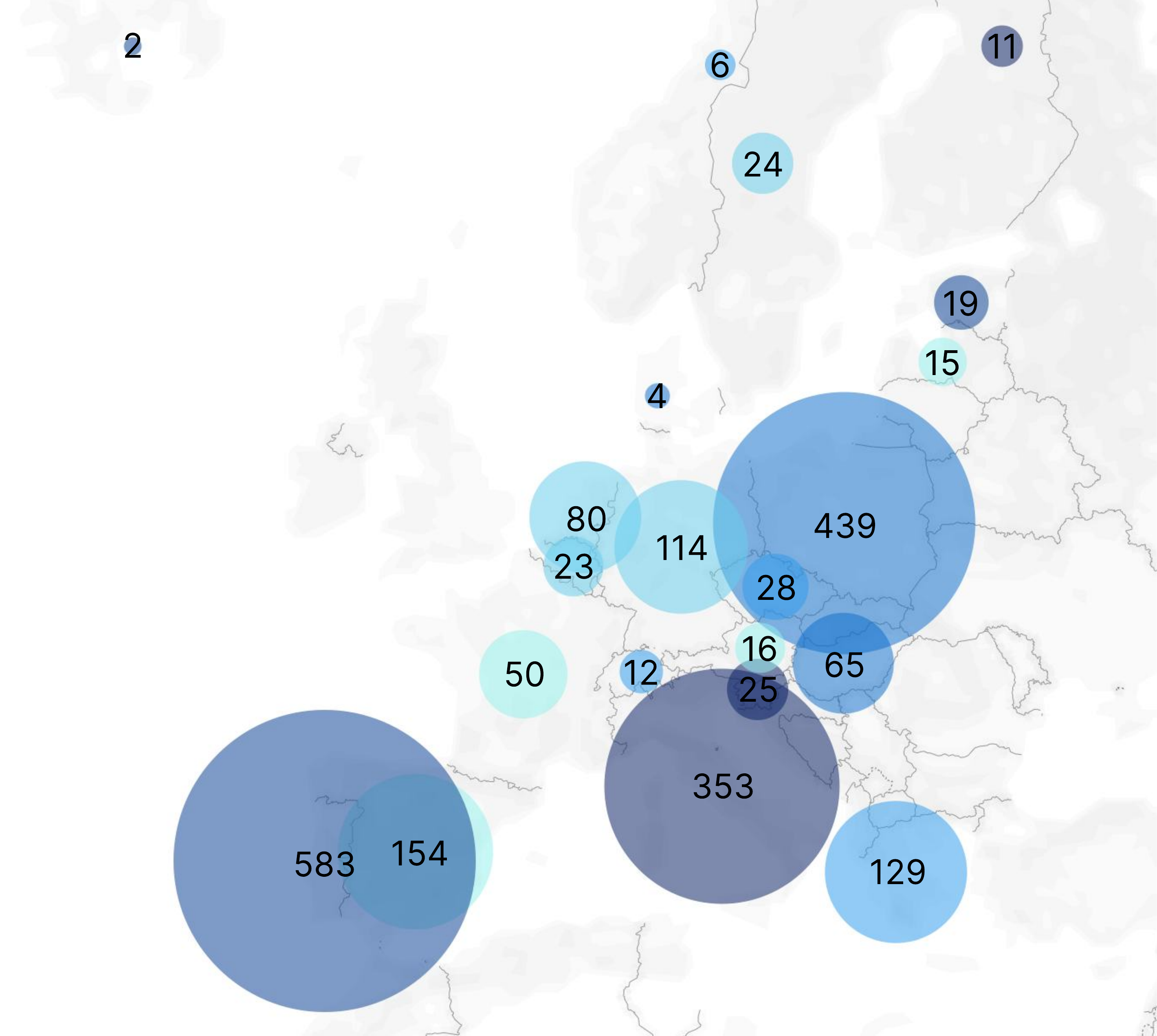}
    \caption{}
    \label{fig_ext:location_map}
    \end{subfigure} 
    \begin{subfigure}{0.4\textwidth}
        \centering
        \includegraphics[width=0.5\textwidth]{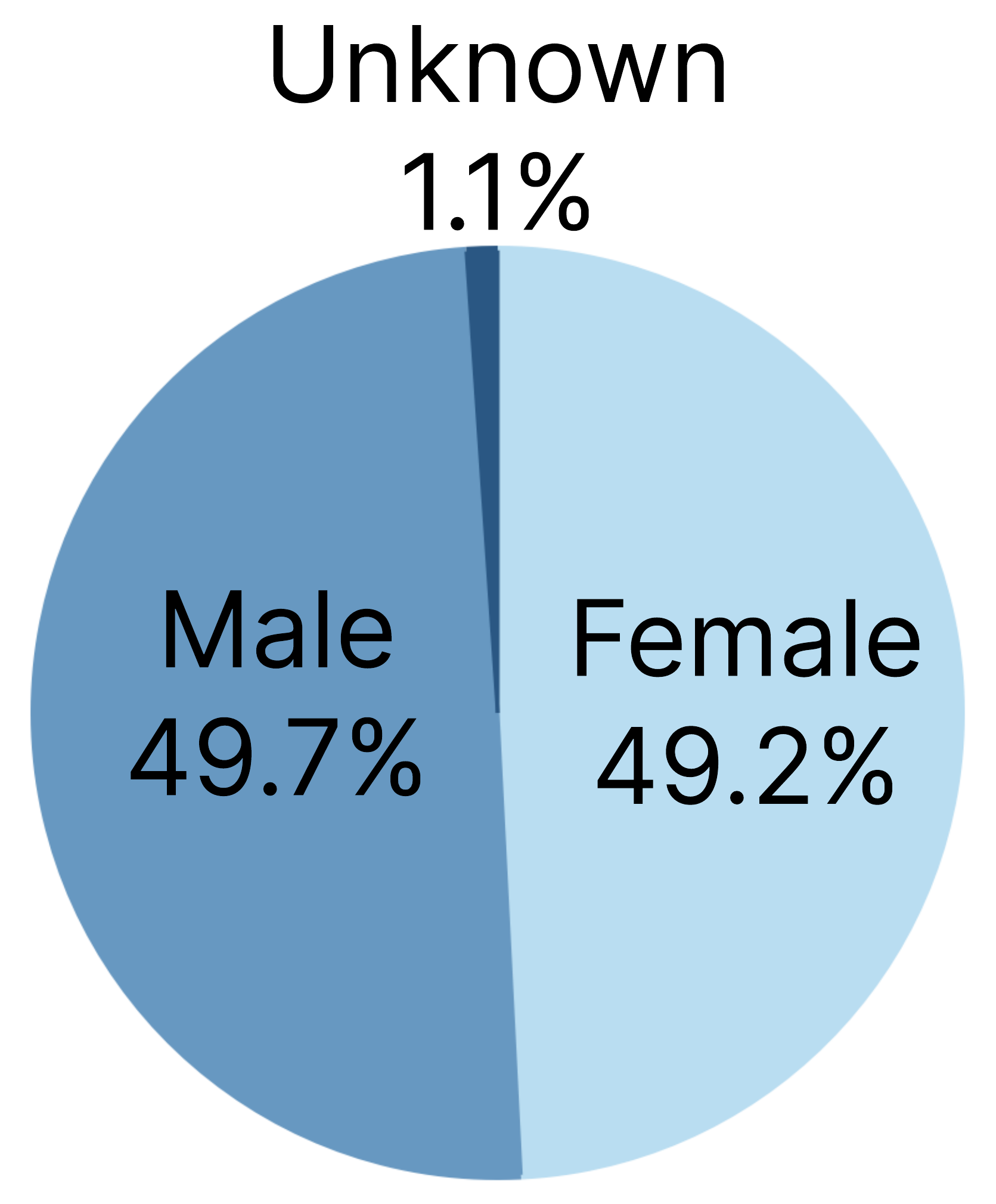}
    \caption{}
    \label{fig_ext: gender}
    \end{subfigure}
    \begin{subfigure}{0.35\textwidth}
        \includegraphics[width=\textwidth]{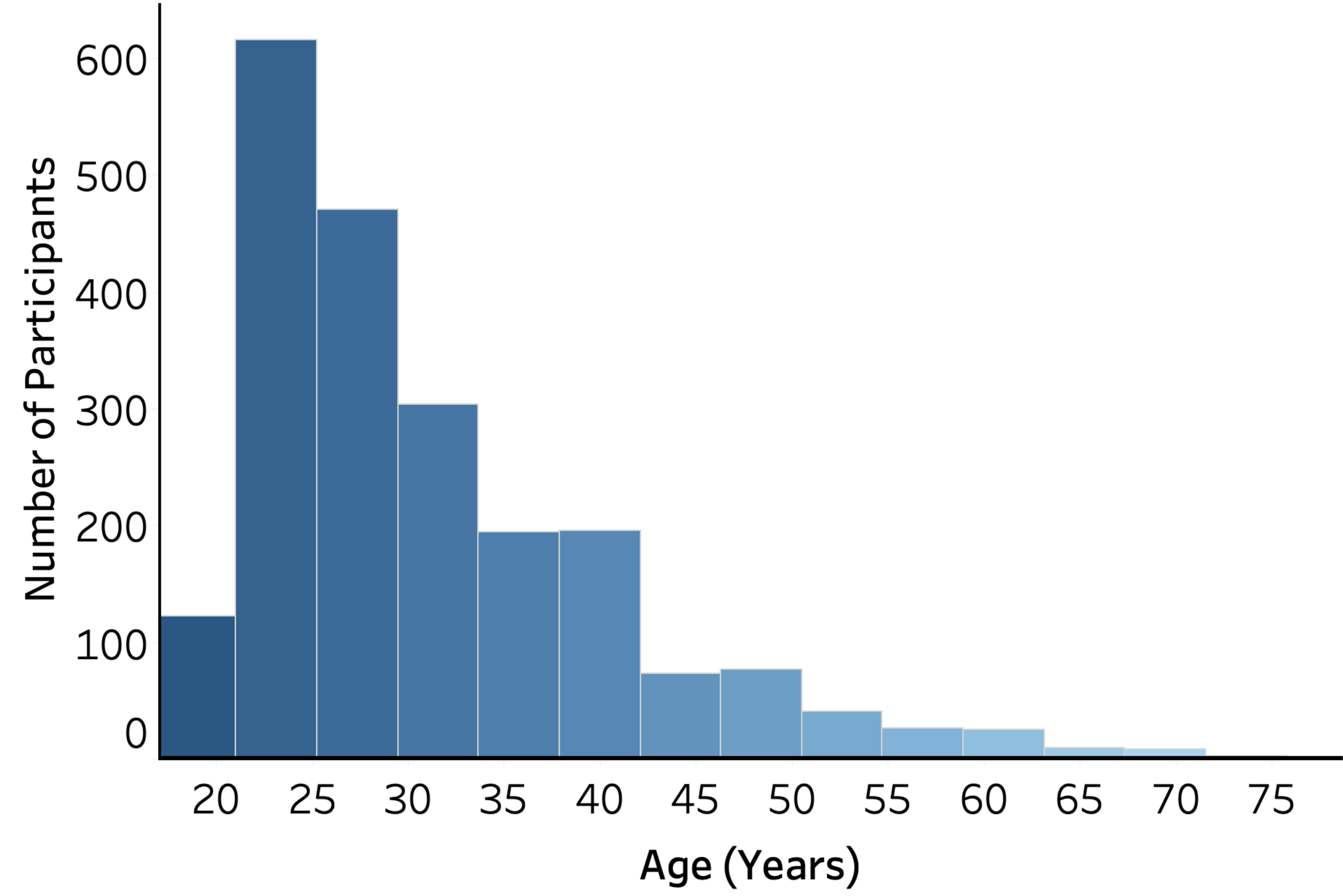}
    \caption{}
    \label{fig_ext: age}
    \end{subfigure}
    \begin{subfigure}{0.35\textwidth}
        \includegraphics[width=\textwidth]{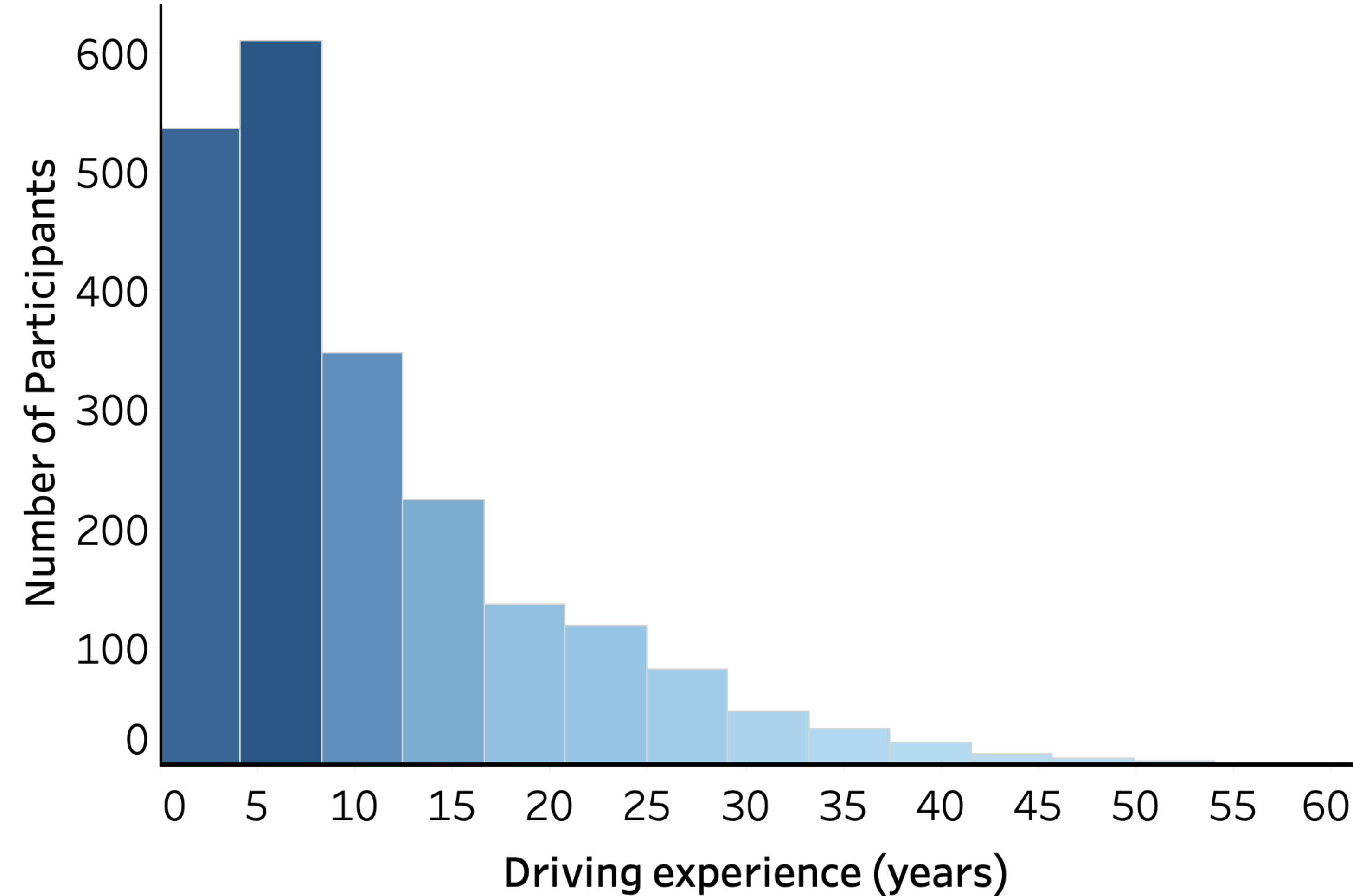}
        \caption{}
        \label{fig_ext: driving experience}
    \end{subfigure}
  \captionof{figure}{Demographic information of the participants. \textbf{a}, geographical distribution by country with counts; \textbf{b}, gender distribution; \textbf{c}, age distribution; \textbf{d}, driving experience distribution.}
\end{figure}

The retained sample was geographically diverse and primarily European (Fig.~\ref{fig_ext:location_map}). Gender distribution was 49.7\% male, 49.2\% female, and 1.1\% preferred not to specify (Fig.~\ref{fig_ext: gender}). Participants ranged in age from 18 to 73 years (mean 31.2 years, standard deviation 9.5 years; Fig.~\ref{fig_ext: age}). All participants held a valid driving licence, with licence duration ranging from 1 to 55 years (mean 11.0 years, standard deviation 9.0 years; Fig.~\ref{fig_ext: driving experience}).
\subsection{Manipulation check of controlled parameters}\label{si_1d_manipulation}
We verified that the scripted grids of controlled parameters produced systematic variation in perceived risk ratings within each interaction type. Ratings were arranged in long format, with one row per clip rating and columns for participant identifier, clip number, controlled parameter values, and the perceived risk rating. For each interaction type, we fitted a mixed effects regression model with participant identifier as a random intercept to account for between participant differences in baseline rating use, while clip number and the controlled parameters were included as fixed effects. Analyses were conducted in IBM SPSS Statistics 29.

\begin{figure}[H]
    \centering
    %--------------
\begin{subfigure}{\linewidth}
    \centering \includegraphics[width=0.3\linewidth]{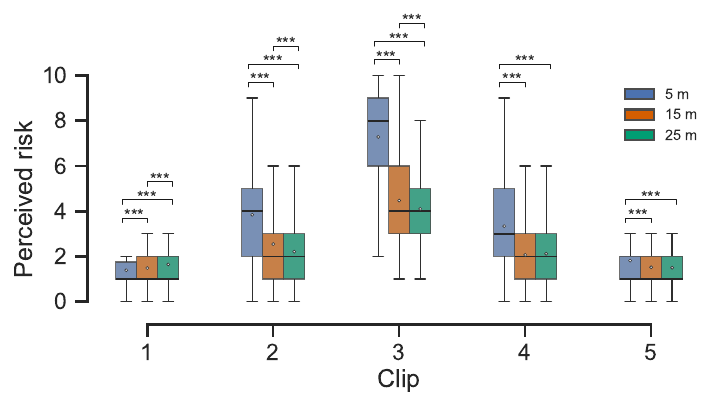}
   %---------------------- 
  \includegraphics[width=0.3\linewidth]{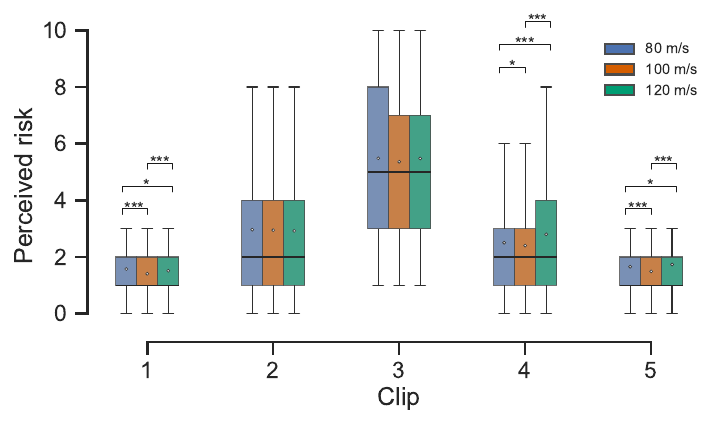}
\includegraphics[width=0.3\linewidth]{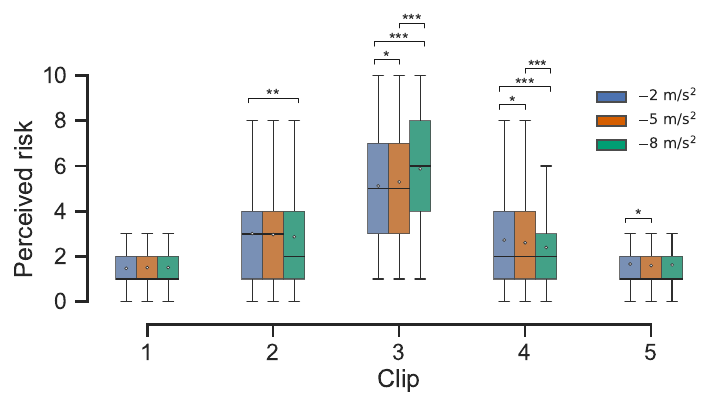}
\caption{Influence of controlled parameters per clip in MB}
\end{subfigure}
% \end{figure}
% \begin{figure}\ContinuedFloat
\begin{subfigure}{\linewidth}
    \centering \includegraphics[width=0.3\linewidth]{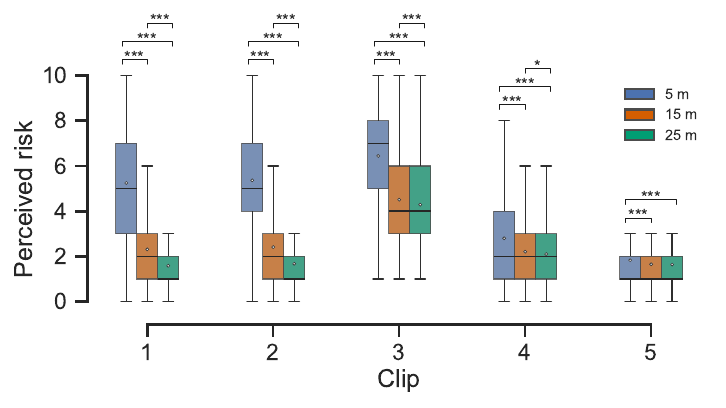}
   %---------------------- 
  \includegraphics[width=0.3\linewidth]{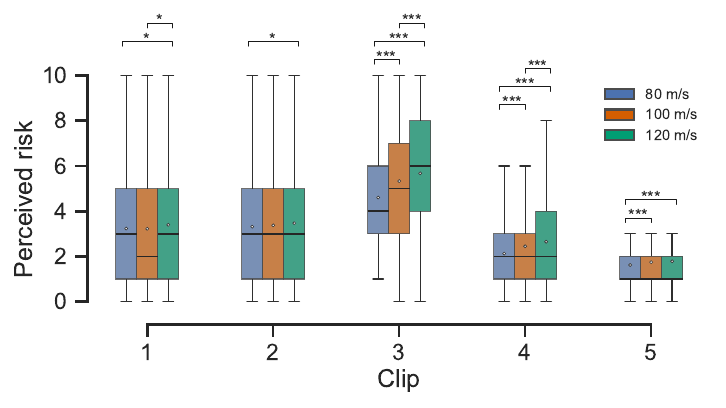}
\includegraphics[width=0.3\linewidth]{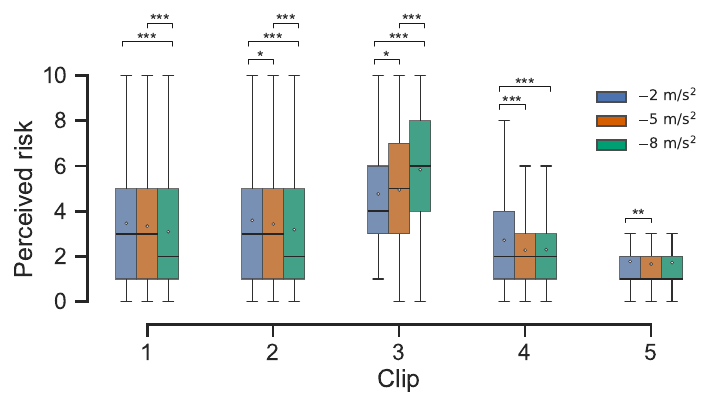}
\caption{Influence of controlled parameters per clip in HB}
\end{subfigure}
\begin{subfigure}{\linewidth}
    \centering
    \includegraphics[width=0.3\linewidth]{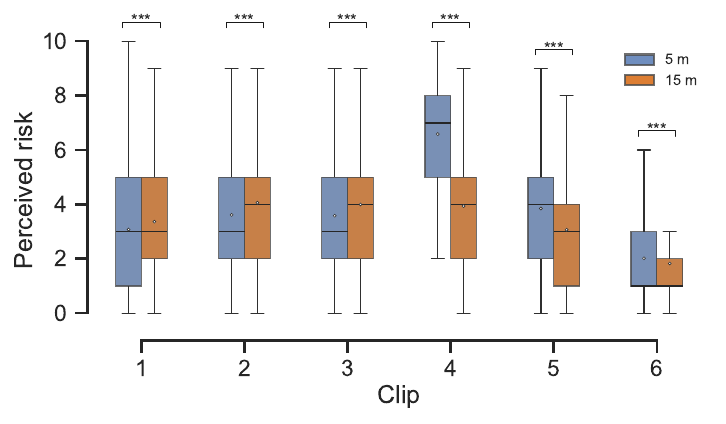}
    \includegraphics[width=0.33\linewidth]{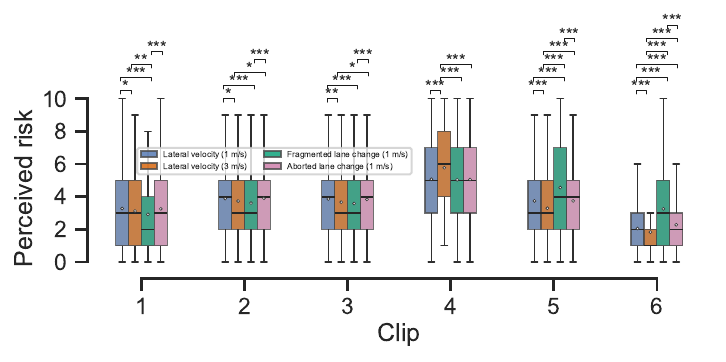}
    \includegraphics[width=0.3\linewidth]{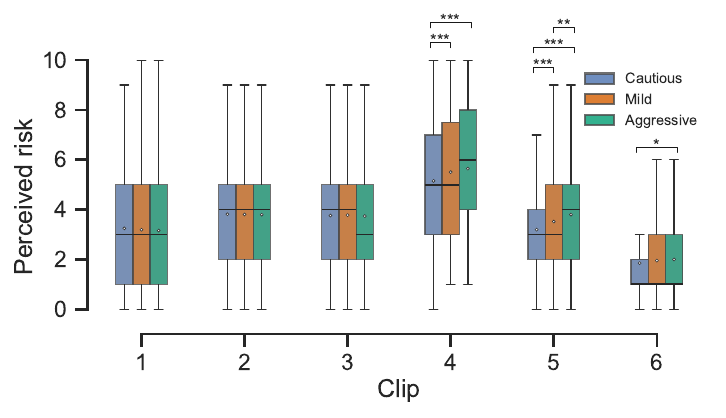}
\caption{Influence of controlled parameters per clip in LC}
\end{subfigure}
\begin{subfigure}{\linewidth}
    \centering \includegraphics[width=0.3\linewidth]{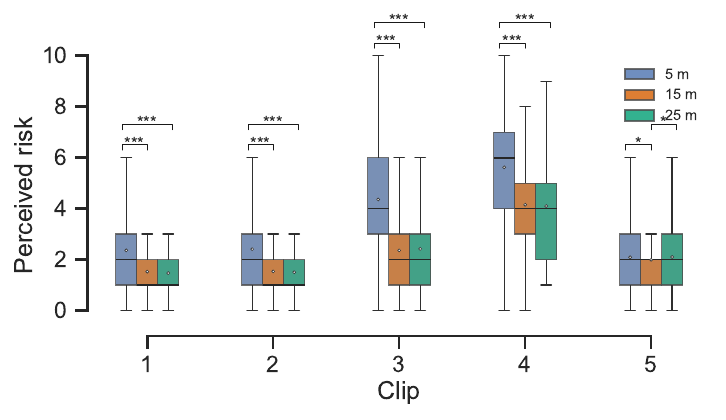}
   %---------------------- 
  \includegraphics[width=0.3\linewidth]{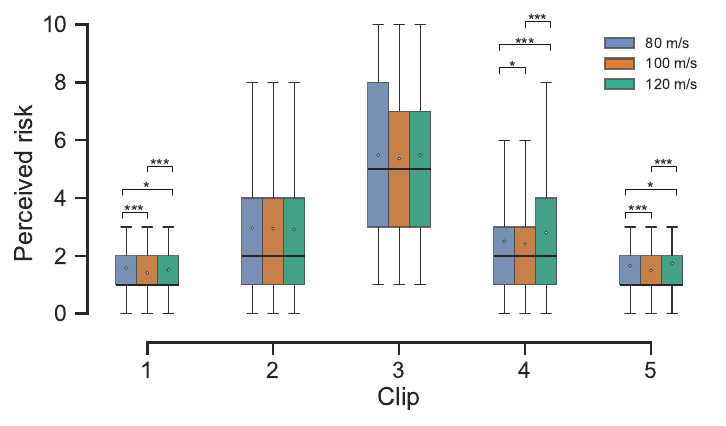}
\includegraphics[width=0.3\linewidth]{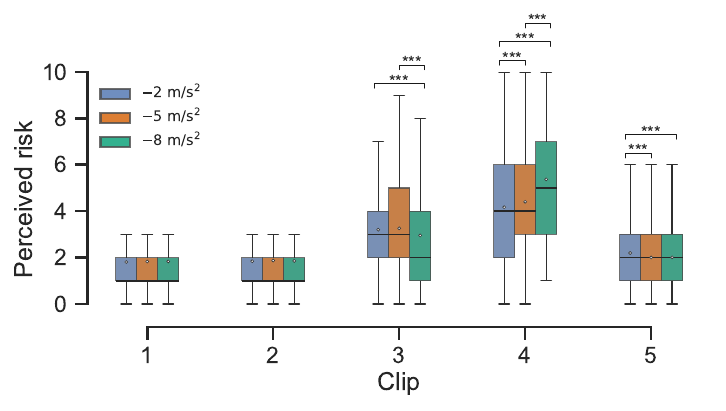}
\caption{Influence of controlled parameters per clip in SVM}
\end{subfigure}
\caption{Influence of controlled parameters per clip across four scenarios}
\label{fig:influence of controlled parameters aross four scenarios}
\end{figure}

Across interaction types, all controlled parameters were significantly associated with perceived risk ratings (all $p<0.001$; Table~\ref{tab:ParametersInfluence}). Clip number also showed a strong effect in each interaction type, consistent with the structured progression induced by the scripted manoeuvre timing. To illustrate how each parameter relates to perceived risk within clips, we additionally report per clip contrasts for each controlled parameter within each interaction type (Fig.~\ref{fig:influence of controlled parameters aross four scenarios}). Together, these results confirm that the designed parameter grids generated reliable perceptual variation across the stimulus set and support the intended coverage from lower to higher perceived risk encounters.

%%%%%%%%% Table: Statistical test of scenario parameters
\makeatletter
\setlength{\@dblfptop}{0pt}    
\setlength{\@fptop}{0pt}   
\makeatother
\begin{table}[H]
    \footnotesize 
    \renewcommand{\arraystretch}{1.3} 
    \setlength{\tabcolsep}{4pt}      
    
    \begin{tabularx}{\textwidth}{L{3cm} c L{5.5cm} c c c}
        \toprule
        \textbf{Scenarios} & \textbf{Sample size} & \textbf{Controlled parameters} & \textbf{$F$} & \textbf{$p$-value} & \textbf{$\eta^2$} \\
        \midrule
        
        % --- MB ---
        \multirow{4}{=}{MB: Merging with hard braking} & \multirow{4}{*}{36,525} 
        & Clip number & 11,475.05 & $<0.001$ & 0.571 \\
        & & Initial merging distance (m) & 3,686.39 & $<0.001$ & 0.169 \\
        & & Desired cruising speed (km/h) & 68.13 & $<0.001$ & 0.004 \\
        & & Braking intensity (\SI{}{m/s^2}) & 29.24 & $<0.001$ & 0.002 \\
        \midrule
        
        % --- HB ---
        \multirow{4}{=}{HB: Hard braking} & \multirow{4}{*}{33,355} 
        & Clip number & 6,141.70 & $<0.001$ & 0.439 \\
        & & Car following distance (m) & 7,753.69 & $<0.001$ & 0.321 \\
        & & Desired cruising speed (km/h) & 388.39 & $<0.001$ & 0.023 \\
        & & Braking intensity (\SI{}{m/s^2}) & 93.56 & $<0.001$ & 0.006 \\
        \midrule
        
        % --- LC ---
        \multirow{4}{=}{LC: Lateral control reaction} & \multirow{4}{*}{39,018} 
        & Clip number & 3,011.40 & $<0.001$ & 0.289 \\
        & & Lane-changing distance (m) & 2,083.87 & $<0.001$ & 0.052 \\
        & & Lateral categories & 63.49 & $<0.001$ & 0.005 \\
        & & Driving style & 133.46 & $<0.001$ & 0.007 \\
        \midrule
        
        % --- SVM ---
        \multirow{4}{=}{SVM: Subject AV merging} & \multirow{4}{*}{32,720} 
        & Clip number & 6,237.33 & $<0.001$ & 0.448 \\
        & & Merging distance to lead (m) & 2,539.26 & $<0.001$ & 0.136 \\
        & & Desired cruising speed (km/h) & 300.18 & $<0.001$ & 0.018 \\
        & & Braking intensity (\SI{}{m/s^2}) & 87.66 & $<0.001$ & 0.005 \\
        \bottomrule
    \end{tabularx}
    \caption{All controlled parameters significantly affect perceived risk.}
    \label{tab:ParametersInfluence}
\end{table}

\section{Simulator dataset, kernel prior and validation protocols}\label{si_sec2}
\subsection{Simulator dataset and kernel prior specification}\label{si_2a_kernel_prior}
\subsubsection{Dataset summary}
To derive physiologically plausible response kernels, we utilised continuous perceived risk ratings from an independent high-fidelity driving simulator study, detailed in Ref.~\cite{He2022_SI}. In brief, $N=25$ participants (mean age $40.6 \pm 16.3$ years) monitored an SAE Level 2 automated vehicle on a simulated motorway. The study included 20 safety-critical events, specifically hard braking scenarios with and without cut-in manoeuvres (decelerations ranging from $-2$ to $-8\,\mathrm{m/s^2}$), which closely match the kinematic dynamics of the online video stimuli. Crucially for our kernel modelling, participants provided time-continuous perceived risk ratings using a force-sensitive handset (pressure sensor) sampled at $60\,\mathrm{Hz}$. They were instructed to apply pressure proportional to their varying sense of risk during the event. These continuous traces captured the temporal evolution of risk perception, specifically the onset delays, rise times, and recovery tails, which were then extracted to parameterise the Gamma-distributed kernels ($\alpha, \beta$) used in our inverse inference model.
\subsubsection{Gamma family kernels and simulator derived shape priors}

\begin{figure}[H]
    \centering
    \includegraphics[width=0.4\linewidth]{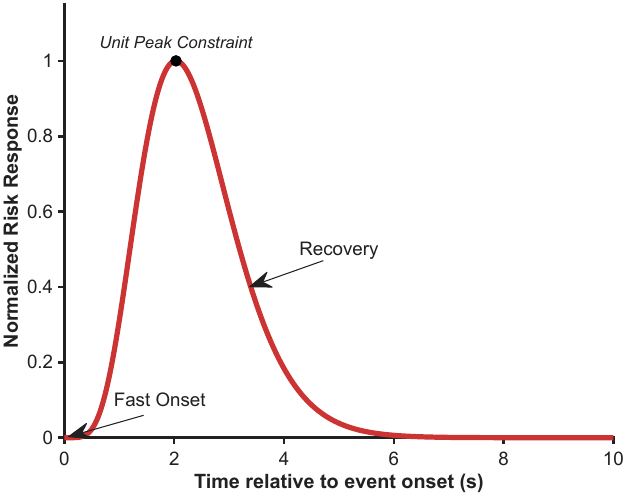}
    \caption{The standard psychophysical response kernel. The inferred temporal perceived risk evolution is modelled as a superposition of Gamma-distributed response kernels ($\phi(t)$), constrained by unit peak amplitude. The figure illustrates the characteristic asymmetry derived from simulator priors \cite{He2022_SI}: the response is anchored to the scripted manoeuvre timing marker ($t=0$), exhibiting a relative rapid reaction (rapid rise) followed by a slow recovery tail (hysteresis). This shape constraint prevents the inference model from producing physically implausible fluctuations (e.g., instant recovery) and ensures psychophysical validity.}
    \label{fig:gamma}
\end{figure}

The core kernel family is a gamma-shaped response with unit peak normalisation, as illustrated in Fig. \ref{fig:gamma}. This profile captures the temporal asymmetry of perceived risk: a relative rapid onset in response to threats and a gradual decay during stability recovery. This functional form is widely established in cognitive science for modelling the temporal dynamics of autonomic and neural arousal, such as skin conductance responses \cite{bach2010time_SI} and the cortical hemodynamic response \cite{boynton1996linear_SI}, which share the fundamental psychophysiological characteristic of a fast stimulus-driven rise and a slow return to baseline.
\begin{equation}
g(t;\alpha,\beta) = \frac{t^{\alpha-1}\exp(-t/\beta)}{\max_{t\ge 0} \big(t^{\alpha-1}\exp(-t/\beta)\big)}\ \ \ \ \text{for}\ t\ge 0,\qquad g(t)=0\ \text{for}\ t<0.
\end{equation}
For components requiring distinct rise and recovery time scales, we use an asymmetric profile constructed from separate rise and fall gamma shapes, blended smoothly across a transition time window,
\begin{equation}
k(t) = s(t)\,g(t;\alpha_{\mathrm{rise}},\beta_{\mathrm{rise}}) + \big(1-s(t)\big)\,g(t;\alpha_{\mathrm{fall}},\beta_{\mathrm{fall}}),
\end{equation}
where $s(t)$ is a logistic transition function controlling the blend between rise dominated and recovery dominated dynamics. The parameters $(\alpha,\beta)$ for vigilance, lateral control, and braking components are fixed by priors estimated from continuous handset based ratings in the simulator study \cite{He2022_SI}.

\subsection{Kernel constrained inverse inference specification}\label{si:2b}
\subsubsection{Inferred evolution representation and anchors}\label{si:2b_representation}

We inferred a risk evolution $r(t)$ on a fixed temporal grid $t \in \{0,\Delta t,2\Delta t,\dots,T\}$ with $\Delta t = 0.1\ \mathrm{s}$ for each participant $p$ and event $e$. The latent evolution is represented as a superposition of anchored response kernels plus a baseline and recovery term,
\begin{equation}
r_{p,e}(t) = b_{p,e}(t;\theta_b) + \mathcal{F}\!\Big(\{w_{p,e,i}\,k_i(t-\tau_{e,i};\theta_i)\}_{i=1}^{M_e}\Big),
\end{equation}
where $\tau_{e,i}$ are scripted manoeuvre timing markers defined by the scripted event timeline, $w_{p,e,i}\ge 0$ are inferred kernel weights, and $\theta_i$ are fixed kernel shape parameters obtained from the simulator calibration.

\subsubsection{Fusion logic for multi cue events}

In events with multiple contemporaneous components, simple addition can yield non physical peaks when kernels overlap. We therefore fuse component contributions using a differentiable SoftMax maximum operator,
\begin{equation}
\mathcal{F}(\{x_i(t)\}_{i=1}^{M}) = \frac{1}{\kappa}\log\Big(\sum_{i=1}^{M}\exp\big(\kappa x_i(t)\big)\Big),
\end{equation}
with $\kappa > 0$ controlling the sharpness of the maximum. This operator approaches $\max_i x_i(t)$ as $\kappa$ increases, while remaining differentiable and preventing unbounded additive amplification.

The lateral control scenario additionally supports an envelope preserving ratchet logic that prevents unphysical dips when transitioning from lateral conflict to braking conflict. In this mode, the fused signal is constrained to remain above a locally defined lower envelope during specified transition windows, implemented as a soft penalty in the objective.

\subsubsection{Observation operator and optimisation}
The online protocol provides one rating $y_{p,e,k}$ per clip interval $[t_k,t_{k+1}]$, corresponding to the perceived risk at the most dangerous moment of that clip. We model this as a constraint on the clip maximum of the latent evolution,
\begin{equation}
y_{p,e,k} \approx \max_{t \in [t_k,t_{k+1}]} r_{p,e}(t).
\end{equation}
Let $\hat y_{p,e,k} = \max_{t\in[t_k,t_{k+1}]} r_{p,e}(t)$ denote the reconstructed clip maximum. Parameters are inferred by minimising a hinge squared reconstruction loss with asymmetric tolerances,
\begin{equation}
J = \sum_{k} \omega_k \, \phi\!\big(\hat y_{p,e,k} - y_{p,e,k}\big)^2 \;+\; \sum_{m}\lambda_m \, \Psi_m(r_{p,e}).
\end{equation}
where $\omega_k$ are clip weights and the error term $\phi(\cdot)$ penalises deviations only outside the tolerance bands $[\delta_k^{-},\delta_k^{+}]$:
\begin{equation}
\phi(\Delta y) = \begin{cases} 
\Delta y - \delta_k^{+} & \text{if } \Delta y > \delta_k^{+} \\
\Delta y - \delta_k^{-} & \text{if } \Delta y < \delta_k^{-} \\
0 & \text{otherwise}
\end{cases}
\end{equation}
This asymmetric hinge loss allows for minor temporal misalignment or measurement noise without penalty. The regularisation terms $\Psi_m(\cdot)$ include penalties enforcing realistic recovery rates via a decay prior, preventing local notches, and discouraging implausible dips in defined transition windows. Boundary constraints at event start and end are implemented as additional tolerance terms on $r_{p,e}(0)$ and $r_{p,e}(T)$.

The optimisation variables include non negative kernel weights $\{w_{p,e,i}\}$ and baseline and recovery parameters $\theta_b$ such as a decay rate. Kernel shape parameters $(\alpha,\beta)$, fusion sharpness $\kappa$, and ratio priors linking component magnitudes are fixed by simulator calibration and held constant across participants and events.
\subsubsection{Scenario specific implementations}

\begin{center}
    \small
    \renewcommand{\arraystretch}{1.5}  
    \setlength{\tabcolsep}{6pt}   
    \begin{tabular}{c >{\raggedright\arraybackslash\hyphenpenalty=10000}m{2.2cm} >{\raggedright\arraybackslash\hyphenpenalty=10000}m{2.4cm} l l >{\raggedright\arraybackslash\hyphenpenalty=10000}m{3.1cm}}
        \toprule
        \textbf{Scenario} & \textbf{Anchors} $\tau$ & \textbf{Components} & \textbf{Fusion} $\mathcal{F}$ & \textbf{Observation} & \textbf{Inferred parameters} \\
        \midrule
        HB & brake onset & braking kernel, sharp or wide & Additive & clip maximum & weight, decay rate, kernel choice \\
        
        MB & vigilance, lane change, brake onset & vigilance, lateral, braking & SoftMax max & clip maximum & weights, braking mixture, decay terms \\
        
        LC & lateral, brake, transition markers & lateral, braking, vigilance & SoftMax & clip maximum & weights, decay terms, nuisance terms \\
        
        SVM & vigilance, lane change, brake onset & vigilance, lateral, braking & SoftMax max & clip maximum & weights, braking mixture, decay terms \\
        \bottomrule
    \end{tabular}
    \captionof{table}{Scenario specific anchors, components, fusion logic, and inferred parameters.}
    \label{tab:scenario_anchors}
\end{center}
\subsubsection{Uncertainty calibration via robust residual mapping in $(r,g)$ state space}
We calibrated uncertainty for the inferred risk evolution by learning a state dependent error scale map $\Sigma(r,g)$ from simulator training events. The calibration emulates the online setting by reconstructing each training event from discrete constraints and measuring how closely this reconstruction reproduces the simulator ground truth. All discretisation and smoothing settings in this section are fixed a priori for numerical stability and robust estimation. They are not tuned to optimise coverage on the held out events.

\paragraph{Residual samples.}
For each simulator training event, we reconstructed $r_{\mathrm{rec}}(t)$ from discrete constraints and computed residuals
\begin{equation}
e(t)=r_{\mathrm{rec}}(t)-r_{\mathrm{gt}}(t),
\end{equation}
where $r_{\mathrm{gt}}(t)$ denotes the simulator ground truth group mean on a common time grid with sampling interval $\Delta t$.

\paragraph{State variables and light smoothing.}
The uncertainty map conditions on both the inferred level and local dynamics of the reconstructed signal, which requires estimating derivatives. Because numerical derivatives amplify high frequency noise, we applied the same short moving average to both $r_{\mathrm{rec}}(t)$ and $e(t)$ using a window length
\begin{equation}
w=\max\!\left(3,\left\lfloor \frac{0.6}{\Delta t}\right\rceil\right),
\end{equation}
yielding $r_s(t)$ and $e_s(t)$. The $0.6\,\mathrm{s}$ window corresponds to $6$ samples on the $10\,\mathrm{Hz}$ grid used in the simulator dataset and is short relative to the rise and recovery time scales captured by the kernel shapes. The lower bound of $3$ samples avoids unstable derivative estimates when the sampling rate or missingness would otherwise yield very small windows. We then defined an effective derivative magnitude
\begin{equation}
g(t)=\sqrt{\left|\frac{\mathrm{d}r_s}{\mathrm{d}t}\right|^2+\left(\kappa\,\left|\frac{\mathrm{d}^2 r_s}{\mathrm{d}t^2}\right|\right)^2},
\end{equation}
where $\kappa$ controls sensitivity to sharp curvature (in our implementation $\kappa=0.8$). This construction makes $g(t)$ responsive both to sustained slopes and to sharp changes that occur at narrow peaks, while remaining stable under light smoothing. Residual samples were pooled as triplets $\{r_s(t),g(t),e_s(t)\}$ across all training events.

\paragraph{Discretised grid.}
We discretised $r$ using fixed edges
\begin{equation}
r_{\mathrm{edges}}=\{0,0.5,1.0,\ldots,10\},
\end{equation}
which matches the online rating scale and uses half unit steps to balance two competing requirements. The grid must be fine enough to capture heteroscedasticity across low, moderate, and high inferred risk, yet coarse enough that each cell receives sufficient residual samples for robust scale estimation. We discretised $g$ using edges spanning the empirical range of derivative magnitudes. Let $g_{0.95}$ be the $95$-th percentile of pooled $g(t)$; we set
\begin{equation}
g_{\mathrm{edges}}=\mathrm{linspace}(0,g_{0.95},12),
\end{equation}
which yields $11$ bins in $g$. The percentile-based upper bound limits sensitivity to rare extreme derivatives, preventing a small number of outliers from expanding the range and leaving most bins sparsely populated. The choice of $11$ bins yields a grid of moderate size, roughly $20\times 11$ cells, which is sufficient to capture systematic dependence of residual scale on both level and dynamics while remaining well populated for the available simulator training data. For each bin pair $(i,j)$, we collected residuals
\begin{equation}
\mathcal{E}_{ij}=\{\,e_s(t)\,:\, r_s(t)\in[r_i,r_{i+1}),\ g(t)\in[g_j,g_{j+1})\,\}.
\end{equation}
Bins with fewer than $n_{\min}=30$ samples were left undefined at this stage. This threshold reduces instability of the within bin scale estimate in sparsely populated cells, and the subsequent completion step propagates information from neighbouring well populated bins.

\paragraph{Robust scale estimation using MAD.}
For each populated bin, we estimated the local uncertainty scale by the median absolute deviation mapped to a Gaussian equivalent standard deviation,
\begin{equation}
\Sigma_{ij}=1.4826\ \mathrm{median}\!\left(\left|x-\mathrm{median}(x)\right|\right),\qquad x\in\mathcal{E}_{ij}.
\end{equation}
The constant $1.4826$ is the standard conversion that makes the MAD consistent with the standard deviation under a normal residual model. This estimator replaces the standard deviation and reduces sensitivity to heavy tails and occasional reconstruction failures.

\paragraph{Completion and smoothing.}
Undefined bins were filled by nearest neighbour propagation along both grid axes, producing a complete map $\Sigma(r,g)$. To enforce continuity over the discretised state space and to avoid non-physical grid artefacts in subsequent interpolation, we applied two-dimensional Gaussian smoothing to the grid with standard deviation $[1.2,1.2]$ bins in the $r$ and $g$ directions. This choice smooths over immediate neighbouring cells without erasing broad state-dependent structure, and yields an uncertainty field that varies smoothly as the reconstructed signal moves through the $(r,g)$ plane. Finally, we imposed a floor $\Sigma_{\min}=0.05$ to avoid degenerate bands,
\begin{equation}
\Sigma(r,g)\leftarrow\max\{\Sigma(r,g),\Sigma_{\min}\}.
\end{equation}
We also stored a conservative scalar fallback
\begin{equation}
\Sigma_{\mathrm{median}}=\mathrm{median}\big(\Sigma(r,g)\big),
\end{equation}
used if the state space map is unavailable.

\paragraph{Applying the calibrated uncertainty during inference.}
For any reconstructed evolution $r_{\mathrm{rec}}(t)$, we computed $r_s(t)$ and $g(t)$ as above, then obtained pointwise uncertainty by interpolating the learned grid. Let $r_{\mathrm{cent}}$ and $g_{\mathrm{cent}}$ denote bin centres. We used bilinear interpolation on $\Sigma(r_{\mathrm{cent}},g_{\mathrm{cent}})$, with values clipped to the grid domain. The resulting $\sigma(t)$ was lightly smoothed by the same moving average window and floored at $\Sigma_{\min}$. The $95\%$ uncertainty band was then
\begin{equation}
r_{\mathrm{rec}}(t)\ \pm\ 1.96\,\sigma(t),
\end{equation}
with bounds truncated to the rating range $[0,10]$.

\subsubsection{Selection of representative events for visualisation}

To ensure the events displayed in Fig.2 and Fig. 4 illustrate characteristic scenario dynamics rather than outlier behaviour, we employed a quantitative selection criterion. For each scenario, we computed a composite ``typicality score'' for every event and selected the top-ranked candidate. The score was defined as a weighted sum of three standardised ($z$-scored) metrics:

\begin{itemize}
    \item Scenario Typicality (Weight $w=0.60$): The root-mean-square deviation (RMSD) between the event's reconstructed trajectory and the scenario-level group mean evolution. Lower values indicate behaviour closer to the population average.
    \item Reconstruction Fidelity (Weight $w=0.30$): The RMSD between the observed clip-wise ratings and the within-clip maxima of the reconstructed curve. Lower values indicate better model agreement with subjective reports.
    \item Uncertainty Width (Weight $w=0.10$): The mean temporal width of the $95\%$ uncertainty band. Lower values indicate higher estimation confidence.
\end{itemize}
The event minimising this weighted score was selected as the representative example for its scenario class. This approach prioritises events that capture the central tendency of the dataset while ensuring high model fitness.
\subsection{Reference measurements and temporal correspondence statistics}\label{si:2c}
\begin{figure}[H]
    \centering
        \includegraphics[width=0.95\textwidth]{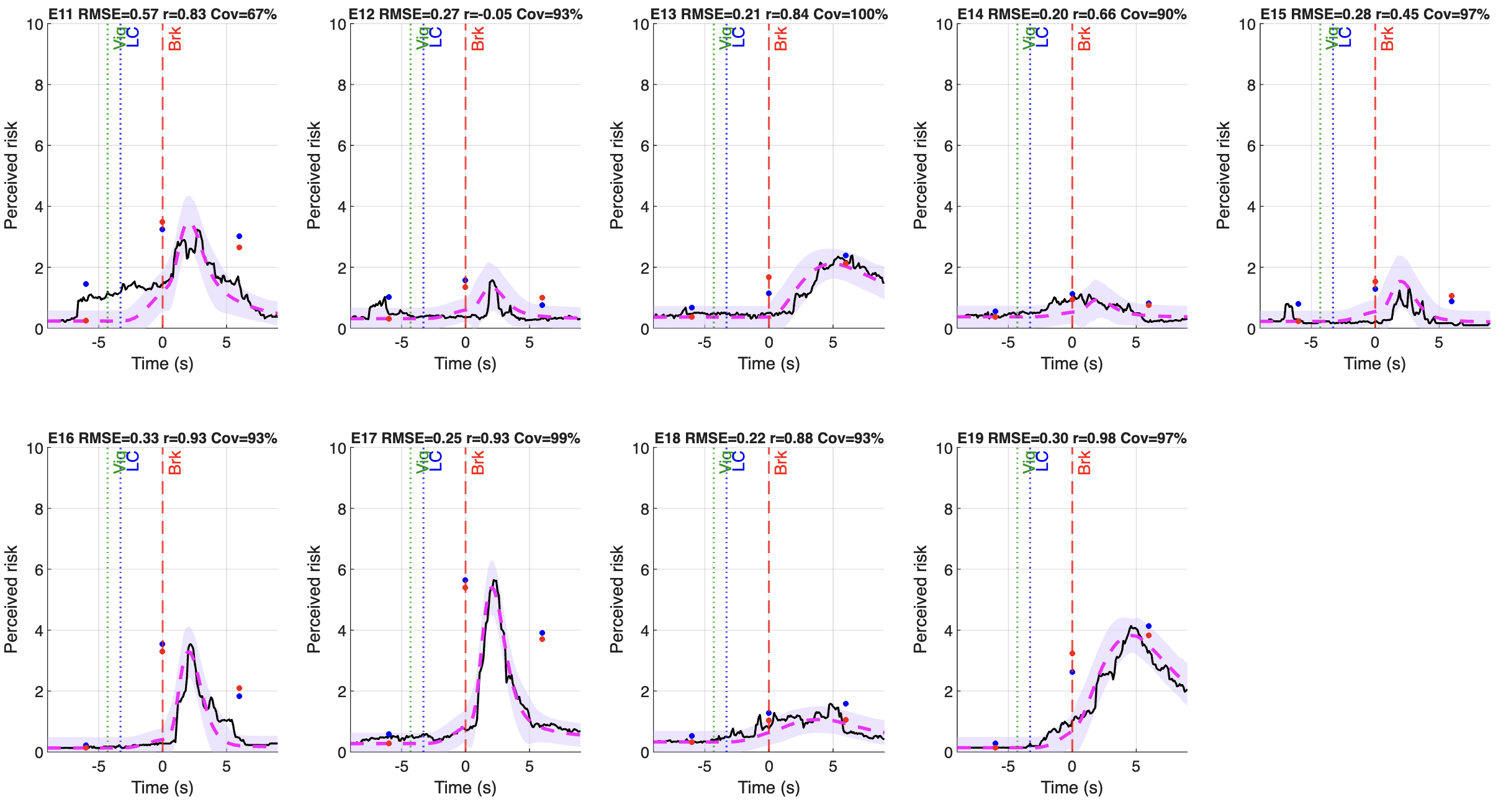} 
    \caption{Inferred perceived risk for all nine simulator events. Solid black curves represent the collected handset-based time-continuous perceived risk. Dashed pink curves represent the inferred evolution of perceived risk.}
    \label{fig:extended_validation_events}
\end{figure}

\paragraph{The inferred perceived risk for simulator events}
The main text highlights the temporal correspondence for two representative events in Fig.~2b. To provide a complete visualisation, the corresponding inferred perceived risk evolutions for all nine simulator events are presented in Fig. \ref{fig:extended_validation_events}. 

%%%%%%%%%%%%%%%%%%%%%%%%%%%%%%%%
\paragraph{Cross-correlation and permutation test.}
All signals were aligned to brake onset, defined as time zero. For each event, pupil size and heart rate were linearly interpolated onto a common $10\,\mathrm{Hz}$ grid spanning $-9\,\mathrm{s}$ to $+9\,\mathrm{s}$ and baseline corrected per participant by subtracting the mean over $-6\,\mathrm{s}$ to $-3\,\mathrm{s}$, after which participant traces were averaged to obtain an event mean physiological trace. The inferred evolution was reconstructed on the same grid and, for temporal correspondence, both the inferred evolution and the event mean physiological trace were restricted to a fixed window from $-3\,\mathrm{s}$ to $+8\,\mathrm{s}$, missing values were filled by linear interpolation with nearest end values, and each series was standardised by z scoring. We computed the normalised cross correlation between the inferred evolution and each physiological trace over lags up to $\pm 3\,\mathrm{s}$, and we report the maximal coefficient $R_{\max}$ and the corresponding lag, where negative lag indicates that physiology precedes the inferred evolution. Statistical significance was assessed by a circular shift permutation test with 5000 permutations, in which the inferred evolution was circularly shifted by a random amount of at least $1.0\,\mathrm{s}$ to disrupt temporal alignment while preserving autocorrelation, and the permutation $P$ value was computed as $(1+\#\{R_{\max}^{\mathrm{null}}\ge R_{\max}\})/(1+N_{\mathrm{perm}})$ together with the 95th percentile of the null distribution as a reference threshold.
\begin{center}
\centering
\begin{tabular}{c | c c c | c c c}
\hline
\multirow{2}{*}{\textbf{Event ID}} & \multicolumn{3}{c|}{\textbf{Pupil Dilation}} & \multicolumn{3}{c}{\textbf{Heart Rate}} \\
\cline{2-7}
 & $R_{\max}$ & Lag [s] & $P$-value & $R_{\max}$ & Lag [s] & $P$-value \\
\hline
E11 & 0.55 & -3.00 & 0.502 & 0.13 & +3.00 & 1.000 \\
E12 & \textbf{0.90} & \textbf{-0.40} & {\boldmath $< 0.001$} & 0.19 & +3.00 & 1.000 \\
E13 & 0.63 & +2.50 & 0.740 & \textbf{0.86} & \textbf{-1.10} & {\boldmath $< 0.001$} \\
E14 & 0.83 & -1.70 & 0.450 & 0.68 & +2.50 & 0.315 \\
E15 & \textbf{0.88} & \textbf{-0.10} & {\boldmath $< 0.001$} & 0.69 & +2.10 & 0.345 \\
E16 & 0.79 & -1.10 & 0.334 & 0.15 & +2.60 & 0.923 \\
E17 & 0.37 & -0.70 & 0.550 & 0.36 & +3.00 & 0.517 \\
E18 & \textbf{0.81} & \textbf{0.00} & {\boldmath $< 0.001$} & \textbf{0.99} & \textbf{0.00} & {\boldmath $< 0.001$} \\
E19 & \textbf{0.62} & \textbf{+1.10} & {\boldmath $< 0.001$} & \textbf{0.94} & \textbf{0.00} & {\boldmath $< 0.001$} \\
\hline
\textbf{Median} & \textbf{0.79} & \textbf{-0.40} & — & \textbf{0.68} & \textbf{+2.50} & — \\
\hline
\end{tabular}
\captionof{table}{\textbf{Event-level physiological validation metrics.} 
Peak cross-correlation coefficients ($R_{\max}$), temporal lags at peak correlation, and permutation-based $P$-values are reported for each validation event ($N=9$). Lags indicate the time shift of the physiological signal relative to the inferred risk; negative values indicate the physiological response \textit{preceded} the inferred risk (e.g., pupil dilation), while positive values indicate the response \textit{followed} the risk. Statistical significance ($P < 0.05$) is denoted in bold.}
\label{tab:physio_stats}
\end{center}
%%%%%%%%%%%%%%%%%%%%%%%%%%%%%%%%
\section{Cross-scenario alignment and clip level metrics}\label{si:secf3}
\subsection{Cross-scenario rescaling algorithm}
To account for scenario-dependent scaling biases, we applied a linear rescaling transformation based on participant-reported relative risk assessments. Participants provided relative severity ratings $w_{i,s}$ for each scenario $s \in \{\text{HB, MB, LC, SVM}\}$, representing the perceived danger of that scenario type in general. We defined a participant-specific scaling factor $\alpha_{i,s}$ using a robust anchoring approach to mitigate outlier effects. The anchor $w_{i, \text{anchor}}$ was defined as the second-largest rating provided by participant $i$ across the four scenarios:
\begin{equation}
    \alpha_{i,s} = \frac{\max(w_{i,s}, \epsilon)}{w_{i, \text{anchor}}}
\end{equation}
where $\epsilon=0.35$ is a lower bound to prevent numerical instability. The global scaling factor for each scenario, $\bar{\alpha}_s$, was computed as the median of $\alpha_{i,s}$ across all $N=2,164$ participants. To apply this scaling to the inferred risk evolution $r(t)$ without shifting the baseline arbitrarily, we used a median-centring transformation. For each event, we first calculated the temporal median $c = \text{median}(r(t))$. The cross-scenario rescaled evolution $r_{\mathrm{cs}}(t)$ is given by

\begin{equation}
    r_{\mathrm{cs}}(t) = c + \bar{\alpha}_s \cdot (r(t) - c).
\end{equation}

This transformation expands or compresses the dynamic range of the risk signal according to the scenario's relative severity, while anchoring the signal to its central tendency. Uncertainty bands were propagated linearly through this transformation.

\subsection{Rescaled inferred evolutions and uncertainty bands (all events)}\label{si:3a_rescaled_all}
The following figures visualise the final cross-scenario rescaled inferred perceived risk evolutions
$r_{\mathrm{cs}}(t)$ and their $95\%$ uncertainty bands for all events in each scenario.
These traces are the inputs used to compute all clip-level metrics in Section~\ref{si:3b} and the DNN targets in Section~\ref{si:fig4_predictability}.
\subsection{Clip level perceived risk metrics}\label{si:3b}
All clip level metrics were computed from the cross scenario rescaled inferred evolution $r_{\mathrm{cs}}(t)$ restricted to the clip interval $[0,T]$ and evaluated on the $10\,\mathrm{Hz}$ grid. Cumulative perceived risk was computed by numerical integration, $A=\Delta t \sum_{n=1}^{N} r_{\mathrm{cs}}(t_n)$ with $\Delta t=0.1\,\mathrm{s}$ and $N=T/\Delta t$, and mean perceived risk was defined as $E=A/T$. For temporal density, we used the baseline removed signal $\tilde r(t)=r_{\mathrm{cs}}(t)-\min_{t\in[0,T]} r_{\mathrm{cs}}(t)$ and computed $F=\tilde E^2/\widetilde{M}_2$ using the same discrete integration for $\tilde E$ and $\widetilde{M}_2$. When $\tilde r(t)$ is constant within a clip, $F$ takes the boundary value $1$ by construction.
\subsection{Derivation and statistical properties of the Temporal Density Index ($F$)}
Here we demonstrate that the Temporal Density Index ($F$) is strictly determined by the evolution's second-order statistics. Let $\mu_r$ be the Mean Risk Intensity ($E$) and $\sigma_r^2$ be the temporal variance of the risk intensity over the interval $[0, T]$. In Fig.~3c, these second-order statistics are computed for the within-clip baseline-removed signal \(\tilde r(t)=r_{\mathrm{cs}}(t)-\min_t r_{\mathrm{cs}}(t)\). By definition:
\begin{equation}
    F = \frac{E^2}{\frac{1}{T} \int_{0}^{T} r(t)^2 dt} = \frac{\mu_r^2}{\sigma_r^2 + \mu_r^2}
\end{equation}
Dividing the numerator and denominator by $\mu_r^2$, we obtain:
\begin{equation}
    F = \frac{1}{1 + \left(\frac{\sigma_r}{\mu_r}\right)^2} = \frac{1}{1 + CV^2}
\end{equation}
where $CV = \sigma_r / \mu_r$ is the Coefficient of Variation of the risk evolution. Thus, $F \in [0, 1]$ serves as an inverse proxy for temporal variability relative to the mean. $F \to 1$ implies constant risk ($CV \to 0$), while $F \to 0$ implies highly concentrated risk excursions ($CV \to \infty$).
\begin{figure}[H]
    \centering
    \includegraphics[width=0.85\textwidth]{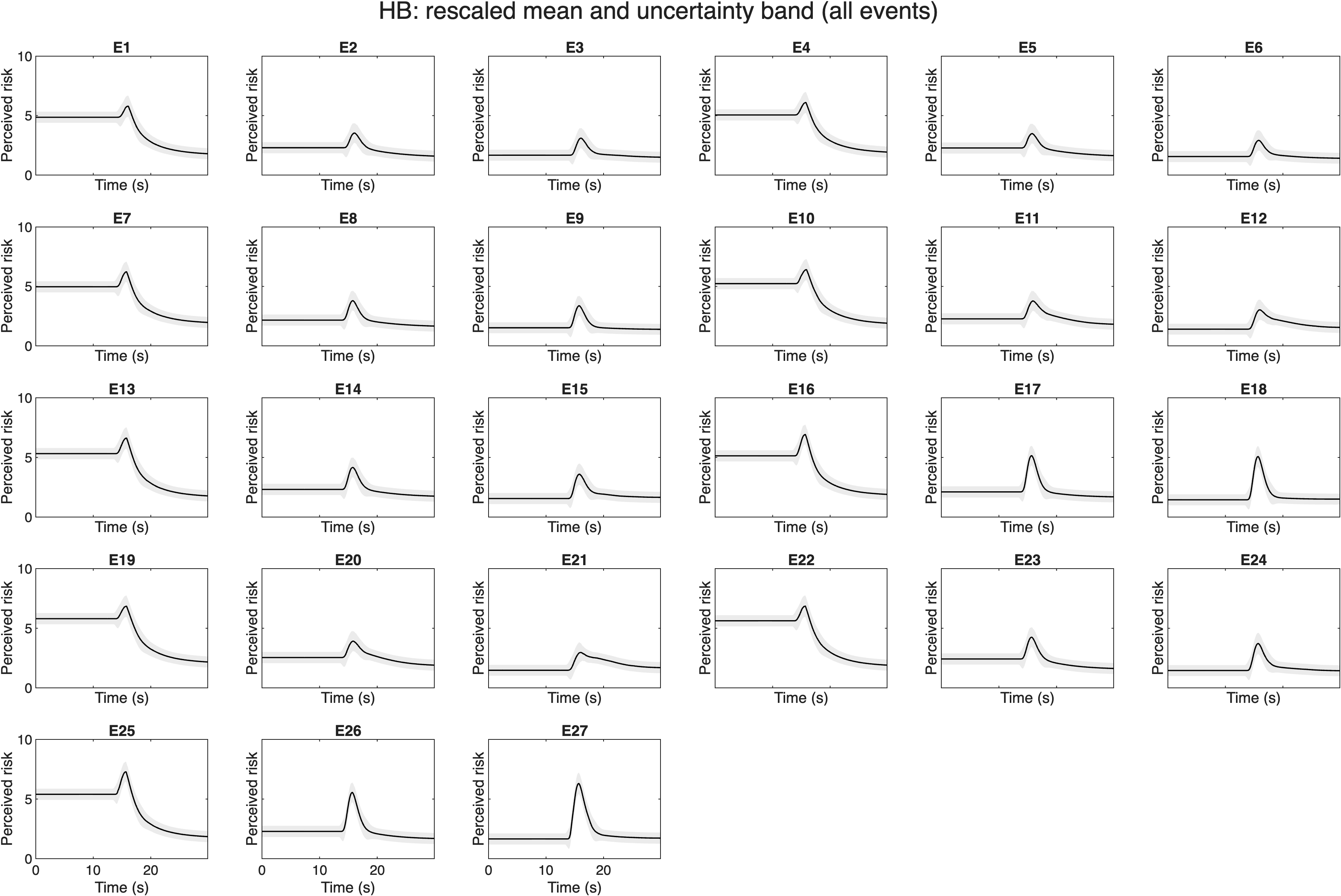}
    \captionof{figure}{\textbf{Cross-scenario rescaled inferred perceived risk evolutions (all events).}
    \textbf{a}, HB; 
    Within each panel, each small subplot corresponds to one event.
    Solid line: rescaled mean inferred evolution $r_{\mathrm{cs}}(t)$; shaded region: $95\%$ uncertainty band propagated through the rescaling transformation. }
    \label{fig:si_rescaled_all_events}
\end{figure}
\begin{figure}[H]\ContinuedFloat
    \centering
    \includegraphics[width=0.85\textwidth]{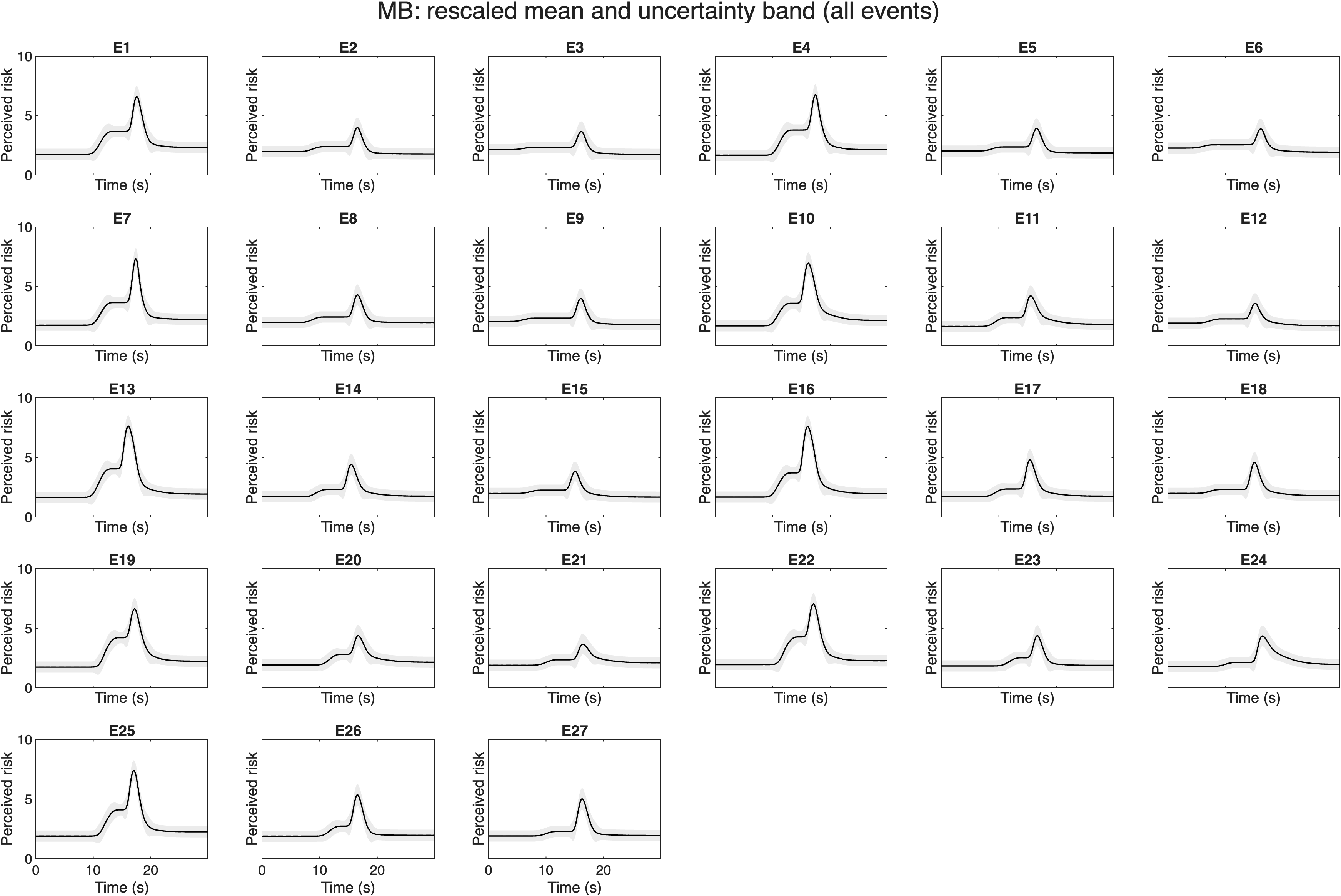}
    \caption{\textbf{Cross-scenario rescaled inferred perceived risk evolutions (all events).}
    \textbf{b}, MB; 
    Within each panel, each small subplot corresponds to one event.
    Solid line: rescaled mean inferred evolution $r_{\mathrm{cs}}(t)$; shaded region: $95\%$ uncertainty band propagated through the rescaling transformation.}
\end{figure}
\begin{figure}[H]\ContinuedFloat
    \centering
    \includegraphics[width=0.85\textwidth]{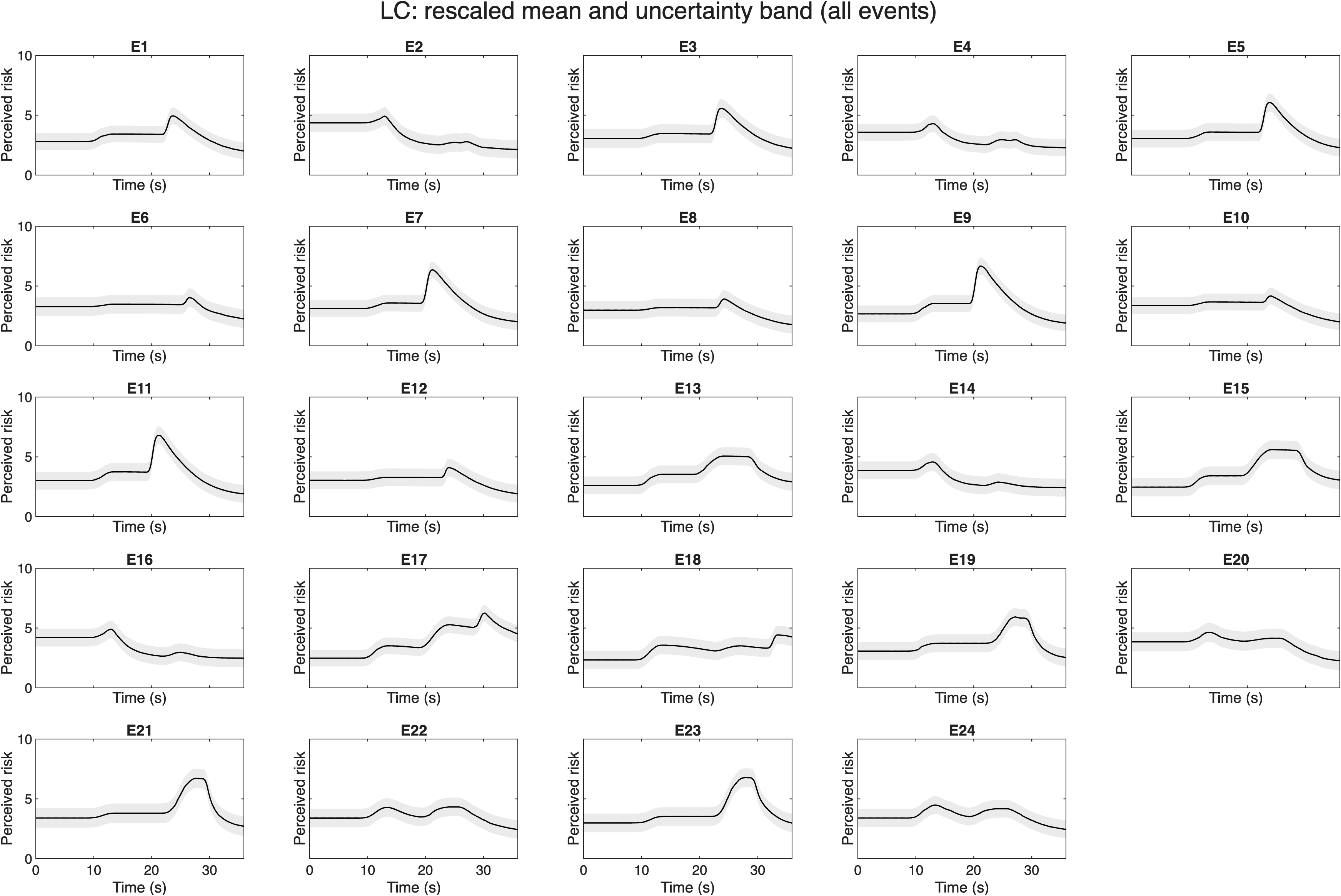}
    \caption{\textbf{Cross-scenario rescaled inferred perceived risk evolutions (all events).}
    \textbf{c}, LC; 
    Within each panel, each small subplot corresponds to one event.
    Solid line: rescaled mean inferred evolution $r_{\mathrm{cs}}(t)$; shaded region: $95\%$ uncertainty band propagated through the rescaling transformation.}
\end{figure}
\begin{figure}[H]\ContinuedFloat
    \centering
    \includegraphics[width=0.85\textwidth]{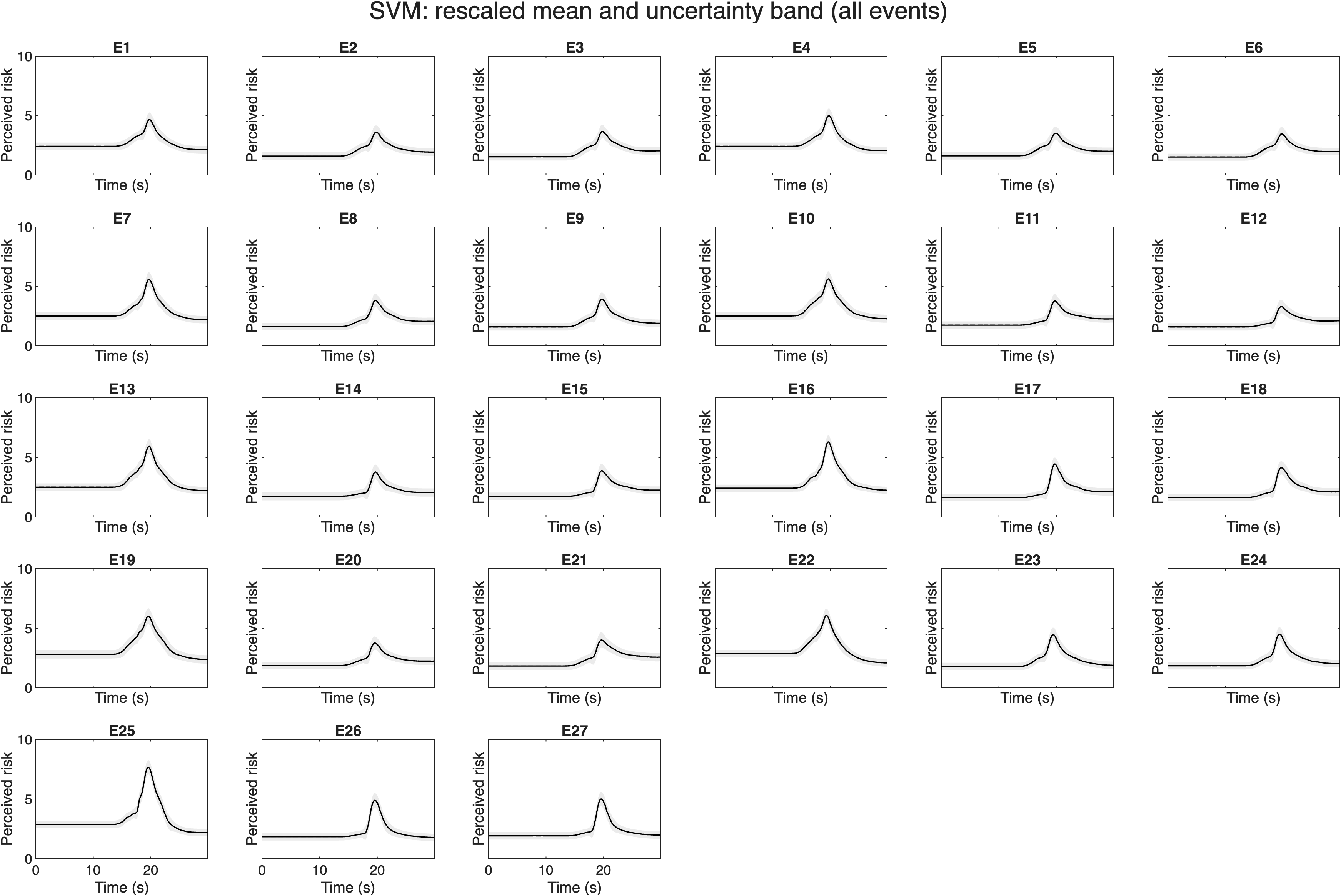}
    \caption{\textbf{Cross-scenario rescaled inferred perceived risk evolutions (all events).}
    \textbf{d}, SVM; 
    Within each panel, each small subplot corresponds to one event.
    Solid line: rescaled mean inferred evolution $r_{\mathrm{cs}}(t)$; shaded region: $95\%$ uncertainty band propagated through the rescaling transformation.}
\end{figure}
\clearpage

\subsection{Statistical details for Fig.~3c}
\begin{center}
\footnotesize
\centering
\begin{tabular}{lcccc}
\hline
\textbf{Scenario} & \textbf{\(n(F_k<1)\)} & \textbf{median \(F_k\) }& \textbf{\(n(F_k=1)\)} & \textbf{\(n_{\mathrm{total}}\)} \\
\hline
HB  & 81  & 0.6281 & 54 & 135 \\
MB  & 117 & 0.4908 & 18 & 135 \\
LC  & 120 & 0.6358 & 24 & 144 \\
SVM & 81  & 0.4417 & 54 & 135 \\
\hline
\end{tabular}

\vspace{0.8em}

\begin{tabular}{lcccc}
\hline
\textbf{Pair} & \textbf{$z$} & \textbf{\(p_{\mathrm{raw}}\)} & \textbf{\(p_{\mathrm{Holm}}\)} \\
\hline
HB vs MB  & 6.336 & \(2.357\times 10^{-10}\) & \(1.414\times 10^{-9}\) \\
HB vs LC  & 0.985 & 0.3244 & 0.3244 \\
HB vs SVM & 4.172 & \(3.018\times 10^{-5}\) & \(1.207\times 10^{-4}\) \\
MB vs LC  & -5.958 & \(2.549\times 10^{-9}\) & \(1.275\times 10^{-8}\) \\
MB vs SVM & -1.801 & 0.07177 & 0.1435 \\
LC vs SVM & 3.573 & \(3.523\times 10^{-4}\) & 0.001057 \\
\hline
\end{tabular}

\vspace{0.8em}

\begin{tabular}{lc}
\hline
\textbf{Test} & \textbf{Result} \\
\hline
Kruskal--Wallis on \(F_k<1\) & \(H=56.6070\), \(df=3\), \(p=3.117\times 10^{-12}\) \\
Chi-square on prevalence \(F_k=1\) & \(\chi^2=43.3916\), \(df=3\), \(p=2.032\times 10^{-9}\) \\
Cram\'er's \(V\) & 0.2811 \\
\hline
\end{tabular}
% \captionsetup{font=normalfont,labelfont=normalfont,textfont=normalfont}
\captionof{table}{Scenario differences in temporal concentration \(F_k\) and prevalence of the boundary case \(F_k=1\) (Fig.~3c). Kruskal--Wallis and Dunn tests are computed on clips with \(F_k<1\). Chi-square tests evaluate the dependence of the boundary-case prevalence \(F_k=1\) on scenario.}
\label{tab:fig3c_stats}
\end{center}

\section{Deep neural network predictability analyses}\label{si:fig4_predictability}

\subsection{Overview and notation}\label{si:fig4_overview}
This section documents the deep neural network analyses used to test kinematic predictability of the inferred perceived risk evolution shown in Fig.~4. Throughout, $r(t)$ denotes the cross scenario rescaled inferred evolution (defined in Methods), and $\hat r(t)$ denotes the DNN prediction of $r(t)$. The smallest unit for splitting and evaluation is the event. Unless stated otherwise, all reported metrics and plots are computed on held out events only.

\subsection{Data alignment, time grid, and target construction}\label{si:fig4_alignment}
For each event, the inferred evolution $r(t)$ is defined on a fixed time grid with sampling interval $\Delta t=\SI{0.1}{s}$ and event duration $\SI{30}{s}$ for HB, MB, and SVM, and $\SI{36}{s}$ for LC. Kinematic time series were aligned to the same grid $\Delta t=\SI{0.1}{s}$, ensuring that each time sample $t$ has a corresponding kinematic feature vector $\mathbf{x}(t)$ and target value $r(t)$.

The cross scenario rescaling was applied to the inferred evolution prior to training and evaluation. Unless otherwise stated, all DNN targets and all baseline computations in this section use the rescaled $r(t)$.

\subsection{Kinematic feature set and window summary construction}\label{si:fig4_features}
We used a unified kinematic feature set across scenarios, derived from the ego vehicle state and the relative state of the interacting vehicle(s) in both longitudinal and lateral components. Table~\ref{tab:inputs_cues_models} summarises the feature set, provides definitions of the base signals used to construct the DNN input vector, and indicates which signals are used by PCAD and DRF.
\subsubsection{Deceleration rate to avoid a crash (DRAC) features}\label{si:fig4_drac}
We include DRAC features as a compact collision-avoidance demand proxy derived from relative motion and spacing \cite{gettman2003surrogate_SI}. For a given direction $j\in\{x,y\}$, let $gap_j(t)$ denote the directional separation between the subject vehicle and the interacting vehicle, and let $\Delta v_j(t)$ denote the directional relative velocity under the sign convention used in feature construction. DRAC in direction $j$ is defined as
\begin{equation}
DRAC_j(t)=
\begin{cases}
\frac{\big(\Delta v_j(t)\big)^2}{gap_j(t)}, & \text{if the vehicles are closing in direction } j,\\
0, & \text{otherwise}.
\end{cases}
\label{eq:drac}
\end{equation}
In Table~\ref{tab:inputs_cues_models}, the $DRAC_{Rj}$ terms are computed from the realised kinematics, whereas the $DRAC_{Ij}$ terms are computed from perceived velocities that incorporate the uncertainty component $\Delta\boldsymbol{v}_{i,u}$ defined in the PCAD model (Eq.~\ref{eq:perceived velocity} and Fig.~\ref{fig_ext:uncertain velocity}). These DRAC features are treated as conflict cues in the attribution analyses.

To incorporate short term temporal context while retaining a pointwise predictor, we augmented each per time sample input vector with precomputed window summary features. For each base signal $z(t)$, window summaries were computed over a past window of duration 5~s. On the common grid. Summary operators is the mean of signal in the past window, yielding a fixed dimensional input vector $\mathbf{x}(t)\in\mathbb{R}^{36}$ at every time sample. The complete list of base signals and derived window summaries is reported in Table~\ref{tab:inputs_cues_models}.

All continuous features were standardised using z scoring based on training events only. Within each fold, feature means and standard deviations were estimated using concatenated time samples from the training events, and the resulting transform was applied to both training and held out events. No information from held out events was used to fit preprocessing transforms. 

\subsection{Event level splits and cross validation protocol}\label{si:fig4_cv}
We evaluated kinematic predictability under event level four fold cross validation across all 105 events, stratified by scenario. Fold assignments were created such that each event belongs to exactly one test fold, and each fold contains approximately equal numbers of events per scenario. The complete list of event IDs and fold assignments is provided in Table~\ref{tab:fold_assign_detailed}.

Within each fold, the training set comprises all events not assigned to the test fold, and the test set comprises the held out fold events. All preprocessing statistics (feature standardisation and any output scaling) were estimated using training events only and then applied unchanged to held out events. The DNN configuration was fixed in a separate pilot grid search described in Section~\ref{si:fig4_hparam_table} and was used unchanged in all four folds.

To obtain full dataset performance, we pooled predictions on held out test events across the four folds. All performance summaries in Fig.~4 are computed from these pooled held out predictions.
\begin{center}
\centering
\fontsize{7pt}{8pt}\selectfont
\begin{tabularx}{\linewidth}{l X l c c l}
\hline
\textbf{Symbol} & \textbf{Definition} & \textbf{Unit} & \textbf{PCAD} & \textbf{DRF} & \textbf{Cue family} \\
\hline
$v_{x,s}$ & Subject vehicle longitudinal speed. & $\mathrm{m\,s^{-1}}$ & $\checkmark$ & $\checkmark$ & Stability \\
$v_{x,n}$ & Neighbour vehicle longitudinal speed. & $\mathrm{m\,s^{-1}}$ & $\checkmark$ &  & Stability \\
$a_{x,s}$ & Subject vehicle longitudinal acceleration. & $\mathrm{m\,s^{-2}}$ & $\checkmark$ &  & Conflict \\
$a_{x,n}$ & Neighbour vehicle longitudinal acceleration. & $\mathrm{m\,s^{-2}}$ & $\checkmark$ &  & Conflict \\
$\Delta x$ & Longitudinal spacing to the neighbour, sign convention as in feature construction. & $\mathrm{m}$ & $\checkmark$ & $\checkmark$ & Stability \\
$\Delta y$ & Lateral offset to the neighbour, sign convention as in feature construction. & $\mathrm{m}$ & $\checkmark$ & $\checkmark$ & Stability \\
$\Delta v_x$ & Relative longitudinal velocity, sign convention as in feature construction. & $\mathrm{m\,s^{-1}}$ &  &  & Conflict \\
$\Delta v_y$ & Relative lateral velocity, sign convention as in feature construction. & $\mathrm{m\,s^{-1}}$ &  &  & Conflict \\
$\Delta a_x$ & Relative longitudinal acceleration, sign convention as in feature construction. & $\mathrm{m\,s^{-2}}$ &  &  & Conflict \\
$\Delta a_y$ & Relative lateral acceleration, sign convention as in feature construction. & $\mathrm{m\,s^{-2}}$ &  &  & Conflict \\
$v_{I_s,x}$ & Subject vehicle inertial frame velocity, $x$ component. & $\mathrm{m\,s^{-1}}$ & $\checkmark$ &  & Stability \\
$v_{I_s,y}$ & Subject vehicle inertial frame velocity, $y$ component. & $\mathrm{m\,s^{-1}}$ & $\checkmark$ &  & Stability \\
$v_{I_n,x}$ & Neighbour vehicle inertial frame velocity, $x$ component. & $\mathrm{m\,s^{-1}}$ & $\checkmark$ &  & Stability \\
$v_{I_n,y}$ & Neighbour vehicle inertial frame velocity, $y$ component. & $\mathrm{m\,s^{-1}}$ & $\checkmark$ &  & Stability \\
$DRAC_{Ix}$ & Collision avoidance deceleration demand, neighbour $I$, longitudinal component. & $\mathrm{m\,s^{-2}}$ &  &  & Conflict \\
$DRAC_{Iy}$ & Collision avoidance deceleration demand, neighbour $I$, lateral component. & $\mathrm{m\,s^{-2}}$ &  &  & Conflict \\
$DRAC_{Rx}$ & Collision avoidance deceleration demand, neighbour $R$, longitudinal component. & $\mathrm{m\,s^{-2}}$ &  &  & Conflict \\
$DRAC_{Ry}$ & Collision avoidance deceleration demand, neighbour $R$, lateral component. & $\mathrm{m\,s^{-2}}$ &  &  & Conflict \\
\hline
\end{tabularx}
\captionof{table}{Kinematic inputs used for the DNN attribution analyses and their cue family assignment. Each listed base signal is included in the DNN both as an instantaneous value and as its mean over the input window $W$, yielding $18\times 2=36$ input features in total. The columns PCAD and DRF indicate whether the corresponding base signal is used by each physics baseline.}\label{tab:inputs_cues_models}
\end{center}

\begin{center}
\centering
\footnotesize
% \fontsize{10pt}{10pt}\selectfont
\begin{tabular}{cllll}
\hline
\textbf{Fold} & \textbf{HB (Event IDs)} & \textbf{MB (Event IDs)} & \textbf{LC (Event IDs)} & \textbf{SVM (Event IDs)} \\ \hline
\textbf{1} & 2, 6, 11, 12, 23, 24, 26 & 1, 4, 5, 16, 18, 21, 25 & 2, 11, 12, 15, 21, 24 & 2, 6, 11, 12, 23, 24, 26 \\ \hline
\textbf{2} & 1, 10, 14, 17, 18, 20, 27 & 8, 13, 15, 17, 19, 20, 24 & 6, 14, 17, 20, 22, 23 & 1, 10, 14, 17, 18, 20, 27 \\ \hline
\textbf{3} & 7, 8, 13, 15, 16, 21, 22 & 7, 9, 10, 11, 14, 23, 27 & 1, 3, 7, 9, 18, 19 & 7, 8, 13, 15, 16, 21, 22 \\ \hline
\textbf{4} & 3, 4, 5, 9, 19, 25 & 2, 3, 6, 12, 22, 26 & 4, 5, 8, 10, 13, 16 & 3, 4, 5, 9, 19, 25 \\ \hline
\end{tabular}
\captionof{table}{Detailed fold assignments for event level four fold cross validation across all scenarios.}
\label{tab:fold_assign_detailed}
\end{center}

\subsection{Unified DNN architecture}\label{si:fig4_dnn_arch}
The unified predictor is a feedforward fully connected network mapping $\mathbf{x}(t)\in\mathbb{R}^{36}$ to a two dimensional output at each time sample. It comprises two hidden layers of width 1000 and a two dimensional linear output head. The predicted perceived risk evolution $\hat r(t)$ is taken as the first output channel. The second output channel is predictive uncertainty.

% \subsubsection{Definition of uncertainty output channel}\label{si:fig4_uncertainty}
% The second output channel represents
% {\color{red}\textbf{[TO FILL: what this uncertainty means and how it is trained]}}.
% It was trained with
% {\color{red}\textbf{[TO FILL: loss definition and $\lambda_{\mathrm{aux}}$ ]}},
% yielding a total objective of the form $\mathcal{L}=\mathcal{L}_{\mathrm{risk}}+\lambda_{\mathrm{aux}}\mathcal{L}_{\mathrm{aux}}$.

\subsection{Training objective and optimisation}\label{si:fig4_training}
The network outputs $(\mu_e(t),\, s_e(t))$, where $\mu_e(t)$ is the predicted evolution and $s_e(t)$ parametrises the predictive variance via $\sigma_e^2(t)=\exp(s_e(t))$. Training minimised the Gaussian negative log likelihood over time samples from training events only,
\[
\mathcal{L}=\frac{1}{\sum_e T_e}\sum_{e\in\mathcal{E}_{\mathrm{train}}}\sum_{t=1}^{T_e}\left(\frac{(r_e(t)-\mu_e(t))^2}{2\exp(s_e(t))}+\frac{s_e(t)}{2}\right).
\]
For all reported figures and metrics we use $\hat r_e(t)=\mu_e(t)$, and the uncertainty channel is not analysed further.

\subsection{Hyperparameter selection}\label{si:hparam_selection}
Architectural choices and training hyperparameters were fixed in a separate pilot grid search conducted prior to the main cross validation experiments. The evaluated configurations and summary results are reported in Section~\ref{si:fig4_hparam_table}. The final chosen configuration is summarised in Table~\ref{tab:dnn_final_config} and was held fixed for all experiments reported in Fig.~4, including the unified predictor, the scenario specific predictors, and the cross scenario stress tests.

\subsection{Independent initialisations and aggregation}\label{si:fig4_seeds}
To assess stability with respect to random seeds, we repeated training with five independent random initialisations. Each run used the same folds and the same preprocessing protocol but a different random seed controlling weight initialisation and minibatch ordering. For reporting and plotting, we aggregated the five runs by averaging predictions at each time sample,
\[
\bar r_e(t)=\frac{1}{5}\sum_{j=1}^{5}\hat r^{(j)}_e(t),
\]
and evaluated all metrics using $\bar r_e(t)$.

\subsection{Physics-based baselines and foldwise calibration protocol}\label{si:baselines_protocol}
PCAD and DRF include free parameters (Table~\ref{tab:pcad_drf_params}) that were calibrated to improve alignment with the inferred evolution $r(t)$. Calibration and evaluation followed the same event level protocol as the unified DNN. Within each fold, parameters were fitted using training events only by minimising the mean squared error between the model output and $r(t)$ over training time samples. Fitted parameters were then fixed and used to generate predictions on held out events only.

To ensure comparability across methods, all baseline outputs were evaluated on the same time grid as $r(t)$. Any output scaling required to place baseline outputs on the scale of $r(t)$ was fitted using training events only and then applied unchanged to held out events.

\subsubsection{Model flexibility and evaluation under held out events.}
The three predictors compared in Fig.~4 differ substantially in flexibility. PCAD and DRF each fit a small number of free parameters per fold (seven for PCAD and four for DRF, plus a fold specific scalar scaling constant for DRF), whereas the unified DNN contains a much larger number of trainable weights given its fully connected architecture (36 inputs, two hidden layers of width 1000, and a two channel output head), corresponding to approximately $1.04\times 10^{6}$ trainable parameters. In this setting, widely used analytic model comparison tools are not straightforward to interpret, because the compared models are not nested and time samples within the same event are correlated, so per sample independence and parameter count based adjustments do not translate cleanly to our evaluation unit. We therefore treat held out event performance as the primary check on generalisation. All fitting steps, including DNN training and hyperparameter selection as well as baseline calibration and any output scaling, are performed using training events only within each fold, and predictive performance is reported only on held out events.

\subsubsection{Calibration objective, output scaling, and fitted parameters for PCAD and DRF}\label{si:baselines_calibration}

Within each event level fold used in Fig.~4, we calibrated the free parameters of PCAD and DRF using training events only. Let $r_e(t)$ denote the cross scenario rescaled inferred evolution for event $e$. Let $u^{\mathrm{PCAD}}_e(t;\boldsymbol{\theta})$ and $u^{\mathrm{DRF}}_e(t;\boldsymbol{\phi})$ denote the raw outputs of PCAD and DRF evaluated on the same aligned grid as $r_e(t)$. For each fold, we pooled all time samples across the training events and estimated parameters by minimising mean squared error,
\begin{equation}
\mathcal{J}_{\mathrm{PCAD}}(\boldsymbol{\theta})=
\frac{1}{N}\sum_{e\in\mathcal{E}_{\mathrm{train}}}\sum_{t}
\Bigl(u^{\mathrm{PCAD}}_e(t;\boldsymbol{\theta})-r_e(t)\Bigr)^2,
\qquad
\mathcal{J}_{\mathrm{DRF}}(\boldsymbol{\phi})=
\frac{1}{N}\sum_{e\in\mathcal{E}_{\mathrm{train}}}\sum_{t}
\Bigl(\tilde u^{\mathrm{DRF}}_e(t;\boldsymbol{\phi})-r_e(t)\Bigr)^2,
\end{equation}
where $N$ is the number of pooled training samples.

For PCAD, we calibrated seven free parameters
$\boldsymbol{\theta}=(\sigma_{x,s},\sigma_{y,s},\sigma_{x,n},\sigma_{y,n},\tau_{p,s},\tau_{p,n},\alpha)$
using a parallelised random search with 500 candidate draws per fold within the bounded ranges in Table~\ref{tab:pcad_drf_params_range}. Prior to evaluation, the longitudinal and lateral accelerations used by PCAD were smoothed with a short moving average over five samples, matching the calibration implementation.

For DRF, we calibrated four free parameters $\boldsymbol{\phi}=(\tau_{la},m,c,s)$ using the same random search procedure. Because the raw DRF magnitude is not on the rating scale, DRF predictions were mapped to the $0$ to $10$ range using a fold specific scalar scaling constant estimated on training events only,
\begin{equation}
\tilde u^{\mathrm{DRF}}_e(t;\boldsymbol{\phi})
=
\frac{10}{\mathrm{scaleTrain}}\,
u^{\mathrm{DRF}}_e(t;\boldsymbol{\phi}),
\qquad \mathrm{scaleTrain} > 0,
\end{equation}
where $\mathrm{scaleTrain}$ is fitted from pooled training samples within the fold and then applied unchanged when generating predictions on held out events.

The fold specific fitted parameter values for PCAD and DRF and the corresponding DRF $\mathrm{scaleTrain}$ values are reported in Table~\ref{tab:pcad_drf_fitted}.

\vspace{0.5em}

\begin{center}
\centering
% \fontsize{7pt}{8pt}\selectfont
\footnotesize
\begin{tabular}{l l c}
\hline
Model & Parameter & Search range \\
\hline
PCAD & $\sigma_{x,s}$ & $[0,6]$ \\
PCAD & $\sigma_{y,s}$ & $[0,6]$ \\
PCAD & $\sigma_{x,n}$ & $[0,6]$ \\
PCAD & $\sigma_{y,n}$ & $[0,6]$ \\
PCAD & $\tau_{p,s}$   & $[0,1]$ \\
PCAD & $\tau_{p,n}$   & $[0,1]$ \\
PCAD & $\alpha$       & $[0,2]$ \\
\hline
DRF  & $\tau_{la}$ & $[0,10]$ \\
DRF  & $m$         & $[0,10^{-3}]$ \\
DRF  & $c$         & $[0,1]$ \\
DRF  & $s$         & $[0,1]$ \\
\hline
\end{tabular}
\captionof{table}{Free parameters and bounded ranges used for fold specific calibration of PCAD and DRF.}
\label{tab:pcad_drf_params_range}
\end{center}

\begin{center}
\centering
\fontsize{7pt}{8pt}\selectfont
% \footnotesize
\begin{tabular}{c m{8cm} m{6cm}}
\hline
Fold &
PCAD fitted parameters $(\sigma_{x,s},\sigma_{y,s},\sigma_{x,n},\sigma_{y,n},\tau_{p,s},\tau_{p,n},\alpha)$ &
DRF fitted parameters $(\tau_{la},m,c,s)$ and $\mathrm{scaleTrain}$ \\
\hline
1 & 3.7709, 3.0297, 4.3228, 2.4997, 0.2016, 0.1895; 0.1686& 0.1596, $6.696\times 10^{-4}$, 0.4563, 0.6959; $4.815\times 10^{5}$ \\
2 & 4.3659, 4.1185, 5.7757, 2.0742, 0.2099, 0.1213; 0.5813& 1.9642, $1.878\times 10^{-4}$, 0.9617, 0.4572; $1.666\times 10^{9}$ \\
3 & 4.7686, 4.1221, 5.8350, 3.0908, 0.8985, 0.9859; 0.2791 & 1.7732, $6.589\times 10^{-4}$, 0.9552, 0.0739; $2.155\times 10^{8}$ \\
4 & 3.9954, 1.8569, 4.5028, 2.9676, 0.2378, 0.4123; 0.2551 & 1.6666, $2.202\times 10^{-4}$, 0.9587, 0.9130; $2.254\times 10^{9}$ \\
\hline
\end{tabular}
\captionof{table}{Fold specific calibrated parameters for PCAD and DRF and the DRF training set scaling constant used for held out evaluation.}
\label{tab:pcad_drf_fitted}
\end{center}

\subsection{Performance metrics and event level summaries}\label{si:fig4_metrics}
For each held out event $e$ with $T_e$ time samples, we computed event level RMSE
\[
\mathrm{RMSE}_e=\sqrt{\frac{1}{T_e}\sum_{t=1}^{T_e}\bigl(\bar r_e(t)-r_e(t)\bigr)^2},
\]
and within event Pearson correlation
\[
\rho_e=\mathrm{corr}\bigl(\bar r_e(t),\,r_e(t)\bigr).
\]
We report the mean and median of $\mathrm{RMSE}_e$ and $\rho_e$ across events, so that each event contributes one value regardless of its number of time samples.
\subsection{Event level prediction traces for the unified DNN and physics based baselines}\label{si:fig4_pred_traces}
To complement the event level summary metrics reported in Fig.~4 and Table~\ref{tab:per_event_metrics}, we provide trace level visualisations of the rescaled inferred evolution $r(t)$ and the corresponding model outputs on held out events. In each scenario, we plot $r(t)$ on the common $\SI{10}{Hz}$ grid and compare it with the unified DNN prediction $\bar r(t)$, computed as the mean across five independent initialisations, together with the calibrated PCAD and DRF baselines. All baseline parameters and any required output scaling were fitted using training events only within each cross validation fold, and were then applied unchanged to held out events, following the protocol in Section~\ref{si:baselines_protocol}. These traces provide a qualitative view of temporal alignment patterns that underlie the quantitative comparisons in Fig.~4.

\begin{figure}[H]
    \centering
    \captionsetup[subfigure]{labelformat=empty, font=bf}
    \begin{subfigure}{\textwidth}
        \centering
        \includegraphics[width=\textwidth]{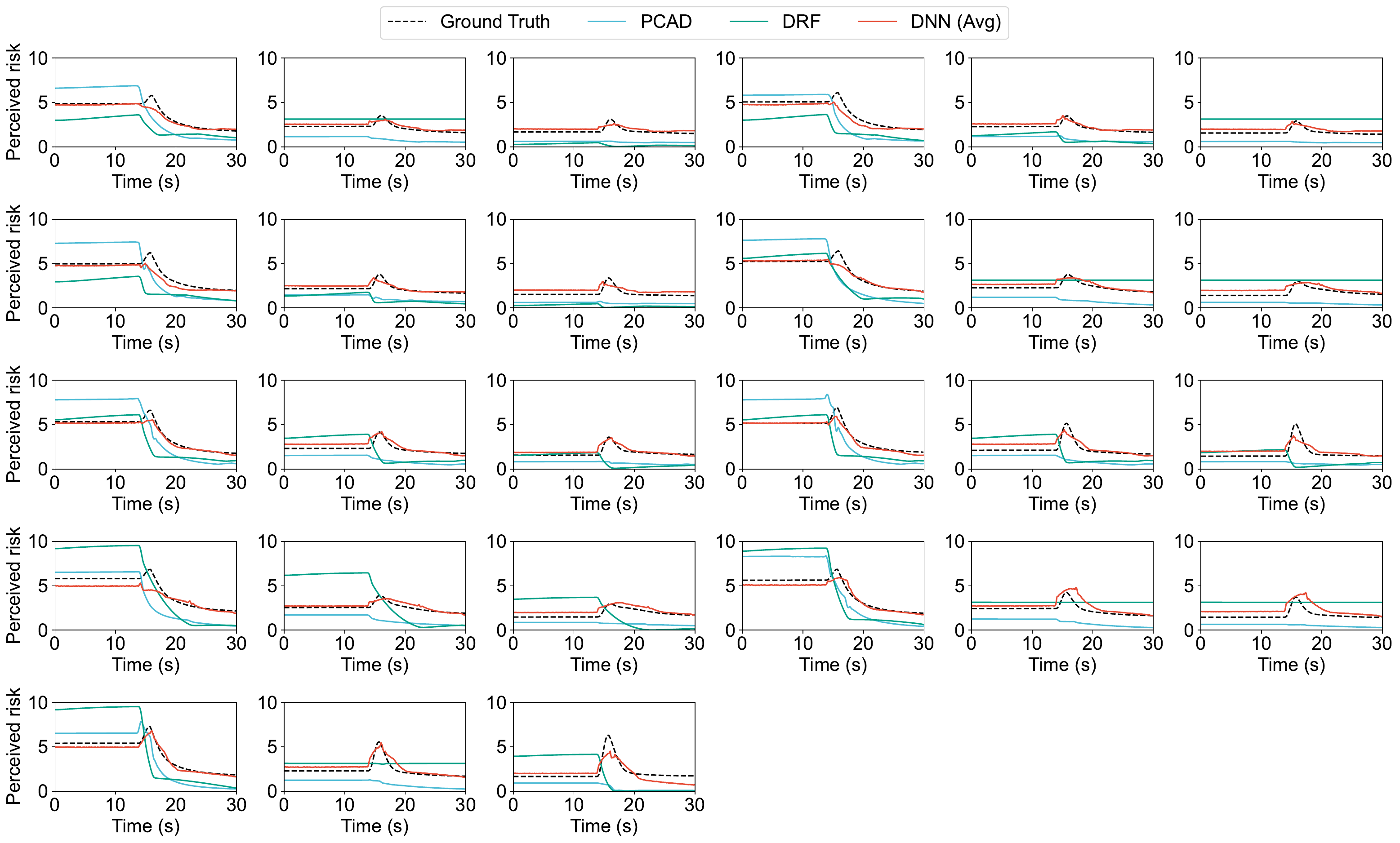}
        \subcaption{a. HB.}
    \end{subfigure}
\end{figure}

\begin{figure}[H]\ContinuedFloat
    \captionsetup[subfigure]{labelformat=empty, font=bf}
    \begin{subfigure}{\textwidth}
        \centering
        \includegraphics[width=\textwidth]{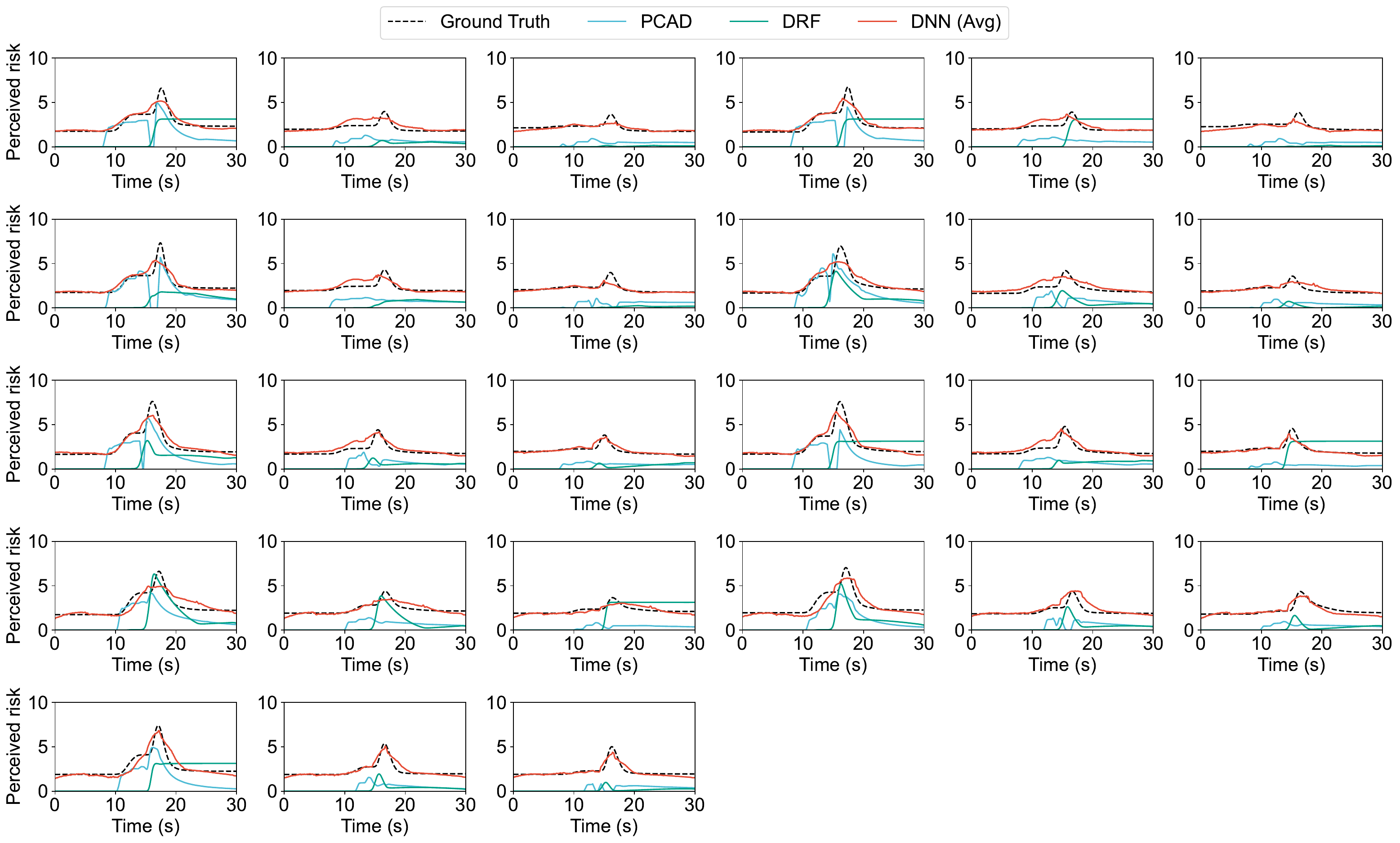}
        \subcaption{b. MB.}
    \end{subfigure}
\end{figure}

\begin{figure}[H]\ContinuedFloat
    \captionsetup[subfigure]{labelformat=empty, font=bf}
    \begin{subfigure}{\textwidth}
        \centering
        \includegraphics[width=\textwidth]{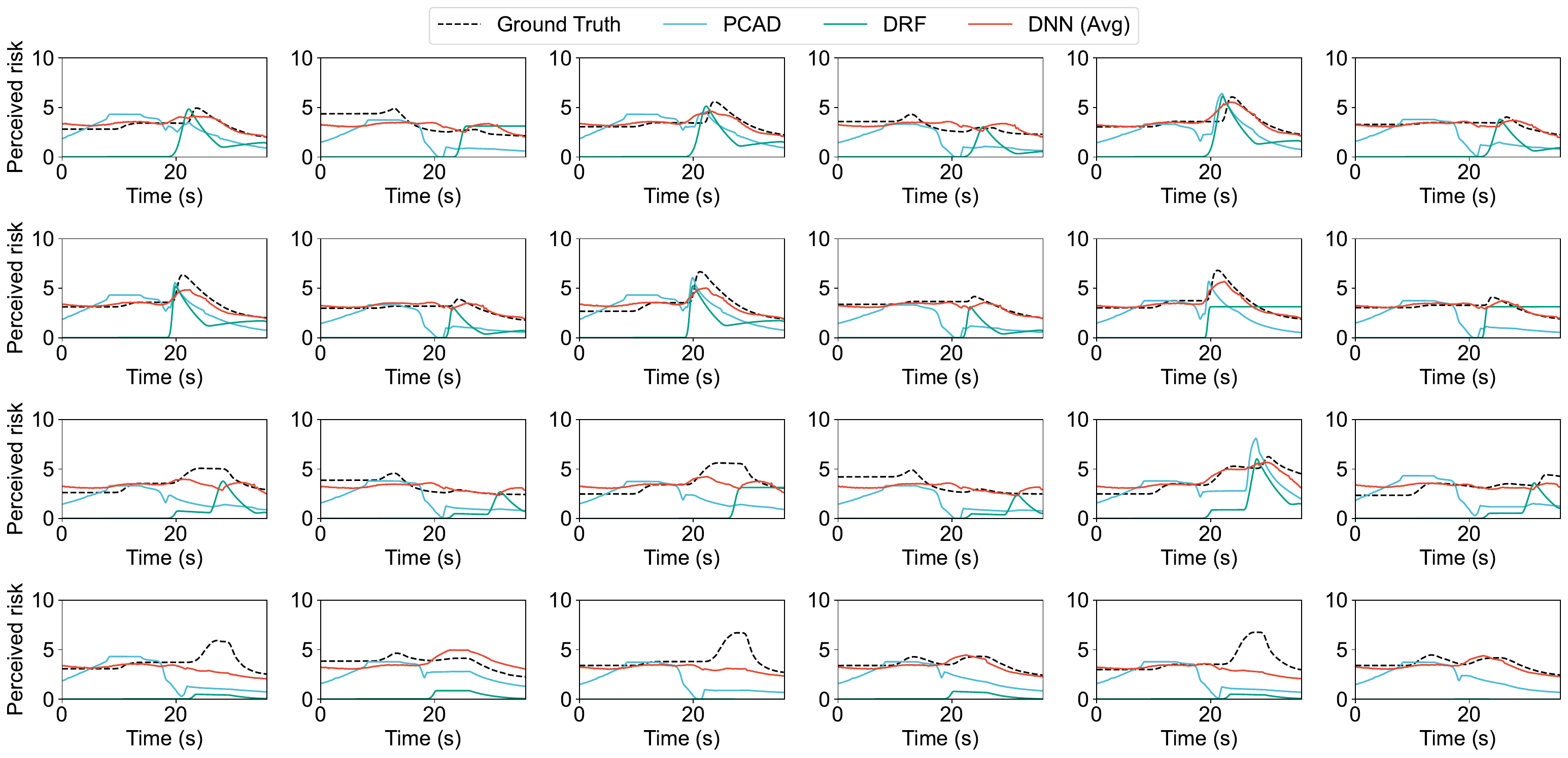}
        \subcaption{c. LC.}
    \end{subfigure}
\end{figure}

\begin{figure}[H]\ContinuedFloat
    \captionsetup[subfigure]{labelformat=empty, font=bf}
    \begin{subfigure}{\textwidth}
        \centering
        \includegraphics[width=\textwidth]{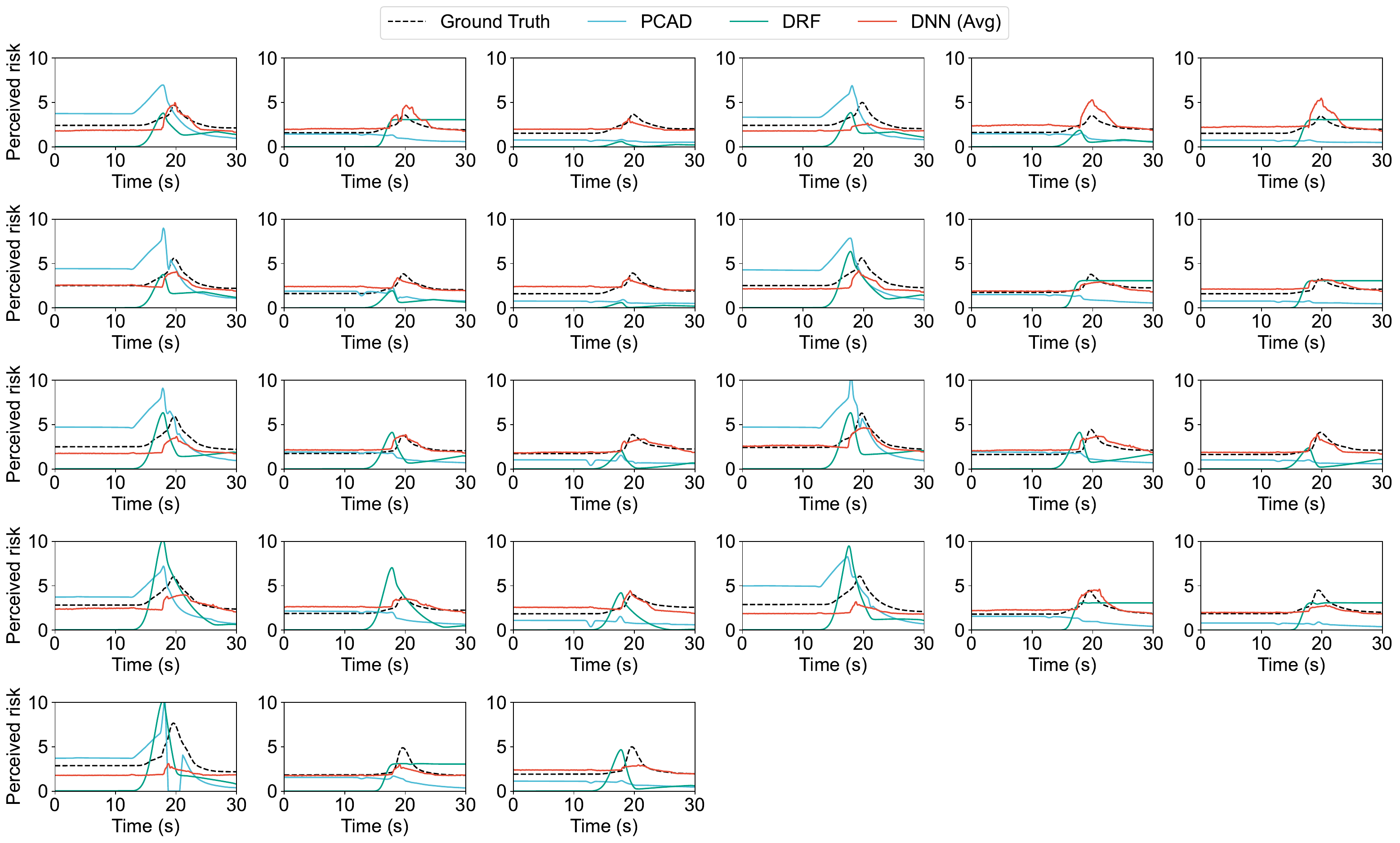}
        \subcaption{d. SVM.}
    \end{subfigure}
    \caption{Prediction traces on held out events for the rescaled inferred evolution $r(t)$ and the corresponding model outputs. The unified DNN prediction $\bar r(t)$ is the mean across five independent initialisations. PCAD and DRF outputs were generated using foldwise calibrated parameters and any required output scaling fitted on training events only within each fold, as described in Section~\ref{si:baselines_protocol}. }
    \label{fig:si_pred_traces}
\end{figure}

\subsection{Per event result tables}\label{si:per_event_tables}
The complete per event table for the DNN and baselines is provided in Table~\ref{tab:per_event_metrics}.
{
\footnotesize
\begin{longtable}{llrrrrrr}
\hline
Scenario & Event & RMSE$_{\mathrm{DNN}}$ & RMSE$_{\mathrm{PCAD}}$ & RMSE$_{\mathrm{DRF}}$ & $\rho_{\mathrm{DNN}}$ & $\rho_{\mathrm{PCAD}}$ & $\rho_{\mathrm{DRF}}$ \\
\hline
\endfirsthead

\multicolumn{8}{c}{{\footnotesize \itshape continued}} \\
\hline
Scenario & Event & RMSE$_{\mathrm{DNN}}$ & RMSE$_{\mathrm{PCAD}}$ & RMSE$_{\mathrm{DRF}}$ & $\rho_{\mathrm{DNN}}$ & $\rho_{\mathrm{PCAD}}$ & $\rho_{\mathrm{DRF}}$ \\
\hline
\endhead

\hline
\endfoot

\hline
\multicolumn{8}{l}{\footnotesize \textbf{Note:} (a) HB Scenario, (b) MB Scenario, (c) LC Scenario, (d) SVM Baseline.} \\
\caption{Per event performance metrics on held out events. The table is divided into four panels corresponding to different scenarios. }
\label{tab:per_event_metrics}
\endlastfoot

% ==========================================
% (a) HB Scenario
% ==========================================
\multicolumn{8}{l}{\textbf{(a) Scenario: HB}} \\
\hline
HB & 1 & 0.3420 & 1.6526 & 1.6953 & 0.9765 & 0.8845 & 0.8358 \\
HB & 2 & 0.2574 & 1.3390 & 1.0214 & 0.9146 & 0.5693 & -0.5905 \\
HB & 3 & 0.3151 & 1.2324 & 1.5588 & 0.8374 & 0.2417 & -0.3379 \\
HB & 4 & 0.4871 & 1.4096 & 1.9237 & 0.9702 & 0.8939 & 0.8365 \\
HB & 5 & 0.2710 & 1.3161 & 1.3030 & 0.8964 & 0.5158 & 0.2841 \\
HB & 6 & 0.3754 & 1.1402 & 1.5163 & 0.8569 & 0.0214 & -0.8475 \\
HB & 7 & 0.5030 & 1.9469 & 1.9215 & 0.9626 & 0.8694 & 0.8036 \\
HB & 8 & 0.2905 & 1.1045 & 1.2358 & 0.8203 & 0.2801 & 0.0618 \\
HB & 9 & 0.4608 & 1.1426 & 1.4508 & 0.7850 & -0.1145 & -0.4426 \\
HB & 10 & 0.3498 & 2.0988 & 1.2601 & 0.9711 & 0.8607 & 0.8880 \\
HB & 11 & 0.2107 & 1.3901 & 1.2514 & 0.9135 & 0.6120 & 0.5312 \\
HB & 12 & 0.4780 & 1.8618 & 1.9827 & 0.8705 & 0.3460 & 0.7886 \\
HB & 13 & 0.4506 & 2.2663 & 1.7510 & 0.9553 & 0.8978 & 0.8118 \\
HB & 14 & 0.4122 & 2.1027 & 2.1044 & 0.9112 & 0.7130 & 0.8055 \\
HB & 15 & 0.4487 & 2.2046 & 2.3754 & 0.9593 & 0.8765 & 0.8231 \\
HB & 16 & 0.3060 & 2.1780 & 1.5795 & 0.9618 & 0.8372 & 0.8765 \\
HB & 17 & 0.4112 & 2.0848 & 1.4941 & 0.9734 & 0.8707 & 0.8853 \\
HB & 18 & 0.3745 & 2.2387 & 2.3142 & 0.9703 & 0.9001 & 0.8718 \\
HB & 19 & 0.5995 & 2.0104 & 2.0324 & 0.9640 & 0.8403 & 0.8400 \\
HB & 20 & 0.4664 & 1.4370 & 1.9452 & 0.9417 & 0.6408 & 0.7381 \\
HB & 21 & 0.4571 & 2.2855 & 1.8030 & 0.9359 & 0.7656 & 0.8879 \\
HB & 22 & 0.5046 & 1.8941 & 2.2390 & 0.9648 & 0.8104 & 0.8324 \\
HB & 23 & 0.3175 & 1.6411 & 1.4312 & 0.9212 & 0.7758 & 0.8284 \\
HB & 24 & 0.6008 & 2.2176 & 2.1129 & 0.9548 & 0.8608 & 0.8439 \\
HB & 25 & 0.5057 & 1.6761 & 1.9853 & 0.9671 & 0.8065 & 0.8139 \\
HB & 26 & 0.4323 & 1.9151 & 1.9175 & 0.8748 & 0.4714 & 0.7626 \\
HB & 27 & 0.5306 & 1.5804 & 1.9341 & 0.9329 & 0.6925 & 0.7815 \\
\\

% ==========================================
% (b) MB Scenario
% ==========================================
\multicolumn{8}{l}{\textbf{(b) Scenario: MB}} \\
\hline
MB & 1 & 0.3747 & 1.4541 & 2.0987 & 0.8743 & 0.6025 & 0.7843 \\
MB & 2 & 0.4270 & 1.6873 & 2.4239 & 0.8255 & 0.4064 & 0.7599 \\
MB & 3 & 0.4257 & 1.8428 & 2.1930 & 0.9301 & 0.7814 & 0.7988 \\
MB & 4 & 0.3920 & 1.5652 & 1.9333 & 0.8606 & 0.5717 & 0.7365 \\
MB & 5 & 0.4587 & 1.6788 & 2.1516 & 0.8213 & 0.2707 & 0.7594 \\
MB & 6 & 0.2801 & 1.8138 & 2.0305 & 0.9444 & 0.8424 & 0.8790 \\
MB & 7 & 0.4075 & 1.5438 & 2.3370 & 0.9009 & 0.7285 & 0.8618 \\
MB & 8 & 0.3558 & 1.6915 & 2.3663 & 0.8398 & 0.4082 & 0.7265 \\
MB & 9 & 0.3266 & 1.4130 & 1.9969 & 0.8822 & 0.5832 & 0.7889 \\
MB & 10 & 0.3060 & 1.6653 & 2.0530 & 0.9157 & 0.8008 & 0.8820 \\
MB & 11 & 0.4097 & 1.7329 & 2.2652 & 0.9239 & 0.7327 & 0.8032 \\
MB & 12 & 0.2441 & 1.3350 & 1.9799 & 0.9109 & 0.7122 & 0.8657 \\
MB & 13 & 0.3492 & 1.6040 & 1.9650 & 0.8770 & 0.6288 & 0.8136 \\
MB & 14 & 0.3840 & 1.6842 & 2.4585 & 0.8801 & 0.6144 & 0.8020 \\
MB & 15 & 0.3056 & 1.6192 & 1.8345 & 0.9103 & 0.7687 & 0.8621 \\
MB & 16 & 0.4268 & 1.8055 & 2.3491 & 0.8829 & 0.6518 & 0.7728 \\
MB & 17 & 0.3131 & 1.6283 & 2.0868 & 0.9238 & 0.8032 & 0.8699 \\
MB & 18 & 0.3491 & 1.4489 & 2.3931 & 0.8479 & 0.4067 & 0.7305 \\
MB & 19 & 0.2391 & 1.7151 & 1.9062 & 0.9559 & 0.8574 & 0.9091 \\
MB & 20 & 0.4307 & 1.4675 & 2.3506 & 0.8017 & 0.0894 & 0.6785 \\
MB & 21 & 0.3598 & 1.6952 & 1.9630 & 0.9108 & 0.7565 & 0.8727 \\
MB & 22 & 0.3797 & 1.7204 & 2.1700 & 0.8907 & 0.6763 & 0.8204 \\
MB & 23 & 0.4940 & 1.8296 & 2.1600 & 0.8662 & 0.5255 & 0.7425 \\
MB & 24 & 0.3030 & 1.6114 & 2.0459 & 0.9254 & 0.8099 & 0.8780 \\
MB & 25 & 0.3710 & 1.6470 & 1.9542 & 0.9079 & 0.7799 & 0.8602 \\
MB & 26 & 0.4196 & 1.7190 & 2.4665 & 0.8985 & 0.7083 & 0.8058 \\
MB & 27 & 0.2843 & 1.5059 & 2.0470 & 0.8949 & 0.7068 & 0.8448 \\
\\

% ==========================================
% (c) LC Scenario
% ==========================================
\multicolumn{8}{l}{\textbf{(c) Scenario: LC}} \\
\hline
LC & 1 & 0.7026 & 2.1014 & 2.5106 & 0.7947 & 0.2265 & 0.6979 \\
LC & 2 & 0.7636 & 2.3368 & 2.7829 & 0.8598 & 0.2315 & 0.6541 \\
LC & 3 & 0.5816 & 1.5787 & 2.2725 & 0.5803 & -0.3098 & 0.5464 \\
LC & 4 & 0.4171 & 1.7163 & 2.1330 & 0.8544 & 0.5866 & 0.8181 \\
LC & 5 & 0.4796 & 1.8541 & 2.0976 & 0.8697 & 0.6056 & 0.7861 \\
LC & 6 & 0.4831 & 1.6948 & 2.1750 & 0.8027 & 0.4644 & 0.8005 \\
LC & 7 & 0.5160 & 1.7086 & 2.2278 & 0.8086 & 0.4581 & 0.7946 \\
LC & 8 & 0.4345 & 1.6672 & 2.1590 & 0.8465 & 0.5291 & 0.8339 \\
LC & 9 & 0.6351 & 1.8045 & 2.2419 & 0.6901 & 0.0929 & 0.6711 \\
LC & 10 & 0.4965 & 1.7792 & 2.1915 & 0.8318 & 0.4738 & 0.8023 \\
LC & 11 & 0.5552 & 1.8706 & 2.1736 & 0.8222 & 0.4426 & 0.7949 \\
LC & 12 & 0.4606 & 1.7740 & 2.0503 & 0.8489 & 0.5867 & 0.8076 \\
LC & 13 & 0.6660 & 1.8496 & 2.3074 & 0.7246 & 0.2005 & 0.7099 \\
LC & 14 & 0.7230 & 2.2603 & 2.5980 & 0.8540 & 0.2697 & 0.6771 \\
LC & 15 & 0.5728 & 1.8686 & 2.1492 & 0.7639 & 0.3262 & 0.7315 \\
LC & 16 & 0.6939 & 1.8607 & 2.3108 & 0.7450 & 0.1602 & 0.6678 \\
LC & 17 & 0.6928 & 1.9332 & 2.2606 & 0.7057 & 0.0531 & 0.6566 \\
LC & 18 & 0.6839 & 1.8464 & 2.3131 & 0.6761 & -0.0481 & 0.6076 \\
LC & 19 & 0.8511 & 2.3115 & 2.8311 & 0.7962 & 0.1679 & 0.6478 \\
LC & 20 & 0.5336 & 1.7650 & 2.1382 & 0.7819 & 0.3548 & 0.7861 \\
LC & 21 & 0.6348 & 1.9297 & 2.3174 & 0.8055 & 0.3126 & 0.7409 \\
LC & 22 & 0.6684 & 2.2154 & 2.4689 & 0.8455 & 0.3360 & 0.7212 \\
LC & 23 & 0.4932 & 1.9821 & 2.1012 & 0.9073 & 0.5848 & 0.7778 \\
LC & 24 & 0.4299 & 1.8074 & 2.0593 & 0.9067 & 0.5658 & 0.8029 \\
\\

% ==========================================
% (d) SVM Baseline
% ==========================================
\multicolumn{8}{l}{\textbf{(d) Baseline: SVM}} \\
\hline
SVM & 1 & 0.4475 & 1.2374 & 1.5736 & 0.9123 & 0.0781 & 0.5822 \\
SVM & 2 & 0.6590 & 1.6116 & 2.0989 & 0.7082 & -0.3170 & 0.5634 \\
SVM & 3 & 0.6291 & 1.4857 & 1.9882 & 0.6737 & -0.4399 & 0.4863 \\
SVM & 4 & 0.6350 & 1.5179 & 2.1325 & 0.7823 & -0.1772 & 0.6159 \\
SVM & 5 & 0.6945 & 1.8350 & 2.2068 & 0.8532 & 0.3244 & 0.6883 \\
SVM & 6 & 0.5572 & 1.8496 & 2.0306 & 0.8732 & 0.4503 & 0.7055 \\
SVM & 7 & 0.5026 & 1.3921 & 1.8895 & 0.7412 & -0.5063 & 0.4882 \\
SVM & 8 & 0.6825 & 1.6860 & 2.2357 & 0.8508 & 0.2727 & 0.7038 \\
SVM & 9 & 0.5873 & 1.6276 & 1.9963 & 0.9139 & 0.3546 & 0.7254 \\
SVM & 10 & 0.6018 & 1.7075 & 2.0439 & 0.7862 & -0.1618 & 0.5556 \\
SVM & 11 & 0.5480 & 1.5020 & 1.9972 & 0.9008 & 0.2977 & 0.7309 \\
SVM & 12 & 0.6665 & 1.8019 & 2.1748 & 0.8671 & 0.2552 & 0.7069 \\
SVM & 13 & 0.7436 & 1.8245 & 2.3237 & 0.8595 & 0.1964 & 0.6908 \\
SVM & 14 & 0.5625 & 1.6405 & 2.0638 & 0.8499 & 0.2129 & 0.6823 \\
SVM & 15 & 0.7028 & 1.5931 & 2.2257 & 0.7907 & 0.0370 & 0.6415 \\
SVM & 16 & 0.5373 & 1.6254 & 1.9192 & 0.9002 & 0.2591 & 0.7250 \\
SVM & 17 & 0.6527 & 1.5080 & 2.0761 & 0.7633 & -0.1178 & 0.6079 \\
SVM & 18 & 0.7077 & 1.7042 & 2.2121 & 0.8063 & 0.0333 & 0.6577 \\
SVM & 19 & 0.7676 & 1.5408 & 2.4664 & 0.7883 & 0.3029 & 0.8256 \\
SVM & 20 & 0.5823 & 1.1110 & 1.8589 & 0.6819 & -0.6147 & 0.6498 \\
SVM & 21 & 0.6009 & 1.5979 & 1.9611 & 0.5088 & -0.5864 & 0.4652 \\
SVM & 22 & 1.3348 & 2.0764 & 2.4575 & 0.8672 & 0.4557 & 0.6587 \\
SVM & 23 & 0.4476 & 1.3256 & 1.4487 & 0.9186 & -0.3703 & 0.6321 \\
SVM & 24 & 0.4482 & 1.6885 & 1.4765 & 0.9035 & -0.0984 & 0.6437 \\
SVM & 25 & 1.7246 & 2.3929 & 2.7386 & 0.8926 & -0.0510 & 0.4992 \\
SVM & 26 & 0.5145 & 1.2256 & 1.5212 & 0.9275 & -0.0111 & 0.4791 \\
SVM & 27 & 0.5765 & 1.5641 & 2.0105 & 0.7012 & -0.3307 & 0.1889 \\

\end{longtable}
}
\subsection{Comparison results supporting Fig.~4b}\label{si:fig4b}
Fig.~4b visualises calibration by pooling time samples from held out events. For each held out time sample we form pairs $\bigl(r(t),\bar r(t)\bigr)$ and plot their density as background shading. To summarise systematic deviations, we bin by the target value $r(t)$ into 20 equally populated bins and compute the conditional mean of the prediction within each bin, yielding the binned curves shown in Fig.~4b. The shaded points represent time sample density and are used for visualisation only, while inference is based on event level metrics.

\subsection{Scenario specific predictors as a performance upper bound}\label{si:scenario_specific_upper}
We trained scenario specific DNNs and evaluated each scenario using strict leave one event out testing within the scenario. This analysis estimates within type predictability and is reported as an upper bound for the unified setting.

To estimate within type kinematic predictability, we trained one scenario specific DNN per scenario and evaluated each model using strict leave one event out testing within that scenario. All metrics are computed on held out events only and then summarised across events so that each event contributes one value. Table~\ref{tab:scenario_specific_summary} reports the corresponding event level RMSE and within event Pearson correlation summaries for each scenario, together with the pooled summary across all scenarios.

\begin{center}
\centering
\footnotesize
% \fontsize{7pt}{8pt}\selectfont
\begin{tabular}{cccccc}
\hline
\textbf{Scenario} & \textbf{Events} & \textbf{Mean RMSE} & \textbf{Median RMSE} & \textbf{Mean} $\rho$ & \textbf{Median} $\rho$\\
\hline
HB  & 27 & 0.26 & 0.24 & 0.92 & 0.92 \\
MB  & 27 & 0.38 & 0.35 & 0.85 & 0.90 \\
LC  & 24 & 0.56 & 0.53 & 0.66 & 0.78 \\
SVM & 27 & 0.24 & 0.23 & 0.95 & 0.96 \\
All & 105 & 0.36 & 0.28 & 0.85 & 0.93 \\
\hline
\end{tabular}
\captionof{table}{Scenario specific DNN predictability under leave one event out testing within each scenario. Metrics are computed on held out events only and summarised across events.}
\label{tab:scenario_specific_summary}
\end{center}

\subsection{Cross-scenario stress test as a performance lower bound and few shot adaptation}\label{si:cross_scenario_stress}
We evaluated robustness under interaction type shift using a leave one scenario out protocol. For each target scenario, we trained a unified model on events from the other three scenarios only and tested it on held out events from the target scenario (zero shot). For few shot adaptation, we added $K$ events from the target scenario to the training set and retrained the model while preserving an independent held out test set from that scenario.

Agreement with the identity relation for the quantile binned conditional means is summarised by the deviation metric $D$ in Table~\ref{tab:fewshot_identity_dev}.

\paragraph*{Quantile binned conditional means and deviation from the identity relation.}
For each target scenario and each reported $K$, we pooled held out time samples from the evaluated target scenario events only. We then formed pairs $\bigl(r(t),\hat r(t)\bigr)$ and binned samples into $B=20$ equally populated bins according to the target value $r(t)$. For each bin $b$, we computed the conditional means
\[
\mu_{\mathrm{true},b}=\mathbb{E}[r(t)\mid b],\qquad
\mu_{\mathrm{pred},b}=\mathbb{E}[\hat r(t)\mid b].
\]
The quantile binned curve plots the pairs $(\mu_{\mathrm{true},b},\mu_{\mathrm{pred},b})$.

To quantify agreement with the identity relation, we computed the mean absolute deviation across bins,
\[
D=\frac{1}{B}\sum_{b=1}^{B}\left|\mu_{\mathrm{pred},b}-\mu_{\mathrm{true},b}\right|.
\]
All reported $D$ values are computed from held out samples only, separately for each target scenario.

\begin{table}[ht]
\centering
\caption{Deviation from the identity relation for the quantile binned conditional means. $D$ is defined in the text above and computed on held out time samples only. Values are computed after pooling held out time samples across all evaluation splits for the target scenario at the specified $K$.}
\label{tab:fewshot_identity_dev}
\footnotesize
% \fontsize{7pt}{8pt}\selectfont
\begin{tabular}{ccc}
\hline
\textbf{Scenario} & \textbf{$D$ for zero shot ($K=0$)} & \textbf{$D$ for few shot 10 ($K=10$)} \\
\hline
HB  & 0.449412 & 0.326527 \\
MB  & 0.969403 & 0.085232 \\
LC  & 0.684057 & 0.346284 \\
SVM & 0.448139 & 0.185299 \\
\hline

\end{tabular}
\end{table}

\subsection{Implementation and reproducibility}\label{si:implementation_repro}

All analyses were implemented in Python using Pytorch together with  numpy, scipy, pandas, scikit learn, shap. The unified predictor, the scenario specific predictors, and the cross scenario stress tests shared the same preprocessing pipeline and event level splitting rules described above.

Randomness was controlled at two levels. The first is the generation of fold assignment, which is held fixed for all experiments reported in Fig.~4. For model training, we used 5 independent random initialisations with different seeds, and we report the mean prediction across these runs. For the cross scenario stress test, the added few shot events was randomly selected and repeated 5 times per target scenario and per $K$. The hyperparameters are detailed in Table~\ref{tab:dnn_final_config}.

Code used to generate all results and figures in Fig.~4, together with the fold assignment table and configuration files, is available at \hyperlink{10.4121/242d9474-e522-4518-8917-8f284fc8a7a8}{10.4121/242d9474-e522-4518-8917-8f284fc8a7a8}. The processed kinematic features and targets used for training and evaluation are available at \hyperlink{10.4121/242d9474-e522-4518-8917-8f284fc8a7a8}{10.4121/242d9474-e522-4518-8917-8f284fc8a7a8} subject to the data sharing constraints described in the Data availability statement.
\subsection{Final unified DNN configuration}\label{si:fig4_final_config}
\begin{center}
\centering
% \fontsize{7pt}{8pt}\selectfont
\footnotesize
\begin{tabular}{p{5cm}p{9cm}}
\hline
\textbf{Item} & \textbf{Value} \\
\hline
Input dimension & 36 \\
Hidden layers and width & $L=2$, $H=1000$ \\
Output head & two dimensional linear head; first channel used as $\hat r(t)$ \\
Window summaries & centred window $W=5$ s; operators: Mean \\
Loss & Gaussian Negative Log Likelihood (GNLL) \\
Optimiser & Stochastic Gradient Descent (SGD) \\
Initial learning rate & 0.001 \\
Number of initialisations & 5 \\
\hline
\end{tabular}
\captionof{table}{Final unified DNN configuration and training settings used in Fig.~4.}
\label{tab:dnn_final_config}
\end{center}

\subsection{Hyperparameter search table}\label{si:fig4_hparam_table}
To ensure that the reported performance of the unified DNN is robust and not an artifact of narrow hyperparameter tuning, we conducted a grid search over learning rate ($lr \in \{10^{-4}, 10^{-3}, 10^{-2}\}$), hidden layer width ($U \in \{200, 500, 1000\}$), dropout rate ($p \in \{0.1, 0.2, 0.3\}$), and window size ($ws \in \{0, 10, \dots, 50\}$).The results, summarised in Table~\ref{tab:hparam_regimes} and Figure~\ref{fig:sensitivity}, reveal distinct performance regimes. An aggressive optimisation schedule ($lr=0.01$) yields the lowest validation errors (RMSE $\approx 0.29$), suggesting a sharp minimum that may be sensitive to event-specific noise. In contrast, a moderate learning rate ($lr=0.001$) converges to a stable solution with an RMSE of $\approx 0.47$. While the aggressive configuration offers numerically superior performance on the validation set, we adopted the conservative configuration ($lr=0.001$) for the main results to prioritise generalisation stability. As shown in Table~\ref{tab:hparam_regimes}, even this conservative choice outperforms the physics-based baselines (PCAD and DRF) by a large margin (reducing RMSE by $>70\%$), demonstrating that the kinematic grounding is robust to hyperparameter variations.

\begin{table}[t]
    \centering
    \caption{\textbf{Comparison of hyperparameter regimes and baselines.} The table contrasts the aggressive optimisation schedule (which minimizes validation error but risks overfitting) against the selected conservative configuration. Even the conservative model significantly outperforms physics-based baselines (PCAD, DRF).}
    \label{tab:hparam_regimes}
    \setlength{\tabcolsep}{5pt} 
    \renewcommand{\arraystretch}{1.1} 
    \footnotesize
    \begin{tabular}{cccccc}
        \toprule
        \textbf{Model / Regime} & \textbf{Learning Rate} & \textbf{Units} & \textbf{Win. Size} & \textbf{Dropout} & \textbf{Mean RMSE} \\
        \midrule
        Aggressive (Min. Error) & $10^{-2}$ & 1000 & 40 & 0.1 & 0.290 \\
        \textbf{Conservative (Selected)} & $\mathbf{10^{-3}}$ & \textbf{1000} & \textbf{50} & \textbf{0.2} & \textbf{0.472} \\
        Under-fitting & $10^{-4}$ & 1000 & 30 & 0.2 & 0.898 \\
        \midrule
        \textit{Baseline: PCAD} & - & - & - & - & 1.622 \\
        \textit{Baseline: DRF} & - & - & - & - & 2.124 \\
        \bottomrule
    \end{tabular}
    \vspace{1em}
\end{table}

\begin{center}
    \centering
    \includegraphics[width=0.7\linewidth]{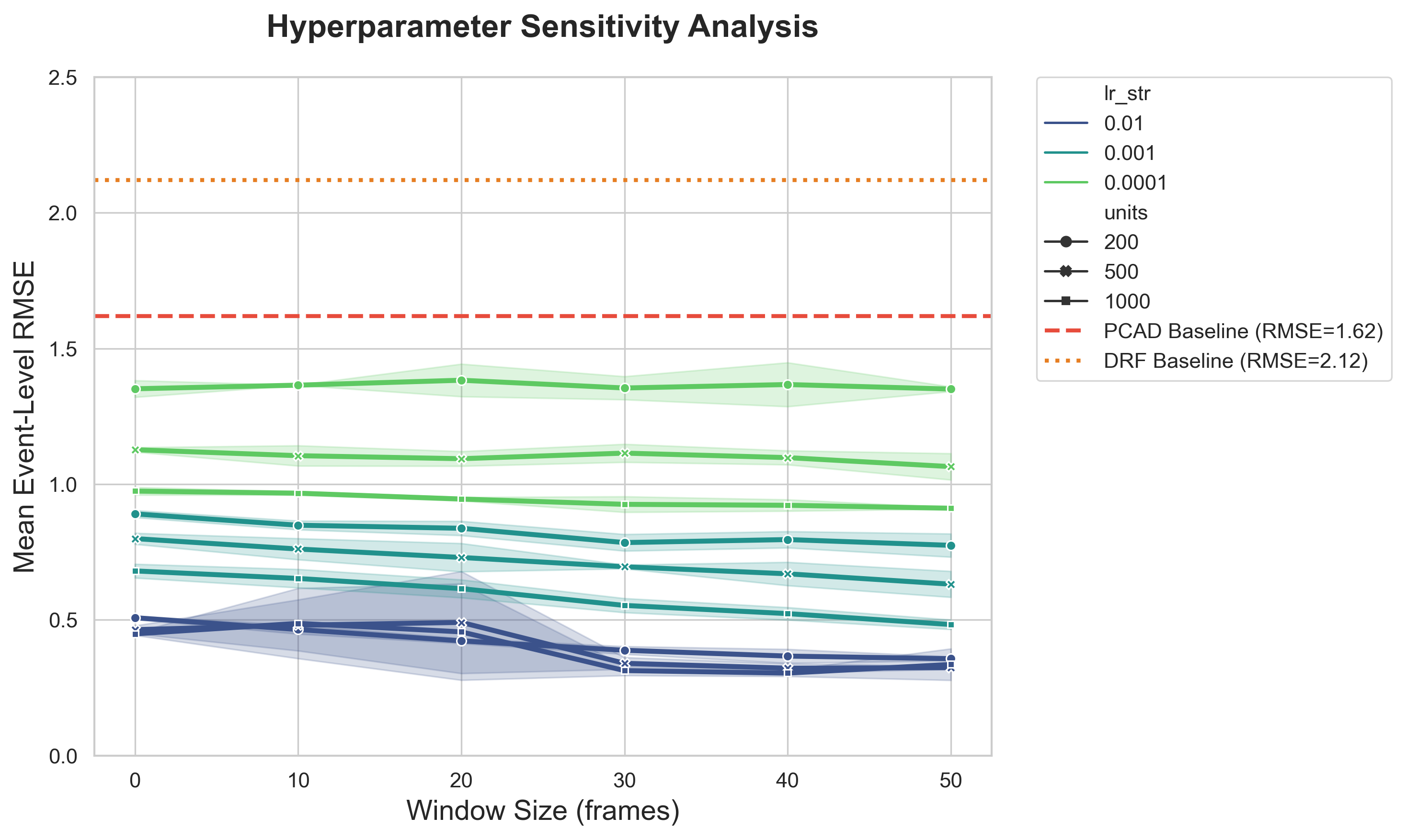}
    \captionof{figure}{\textbf{Hyperparameter sensitivity analysis.} The plot shows the Mean Event-Level RMSE across varying window sizes (x-axis) and hidden layer width (line styles), grouped by learning rate regimes (colours). 
    The aggressive regime ($lr=10^{-2}$, purple lines) achieves the lowest error but exhibits a sharp minimum. 
    The selected conservative regime ($lr=10^{-3}$, teal/green lines) demonstrates robust performance with RMSE consistently between 0.4 and 0.6, significantly outperforming the PCAD baseline (red dashed line, RMSE=1.62) regardless of structural parameters.}
    \label{fig:sensitivity}
\end{center}
\section{Surrogate perceived risk models}
\subsection{Potential collision avoidance difficulty (PCAD) model}
The Potential Collision Avoidance Difficulty (PCAD) model \cite{He2023_SI} is a computational driver perceived risk model grounded in Fuller’s Risk Allostasis Theory, which posits that drivers strive to keep perceived risk within a preferred range and that the feeling of risk corresponds to driving task difficulty. Accordingly, PCAD quantifies perceived risk as the difficulty of avoiding a potential collision, measured by the required change in the vehicle’s velocity to reach a safe (collision-free) state. In essence, it evaluates how hard collision avoidance would be in the current situation. The model explicitly accounts for key human factor elements including visual looming cues, motion uncertainty, and crash severity.

At its core, PCAD defines perceived risk $R_{\text{PCAD}}(t)$ at time $t$ as the product of an \textit{avoidance difficulty} function $\mathcal{A}$ and a speed-dependent \textit{weighting} function $\mathcal{W}$:
\begin{equation*}
    R_{\text{PCAD}}(t) = \mathcal{A}(\boldsymbol{p}_s, \boldsymbol{p}_n, \mathcal{V}_s(\boldsymbol{\textbf{X}}_s, \boldsymbol{\textbf{X}}_n), \mathcal{V}_n(\boldsymbol{\textbf{X}}_s, \boldsymbol{\textbf{X}}_n))\cdot \mathcal{W}(\boldsymbol{v}_s)
\label{eq:general structure}
\end{equation*}
where $\boldsymbol{p}_s$ and $\boldsymbol{p}_n$ are the positions of the subject vehicle $s$ and a neighbouring vehicle $n$, and $\mathcal{V}_s(\boldsymbol{\textbf{X}}_s, \boldsymbol{\textbf{X}}_n)$ and $\mathcal{V}_n(\boldsymbol{\textbf{X}}_s, \boldsymbol{\textbf{X}}_n))$ are the \textit{perceived velocities} of the subject and neighbour, respectively. 

The function $\mathcal{A}$ quantifies the minimal 2D velocity change required for the subject vehicle to reach a ``safe'' velocity such that a collision with the neighbour is avoided. It formalises the intuitive concept of a \textit{safe velocity set} $\boldsymbol{V}$ that is the set of subject velocities under which the neighbouring vehicle is not on a collision course (no looming occurs). If $\boldsymbol{v}_{s}$ is the current velocity (vector) of the subject and $\boldsymbol{v}_{s,\boldsymbol{V}}$ is the nearest velocity within the safe region $\boldsymbol{V}$, then the collision avoidance difficulty is defined as the magnitude of the velocity gap $||\boldsymbol{v}_g||$ between these two velocities:
\begin{equation*}
% \begin{gathered}
        ||\boldsymbol{v}_g|| = \mathcal{A} = ||\boldsymbol{v}_{s,\boldsymbol{V}} - \boldsymbol{v}_s||
% \end{gathered}
\label{eq:velocity gap}
\end{equation*}
where $\boldsymbol{v}_{s,\boldsymbol{V}}$ is the vector in the safe velocity set $\boldsymbol{V}$, the end point of which is closest to the subject velocity vector $\boldsymbol{v}_s$. This equation essentially represents the required total speed change (combining braking and/or steering) to avoid a collision. By definition, if $\boldsymbol{v}_{s}$ already lies inside the safe velocity set $\boldsymbol{V}$, then $\mathcal{A} = 0$ (no avoidance manoeuvre needed). PCAD uses a looming-based detection mechanism to determine the safe set $\boldsymbol{V}$ at any moment (i.e. to identify whether the neighbour is looming and what subject velocities would eliminate looming), aligning the model with human visual collision-detection cues.

Human drivers do not rely only on instantaneous velocities, but also anticipate near-future motion. PCAD captures this via a perceived velocity function $V_i$ for each vehicle $i$, which adjusts the actual velocity for acceleration and uncertainty. Formally, for $i \in \{s,n\}$:
\begin{equation*}
    \boldsymbol{v}'_i = \mathcal{V}_i(\boldsymbol{\textbf{X}}_s,\boldsymbol{\textbf{X}}_n)=\boldsymbol{v}_i+\Delta\boldsymbol{v}_{i,a}+\Delta\boldsymbol{v}_{i,u}
    \label{eq:perceived velocity}
\end{equation*}
where $\mathcal{V}_i$ is the functional operator to compute the perceived velocity $\boldsymbol{v}'_i$ of the vehicle $i \in \{s,n\}$ by human drivers for perceived risk computation.  The perceived velocity combines three components: the velocity $\boldsymbol{v}_i$ ($i \in \{s,n\}$), an acceleration-based velocity $\Delta \boldsymbol{v}_{i,a} (i \in \{s,n\})$ that accounts for the influence of the known acceleration and an uncertain velocity $\Delta\boldsymbol{v}_{i,u}$ that accounts for uncertainties in vehicle motion. For example, consider a driver who notices that a car ahead is braking rapidly. The driver might perceive the car's velocity to be lower than it actually is because the driver anticipates its future motion based on the acceleration. The uncertain velocity component captures uncertainties in vehicle motion, such as a neighbouring vehicle suddenly swerving or the subject vehicle's imprecise control. In our implementation, the uncertainty component exists in both longitudinal and lateral directions for both the subject vehicle and the interacting vehicle(s). We assume that the uncertainty direction that most reduces the inter-vehicle spacing contributes most strongly to perceived risk, as illustrated in Fig.~\ref{fig_ext:uncertain velocity}. 
\begin{center}
\centering
\includegraphics[width=0.5\textwidth]{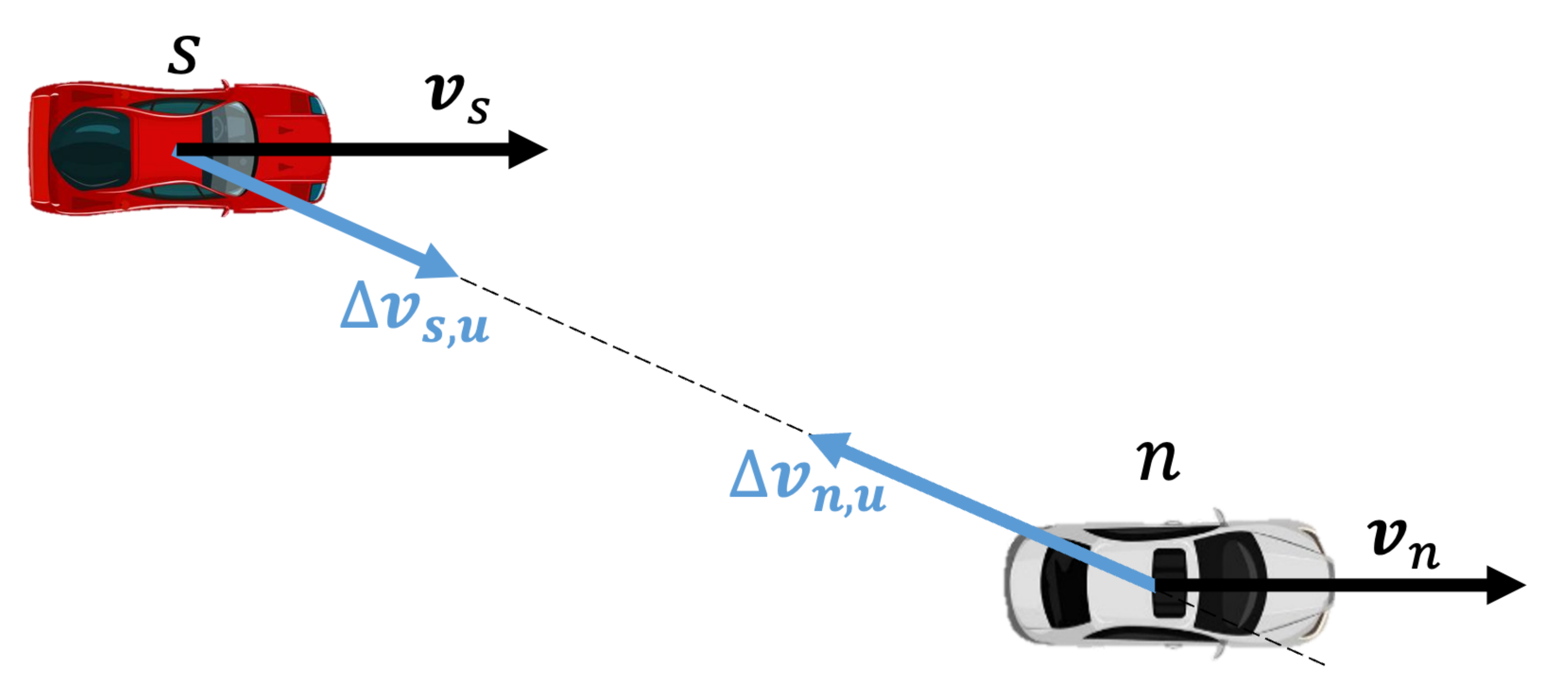}
\captionsetup{type=figure}
\captionof{figure}{The uncertain velocity $\Delta\boldsymbol{v}_{s,u}$ and $\Delta\boldsymbol{v}_{n,u}$ of the subject vehicle $s$ and the neighbouring vehicle $n$. In this case, the subject vehicle (red) is passing by a neighbouring vehicle (white). The uncertain velocities $\Delta\boldsymbol{v}_{s,u}$ and $\Delta\boldsymbol{v}_{n,u}$ are pointing to each other. }
\label{fig_ext:uncertain velocity} 
\end{center}
The weighting function $W(v_s)$ scales the difficulty $\mathcal{A}$ according to the subject vehicle’s speed, reflecting the influence of collision severity at different speeds. Higher speeds yield a larger weight (since a collision would be more severe), while lower speeds reduce the weight. In the PCAD formulation, $W$ is expressed as a normalised power-law of $v_s$ (subject vehicle speed):
\begin{equation*}
\mathcal{W}(\boldsymbol{v}_s) \;=\; \left(\frac{v_s}{v_{\text{lim}}}\right)^{\alpha}~,
\end{equation*}
where $v_{\text{lim}}$ is a reference speed (e.g. the speed limit or another context-specific maximum) and $\alpha$ is a positive exponent determining how sharply risk grows with speed. This form ensures $\mathcal{W}$ ranges between 0 and 1 for $0 \leqslant v_s \leqslant v_{\text{lim}}$. The inclusion of $\mathcal{W}$ increases perceived risk for higher $v_s$ to account for the greater potential crash energy at higher speeds, even if the avoidance difficulty $\mathcal{A}$ (purely in terms of required manoeuvre) is the same. In summary, $\mathcal{W}$ modulates the base difficulty by the severity of outcome, completing the PCAD risk estimate.

\subsection{Driving risk field (DRF) model}
The DRF represents human drivers' risk perception as a 2D field, combining the probability (probability field) and consequence (severity field) of an event, the product of which provides an estimation of driver’s perceived risk. The DRF model was derived from a simulator experiment involving obstacle avoidance with 77 obstacles distributed on a 2D plane in front of the subject vehicle. During each drive, one obstacle was randomly chosen and suddenly appeared, after which participants needed to steer to avoid the obstacle and gave a non-negative number indicating the required steering effort. Based on the position information of the obstacles, the maximum steering angle, and the subjective ratings, the DRF model was fitted to the data, and thereby it is essentially an empirical model \cite{Kolekar2020_SI}. 

The DRF model quantifies overall perceived risk as 
\begin{equation}
    R_{\text{DRF}}(t)=\sum p(x(t), y(t)) \cdot {sev}(t)
\label{DRF_RDRF}
\end{equation}
where $p(x(t), y(t))$ is the probability of an event happening at position $(x(t),y(t))$;  ${sev}(t)$ is the severity field of events. Specifically, in straight drive, the probability field can be simplified as\\
\vspace{-10pt}
\begin{equation}
    p(x(t), y(t))=h \cdot \exp \left(\frac{-y(t)^2}{2 \sigma^{2}}\right)
\end{equation}%
\vspace{-10pt}
\begin{equation}
    h=s \cdot \left(x(t)-v_{s,X}(t) \cdot t_{l a}\right)^{2}
\end{equation}%
\vspace{-12pt}
\begin{equation}
    \sigma = m \cdot x(t)+c
\end{equation}%
where the subject vehicle is at the origin $(0,0)$ with $h$ and $\sigma$ representing the height and the width of the Gaussian at longitudinal position $x(t)$; $s$ defines the steepness of the height parabola; $t_{la}$ is the human driver's preview time (s); $m$ defines the widening rate of the 2D probability field; $c$ is the quarter width of the subject vehicle (m). $v_{s,X}(t)$ is the subject vehicle's velocity (m/s). The lateral cross-section of the 2D probability field is a Gaussian. Note that the height of the Gaussian $h$ and the width $\sigma$ are separately modelled as a parabola and linear function of longitudinal distance $x$ in front of the subject vehicle. 

The severity field of the events in this study can be defined as 
\begin{equation}
    {sev}(t) = \begin{cases}C_{sev}, & (x(t), y(t)) \in A^O, \\ 0, & (x(t), y(t)) \notin A^O.\end{cases}
\label{DRF_costmap}
\end{equation}
where $C_{sev}$ is the severity value that is set empirically and $A^O$ represents a neighbouring vehicle's spatial area. 
\subsection{Foldwise calibration of PCAD and DRF}
PCAD and DRF include free parameters (Table~\ref{tab:pcad_drf_params}) that were calibrated to improve alignment with the inferred evolution $r(t)$ used throughout the paper. Calibration and evaluation followed the same event level cross validation protocol as the DNN. Within each fold, parameters were fitted using training events only by minimising the mean squared error between the model output and $r(t)$ over training time samples. The fitted parameters were then fixed and used to generate predictions on held out events only.

To ensure comparability of metrics across methods, all baseline outputs were evaluated on the same time grid as $r(t)$. Any output scaling required to place model outputs on the same numerical scale as $r(t)$ was fitted using training events only within each fold and then applied unchanged to held out events.
%%%%%%%%%%%%%% Table: model parameters to be calibrated
\begin{center}
\centering
\footnotesize
\begin{tabular}{@{}cc>{\raggedright\arraybackslash}p{12.8cm}@{}}
\toprule
\textbf{Model} & \textbf{Parameters} &  \textbf{Explanation} \\ \midrule

\multirow{7}{*}{PCAD} & \(\sigma_{n,X}\) & The standard deviation in \(X\) of the velocity Gaussian of a neighbouring vehicle \\
 & \(\sigma_{n,Y}\) & The standard deviation in \(Y\) of the velocity Gaussian of a neighbouring vehicle  \\
 & \(\sigma_{s,X}\) & The standard deviation in \(X\) of the velocity Gaussian of the subject vehicle  \\
 & \(\sigma_{s,Y}\) & The standard deviation in \(Y\) of the velocity Gaussian of the subject vehicle  \\
 & \(t_{s,a}\) & The accumulation time for the acceleration-based velocity of the subject vehicle  \\
 & \(t_{n,a}\) & The accumulation time for the acceleration-based velocity of a neighbouring vehicle  \\
 & \(\alpha\) & The exponent of the power function in weighting function \\ \midrule
\multirow{4}{*}{DRF} & \(D\) & The steepness of descent of the potential field  \\
 & \(s\) & The steepness of the height parabola of the risk field \\
 & \(t_{la}\) & Human driver's preview time  \\
 & \(m\) & The rate of the risk field width expanding \\
 & \(c\) & The initial width of the DRF \\ \bottomrule
\end{tabular}
\captionsetup{type=table}
\captionof{table}{Parameters to be calibrated of computational perceived risk models}
\label{tab:pcad_drf_params}
\end{center}

\section{State stratification and SHAP based cue utilisation analyses}
\subsection{Definition of the nine state labels}\label{si:state9_definition}
State labels were constructed from the inferred evolution $r(t)$ and its local rate of change $\dot r(t)$ on the $10\,\mathrm{Hz}$ grid. Two intensity thresholds $q_1$ and $q_2$ partitioned $r(t)$ into low, medium, and high intensity across pooled time samples. A symmetric derivative dead band $\varepsilon$ partitioned $\dot r(t)$ into falling, stable, and rising. The mapping to $state9\in\{1,\dots,9\}$ follows the order low falling, low stable, low rising, medium falling, medium stable, medium rising, high falling, high stable, high rising, consistent with the axis labels used in Fig.~5. 
\paragraph{Smoothing, derivative, and thresholds.}
The inferred evolution was first smoothed using the deterministic procedure. The local rate of change $\dot r(t)$ was then estimated by a centred finite difference on the $10\,\mathrm{Hz}$ grid. Within each scenario, $q_1$ and $q_2$ were defined as the $1/3$ and $2/3$ quantiles of the smoothed $r(t)$ pooled across events and time samples. The derivative dead band $\varepsilon$ defining falling, stable, and rising was computed within each scenario by the same implementation. Numerical values of $q_1$, $q_2$, and $\varepsilon$ are reported in Table~\ref{tab:state9_thresholds}.
\begin{center}
\centering
\footnotesize
\begin{tabular}{lccc}
\hline
\textbf{Scenario} & \textbf{$q_1$} & \textbf{$q_2$} & \textbf{$\varepsilon$} \\
\hline
HB  & 1.849989 & 2.505103 & 0.0291995 \\
MB  & 1.915542 & 2.275726 & 0.0296368 \\
LC  & 3.050245 & 3.573933 & 0.0250383 \\
SVM & 1.882308 & 2.507145 & 0.0305503 \\
\hline
\end{tabular}
\captionof{table}{Scenario specific thresholds used to define the nine state labels.}\label{tab:state9_thresholds}
\end{center}

\subsection{Feature set and cue family assignment}\label{si:cue_family_def}
Table~\ref{tab:inputs_cues_models} lists all DNN input features used for attribution analyses, their physical meaning, units, and cue family assignment. Conflict cues include DRAC components, longitudinal acceleration terms, and relative motion transients, whereas stability cues include spacing margin and speed related terms and any remaining kinematic quantities not classified as conflict cues. The cue family assignment follows the implementation used to generate Fig.~5d.

\subsection{Eventwise paired contrasts for state dependent reweighting}\label{si:paired_contrasts}
We computed within event paired contrasts from eventwise state means of cue family shares. Let $C_{e,s}$ denote the conflict share for event $e$ in state $s$, and $S_{e,s}$ denote the stability share. For conflict shift, define the set of available matched intensity levels
\[
\mathcal{L}^{C}_e=\Bigl\{\ell\in\{\mathrm{low},\mathrm{med},\mathrm{high}\}\;:\; C_{e,s_r(\ell)} \text{ and } C_{e,s_s(\ell)} \text{ are both available}\Bigr\},
\]
where $(s_r,s_s)$ are the rising and stable state indices at the same intensity, namely $(3,2)$, $(6,5)$, and $(9,8)$ for low, medium, and high. The eventwise conflict shift was then computed as the mean over available matched levels,
\[
\Delta C_e=\frac{1}{|\mathcal{L}^{C}_e|}\sum_{\ell\in\mathcal{L}^{C}_e}\Bigl(C_{e,s_r(\ell)}-C_{e,s_s(\ell)}\Bigr).
\]
For stability rebound, define
\[
\mathcal{L}^{S}_e=\Bigl\{\ell\in\{\mathrm{low},\mathrm{med},\mathrm{high}\}\;:\; S_{e,s_f(\ell)} \text{ and } S_{e,s_r(\ell)} \text{ are both available}\Bigr\},
\]
where $(s_f,s_r)$ are the falling and rising indices at the same intensity, namely $(1,3)$, $(4,6)$, and $(7,9)$ for low, medium, and high. The eventwise stability rebound was computed as
\[
\Delta S_e=\frac{1}{|\mathcal{L}^{S}_e|}\sum_{\ell\in\mathcal{L}^{S}_e}\Bigl(S_{e,s_f(\ell)}-S_{e,s_r(\ell)}\Bigr).
\]
Events with no available matched level for a given contrast were excluded from that contrast.

\subsection{Attribution concentration metrics and matched event count resampling}\label{si:concentration_metrics}

This section reports the concentration summaries used for Fig.~5f and the associated robustness checks.

\subsubsection{Normalised attribution distribution}
For each event and each $10\,\mathrm{Hz}$ time step, we obtained Shapley values $\phi_i(t)$ for all SHAP feature columns. Let $K$ denote the number of feature columns retained after excluding non feature columns (e.g., \texttt{t\_index}, \texttt{scenario}, \texttt{event\_id}, \texttt{state9} and any bookkeeping columns). We formed a scale free attribution distribution by normalising absolute Shapley magnitudes within each time step,
\begin{equation}
p_i(t)=\frac{|\phi_i(t)|}{\sum_{j=1}^{K}|\phi_j(t)|+\epsilon},
\end{equation}
where $\epsilon=10^{-12}$ ensures numerical stability.

\subsubsection{Concentration metrics.}
From $p_i(t)$ we computed three complementary metrics. Attribution entropy was
\begin{equation}
H(t)=-\sum_{i=1}^{K} p_i(t)\log_2\!\bigl(p_i(t)+\epsilon\bigr),
\end{equation}
the top one share was
\begin{equation}
Top1(t)=\max_{i} \, p_i(t),
\end{equation}
and the effective number of contributing cues was
\begin{equation}
K_{\mathrm{eff}}(t)=2^{H(t)}.
\end{equation}
Under these definitions, greater concentration corresponds to larger $Top1(t)$ and smaller $H(t)$ and $K_{\mathrm{eff}}(t)$.

\subsubsection{Binning by inferred risk intensity and eventwise aggregation.}
Each time sample was paired with its inferred risk intensity $r(t)$ using the smoothed inferred evolution provided in the exported risk tables. When a smoothed trace was not available, we used the corresponding unsmoothed inferred evolution. This choice was implemented as a deterministic rule in the released analysis scripts, ensuring that all binning and slope tests used a single, consistently defined intensity trace within each dataset.

We summarised trends using two binning schemes as labelled in Fig.~5f: quantile binning and equal width binning, each with $B=20$ bins. Within each bin, we first averaged the metric over time steps within each event, yielding one event level value per bin. This prevents events with longer dwell time in a bin from dominating the aggregate.

\subsubsection{Matched event count resampling and uncertainty intervals.}
\begin{figure}[H]
    \centering
    \begin{subfigure}{0.48\textwidth}
        \centering
        \includegraphics[width=\linewidth]{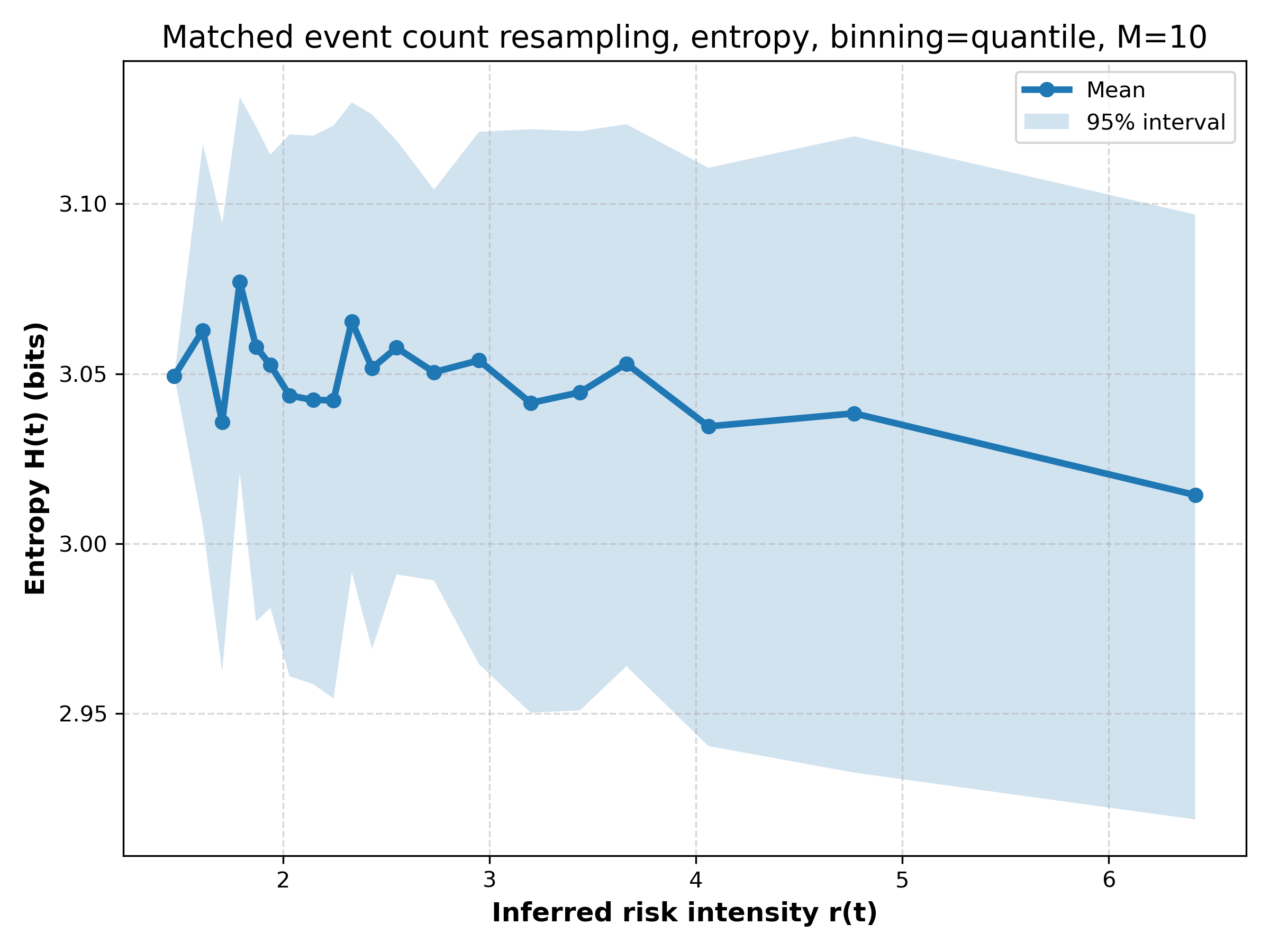}
        \caption{Entropy under Quantile Binning}
    \end{subfigure}
    \hfill
    \begin{subfigure}{0.48\textwidth}
        \centering
        \includegraphics[width=\linewidth]{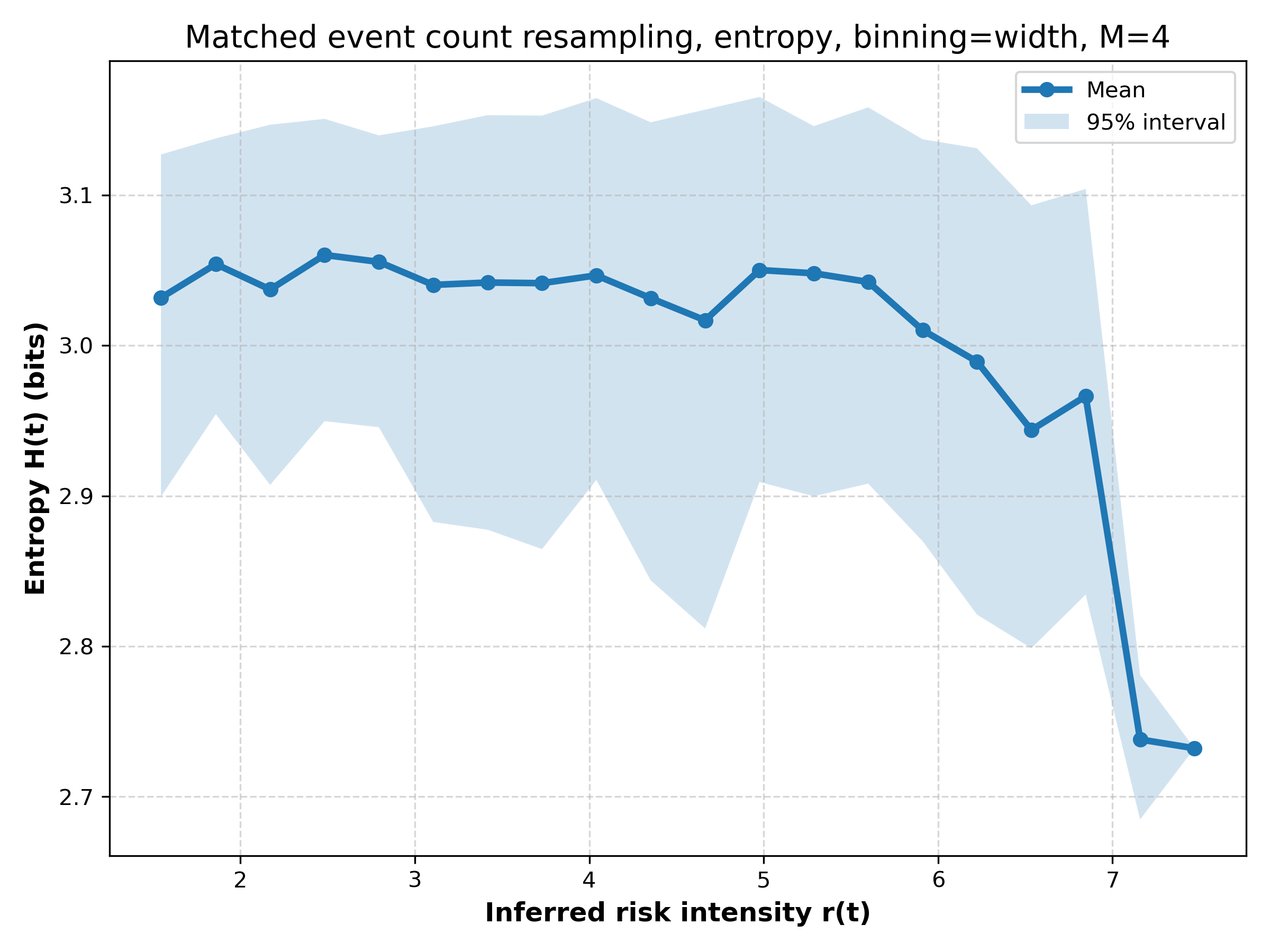}
        \caption{Entropy under Equal-width Binning}
    \end{subfigure}
    \caption{Attribution entropy across inferred risk levels using matched event count resampling. The mild downward trend demonstrates that the global distribution of cue utilization narrows slightly but avoids absolute collapse at high risk levels.}
    \label{fig:si_entropy_curves}
\end{figure}

Event coverage varies across risk bins because not every event visits every risk range. For each bin $b$, let $E_b$ denote the set of events that contribute at least one time step to $b$. We defined
\begin{equation}
M=\min_{b:\,|E_b|>0} |E_b|,
\end{equation}
and retained only bins with $|E_b|\ge M$ for plotting the matched curves. For each retained bin, we performed matched event resampling by repeatedly sampling $M$ event level values without replacement and taking their mean. We used $R=800$ resamples with a fixed random seed, and reported the resampling mean as the curve and the $2.5$th and $97.5$th percentiles as the $95\%$ interval. This procedure constrains each bin estimate to the same number of contributing events in each resample while allowing all eligible events to contribute across resamples. Consequently, end to end contrasts for entropy and top one share are estimated under a conservative effective sample size of $M$ events per retained bin.
Building on this rigorous resampling protocol, we evaluated the global dispersion of feature importance. While the Top-1 share (discussed in the main text) captures the dominance of the primary cue, the attribution entropy provides a comprehensive measure of whether the remaining attribution mass collapses or maintains a distributed long tail. As shown in Fig.~\ref{fig:si_entropy_curves}, the attribution entropy exhibits a mild downward trend as inferred risk increases. This divergence—a rising primary cue dominance coupled with a gently declining entropy—indicates a mild focusing mechanism. The primary conflict cue claims a larger proportion of the model's attention, yet the background stability cues continue to receive distributed, non-zero weights, preventing a complete collapse of global situational awareness. The consistent trajectory observed across both quantile and equal-width binning schemes confirms that this global concentration pattern is structurally robust.

\subsection{Sensitivity checks for the focusing claim}\label{si:focusing_sensitivity}
To verify that the modest increase in attribution concentration is not driven by a particular modelling choice in the concentration metric, we performed a sensitivity grid in which small attribution components were optionally suppressed before normalisation, using a threshold proportional to the per time sample total attribution mass. For each configuration, we recomputed the end minus start change of entropy and top one share under both binning schemes. 

\begin{figure}[H]
    \centering
    \includegraphics[width=0.5\textwidth]{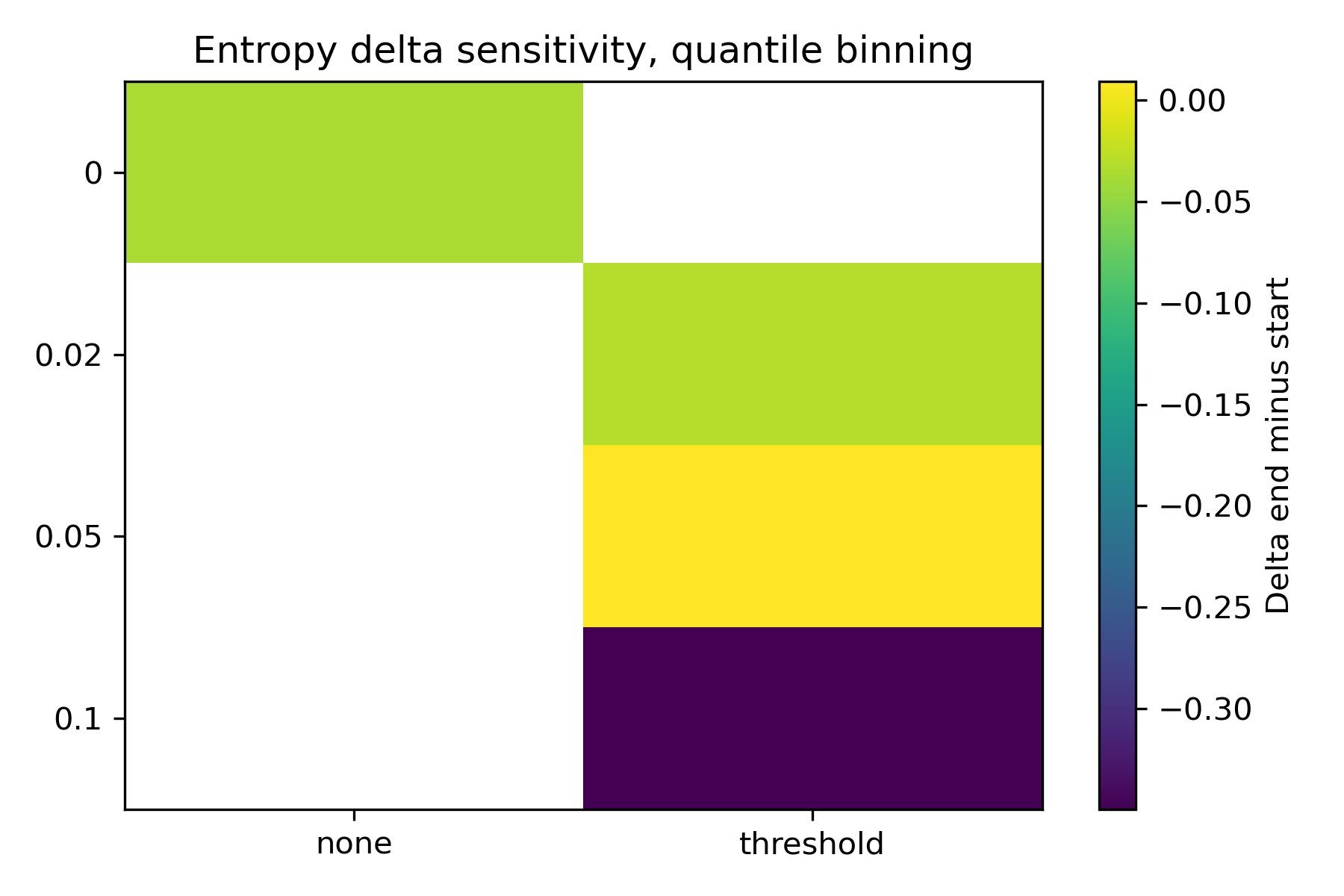}
    \caption{Sensitivity heatmap illustrating the end-minus-start delta for attribution entropy. The matrix compares absolute changes across different threshold ratios and binning methods, demonstrating persistent structural stability in the concentration trend.}
    \label{fig:si_sensitivity_heatmap}
\end{figure}

Fig.~\ref{fig:si_sensitivity_heatmap} presents the resulting end-minus-start delta of the attribution entropy across these thresholds. The delta values remain persistently negative under different thresholding ratios and binning methodologies. This consistency verifies that the reduction in entropy represents a robust underlying property of the risk attribution, rather than a byproduct of retaining long-tail computational noise.

To completely decouple the analysis from discrete binning choices, we further computed ordinary least squares slopes for the unbinned, time-continuous entropy traces within each independent event. As detailed in Table~\ref{tab:si_eventwise_slopes}, the continuous eventwise slopes for attribution entropy are predominantly negative across all scenarios. This strict within-event continuous evaluation corroborates the aggregate binned results, confirming that the narrowing of cue distribution during risk escalation is an event-level characteristic.

\begin{center}
\footnotesize
\begin{longtable}{lccc}
\\
\toprule
\textbf{Event ID} & \textbf{Sample Count ($n$)} & \textbf{Entropy Slope} & \textbf{Top-1 Share Slope} \\
\midrule
\endfirsthead
\multicolumn{4}{c}{{\tablename\ \thetable{} -- continued from previous page}} \\
\toprule
\textbf{Event ID} & \textbf{Sample Count ($n$)} & \textbf{Entropy Slope} & \textbf{Top-1 Share Slope} \\
\midrule
\endhead
\midrule
\multicolumn{4}{r}{{Continued on next page}} \\
\endfoot
\bottomrule
\endlastfoot
HB\_1 & 300 & -0.0020 & 0.0030 \\
HB\_2 & 300 & -0.1123 & 0.0571 \\
HB\_3 & 300 & -0.2444 & 0.1240 \\
HB\_4 & 300 & -0.0037 & 0.0132 \\
HB\_5 & 300 & -0.0339 & 0.0436 \\
HB\_6 & 300 & 0.0435 & 0.0256 \\
HB\_7 & 300 & -0.0258 & -0.0022 \\
HB\_8 & 300 & 0.2006 & -0.0418 \\
HB\_9 & 300 & 0.0915 & 0.0088 \\
HB\_10 & 300 & 0.0110 & -0.0020 \\
HB\_11 & 300 & -0.0180 & 0.0599 \\
HB\_12 & 300 & -0.0453 & 0.0525 \\
HB\_13 & 300 & 0.0259 & -0.0033 \\
HB\_14 & 300 & 0.0249 & -0.0092 \\
HB\_15 & 300 & 0.0918 & -0.0133 \\
HB\_16 & 300 & 0.0364 & -0.0084 \\
HB\_17 & 300 & 0.0774 & -0.0097 \\
HB\_18 & 300 & 0.0264 & -0.0010 \\
HB\_19 & 300 & -0.0039 & 0.0113 \\
HB\_20 & 300 & 0.0669 & -0.0030 \\
HB\_21 & 300 & 0.1766 & -0.0152 \\
HB\_22 & 300 & 0.0260 & -0.0004 \\
HB\_23 & 300 & 0.0487 & 0.0153 \\
HB\_24 & 300 & -0.0707 & 0.0476 \\
HB\_25 & 300 & -0.0003 & 0.0121 \\
HB\_26 & 300 & 0.0454 & -0.0106 \\
HB\_27 & 300 & 0.0619 & -0.0176 \\
LC\_1 & 360 & 0.0138 & -0.0034 \\
LC\_2 & 360 & 0.0061 & -0.0020 \\
LC\_3 & 360 & 0.0211 & -0.0050 \\
LC\_4 & 360 & 0.0441 & -0.0133 \\
LC\_5 & 360 & -0.0286 & 0.0265 \\
LC\_6 & 360 & -0.0795 & 0.0269 \\
LC\_7 & 360 & 0.0025 & 0.0038 \\
LC\_8 & 360 & -0.0878 & 0.0355 \\
LC\_9 & 360 & 0.0066 & 0.0007 \\
LC\_10 & 360 & -0.0834 & 0.0368 \\
LC\_11 & 360 & -0.0004 & 0.0098 \\
LC\_12 & 360 & -0.0389 & 0.0229 \\
LC\_13 & 360 & 0.0005 & -0.0042 \\
LC\_14 & 360 & -0.0045 & -0.0039 \\
LC\_15 & 360 & 0.0000 & -0.0005 \\
LC\_16 & 360 & 0.0336 & -0.0150 \\
LC\_17 & 360 & -0.0098 & 0.0044 \\
LC\_18 & 360 & 0.0280 & -0.0109 \\
LC\_19 & 360 & 0.0061 & -0.0131 \\
LC\_20 & 360 & 0.0522 & -0.0012 \\
LC\_21 & 360 & -0.0090 & 0.0009 \\
LC\_22 & 360 & 0.0195 & -0.0057 \\
LC\_23 & 360 & 0.0114 & -0.0049 \\
LC\_24 & 360 & 0.0532 & -0.0069 \\
MB\_1 & 300 & -0.0742 & 0.0310 \\
MB\_2 & 300 & -0.2221 & 0.0878 \\
MB\_3 & 300 & -0.1581 & 0.0738 \\
MB\_4 & 300 & -0.0630 & 0.0224 \\
MB\_5 & 300 & -0.1618 & 0.0768 \\
MB\_6 & 300 & -0.1717 & 0.0859 \\
MB\_7 & 300 & -0.0685 & 0.0306 \\
MB\_8 & 300 & -0.1366 & 0.0744 \\
MB\_9 & 300 & -0.0803 & 0.0512 \\
MB\_10 & 300 & -0.0530 & 0.0290 \\
MB\_11 & 300 & 0.0040 & 0.0030 \\
MB\_12 & 300 & -0.0063 & 0.0199 \\
MB\_13 & 300 & -0.0344 & 0.0115 \\
MB\_14 & 300 & -0.1065 & 0.0475 \\
MB\_15 & 300 & -0.1934 & 0.0968 \\
MB\_16 & 300 & -0.0668 & 0.0178 \\
MB\_17 & 300 & -0.1280 & 0.0618 \\
MB\_18 & 300 & -0.2321 & 0.0933 \\
MB\_19 & 300 & -0.0529 & 0.0256 \\
MB\_20 & 300 & -0.0235 & 0.0110 \\
MB\_21 & 300 & 0.0916 & -0.0348 \\
MB\_22 & 300 & -0.0719 & 0.0288 \\
MB\_23 & 300 & -0.0266 & 0.0186 \\
MB\_24 & 300 & -0.0992 & 0.0622 \\
MB\_25 & 300 & -0.0510 & 0.0169 \\
MB\_26 & 300 & -0.1648 & 0.0715 \\
MB\_27 & 300 & -0.0578 & 0.0347 \\
SVM\_1 & 300 & 0.0061 & 0.0141 \\
SVM\_2 & 300 & -0.0580 & 0.0280 \\
SVM\_3 & 300 & -0.1759 & 0.0843 \\
SVM\_4 & 300 & -0.0684 & 0.0461 \\
SVM\_5 & 300 & -0.1577 & 0.0950 \\
SVM\_6 & 300 & -0.0266 & 0.0063 \\
SVM\_7 & 300 & 0.0185 & -0.0108 \\
SVM\_8 & 300 & -0.0092 & -0.0048 \\
SVM\_9 & 300 & -0.1078 & 0.0491 \\
SVM\_10 & 300 & -0.0057 & 0.0047 \\
SVM\_11 & 300 & 0.0256 & -0.0042 \\
SVM\_12 & 300 & 0.0446 & -0.0133 \\
SVM\_13 & 300 & 0.0221 & -0.0058 \\
SVM\_14 & 300 & -0.0575 & 0.0231 \\
SVM\_15 & 300 & 0.0274 & -0.0302 \\
SVM\_16 & 300 & 0.0121 & -0.0068 \\
SVM\_17 & 300 & -0.0274 & 0.0142 \\
SVM\_18 & 300 & 0.0091 & -0.0065 \\
SVM\_19 & 300 & 0.0092 & 0.0044 \\
SVM\_20 & 300 & -0.0717 & 0.0417 \\
SVM\_21 & 300 & -0.0548 & 0.0330 \\
SVM\_22 & 300 & 0.0291 & 0.0053 \\
SVM\_23 & 300 & -0.0075 & 0.0099 \\
SVM\_24 & 300 & -0.0335 & 0.0133 \\
SVM\_25 & 300 & -0.0448 & 0.0190 \\
SVM\_26 & 300 & -0.0350 & 0.0067 \\
SVM\_27 & 300 & -0.0172 & 0.0006 \\
\end{longtable}
\captionof{table}{Eventwise continuous ordinary least squares slopes for attribution entropy and Top-1 share on unbinned time samples.} \label{tab:si_eventwise_slopes} 
\end{center}
\subsection{Brief introduction of SHapley Additive exPlanations (SHAP)}
SHAP is a method to explain the output of machine learning models by computing the contribution of each feature to the prediction for each instance. It uses Shapley values, a concept from game theory, to assign an importance value to each feature, showing how much each feature contributes to the prediction \cite{lundberg2017unified_SI,lundberg2018explainable_SI}. Based on neural network models that have been well-trained and demonstrate good performance, SHAP can serve to analyse the contribution of various features to the prediction. Specifically, some features may have a higher impact on risk prediction (e.g., longitudinal distance), while others may have a lower impact (e.g., acceleration). 

For a sample $\mathbf{X}=\{x_i\}_{i=1}^{D_\text{feature}}$ feed into model $f$, resulting in prediction $\hat{Y}=f(\mathbf{X})$, where $x_i$ represents $i^{\text{th}}$ features collected from the scenario, this process of analysing the feature impact can be formulated as follows:
\begin{equation}
\phi_i : (x_i, f, \hat{Y}) \rightarrow \mathbb{R}.
\end{equation}

The SHAP value is often used to analyse the contribution of each input feature in predictive models, which employs game theory and is theoretically justified. To illustrate using the computation of the $i^{\text{th}}$ feature as an example, the fundamental principle of the Shapley value involves traversing all permutations of feature subset coalitions $S$ without considering the feature $x_i$:
\begin{equation}
\phi_i^{\text{Shapley}} = \sum_{S \subseteq F \setminus \{i\}} \frac{|S|! \cdot (|F| - |S| - 1)!}{|F|!} \left[ f(x_{S \cup \{i\}}) - f(x_S) \right]
\end{equation}
where $|F|=D_\text{feature}$. $x_{S \cup \{i\}}$ and $x_S$ are the features with and without in the set ${S \cup \{i\}}$ and $S$. $f(x_{S \cup \{i\}})$ and $f(x_S)$ are neural network models trained in corresponding feature set.

The SHAP method is a unified approach that offers both global and local interpretability for inputs, which is the Shapley value of a conditional expectation function of the original model. It attributes to each feature the change in the expected model prediction when conditioning on that feature. A surprising characteristic of SHAP is the presence of a single unique solution in this class with three desirable properties: local accuracy, missingness, and consistency \cite{lundberg2020local2global_SI}. 

\begin{itemize}
    \item \textbf{Local accuracy}. Local accuracy in the context of SHAP refers to the property that ensures the explanation model’s output matches exactly with the original model’s $f(x)$ output for individual predictions. This means that for any specific instance, the sum of SHAP values assigned to each feature, along with the base value, will accurately equal the prediction of the original model. This characteristic ensures that SHAP provides precise and faithful explanations at the individual sample level.
    Local accuracy ensures that the sum of SHAP values and the base value exactly matches the model's output for any given input.
\begin{equation}
\begin{aligned}
f(x)=g(x') &= \phi_0 + \sum_{i=1}^{D_\text{feature}} \phi_i x_i' 
        &= \text{bias} + \sum \text{contribution of each feature}
\end{aligned}
\end{equation}
where $x$ is related to its original feature by $x=h_x(x')$ by mapping function $h_x(\cdot)$, and $\phi_0=h_x(0)$ denotes that all simplified features are toggled off (i.e. missing). 
    \item \textbf{Missingness}. 
The concept of missingness in SHAP relates to how the method handles features that are absent or missing in a given data instance. SHAP accounts for missing features by allocating a Shapley value of zero to them, reflecting their non-contribution to the model's prediction for that specific instance. This approach acknowledges the absence of data and ensures that only the present features contribute to the explanation of a model's output. The property of missingness can be formulated as: If
\begin{equation}
f_x(S \cup i) = f_x(S)
\end{equation}
for all subsets of features $S\subseteq F$, then $\phi_i(f,x)=0$. 
\item \textbf{Consistency}. The consistency characteristic refers to the principle that if a model changes so that the contribution of a feature increases or stays the same. Meanwhile, regardless of other features, the SHAP value for that feature should not decrease. This ensures that the explanation model remains consistent with changes in the feature's impact on the prediction. In other words, SHAP values faithfully represent the proportional impact of each feature on the model's output, adhering to changes in feature importance. The property of consistency can be formulated as: For any two models $f$ and $f'$, if
\begin{equation}
f'_x(S) - f'_x(S \setminus i) \geq f_x(S) - f_x(S \setminus i)
\end{equation}
for all ${S \subseteq F \setminus \{i\}}$, where $F$ is the set of $D_\text{feature}$ input features, then $\phi_i(f',x)\geq\phi_i(f,x)$.
\end{itemize}

According to Ref.~\cite{lundberg2017unified_SI}, only one possible explanation satisfies all three properties: 
\begin{equation}
\phi_i(f, x) = \sum_{z' \subseteq x'} \frac{|z'|!(D_\text{feature} - |z'| - 1)!}{D_\text{feature}!} [f_x(z') - f_x(z' \setminus i)]
\end{equation}
where $|z'|$ is the number of non-zero entries in $z'$, and $z' \subseteq x'$ represents all $z'$ vectors where the non-zero entries are a subset of the non-zero entries in $x'$.
\begin{equation}
\begin{aligned}
f(h_x(z')) &= \mathbb{E}[f(z) | z_S] & \text{simplified input mapping} \\
&= \mathbb{E}_{z_{\overline{S}}|z_S}[f(z)] & \text{expectation over} z_{\overline{S}}|z_S  \\
&\approx \mathbb{E}_{z_{\overline{S}}}[f(z)] & \text{assumption (as in Refs.~\cite{vstrumbelj2014explaining_SI,ribeiro2016should_SI,shrikumar2017learning_SI,datta2016algorithmic_SI})}  \\
&\approx f((z_S, \mathbb{E}[z_{\overline{S}}])). & \text{assume model linearity} 
\end{aligned}
\end{equation}
Note that $z_{\overline{S}}$ is the set of features not in $S$.
\subsubsection{Shapley results availability}
\begin{figure*}[t]
    \centering

    \begin{subfigure}[t]{0.49\textwidth}
        \centering
        \includegraphics[width=\linewidth]{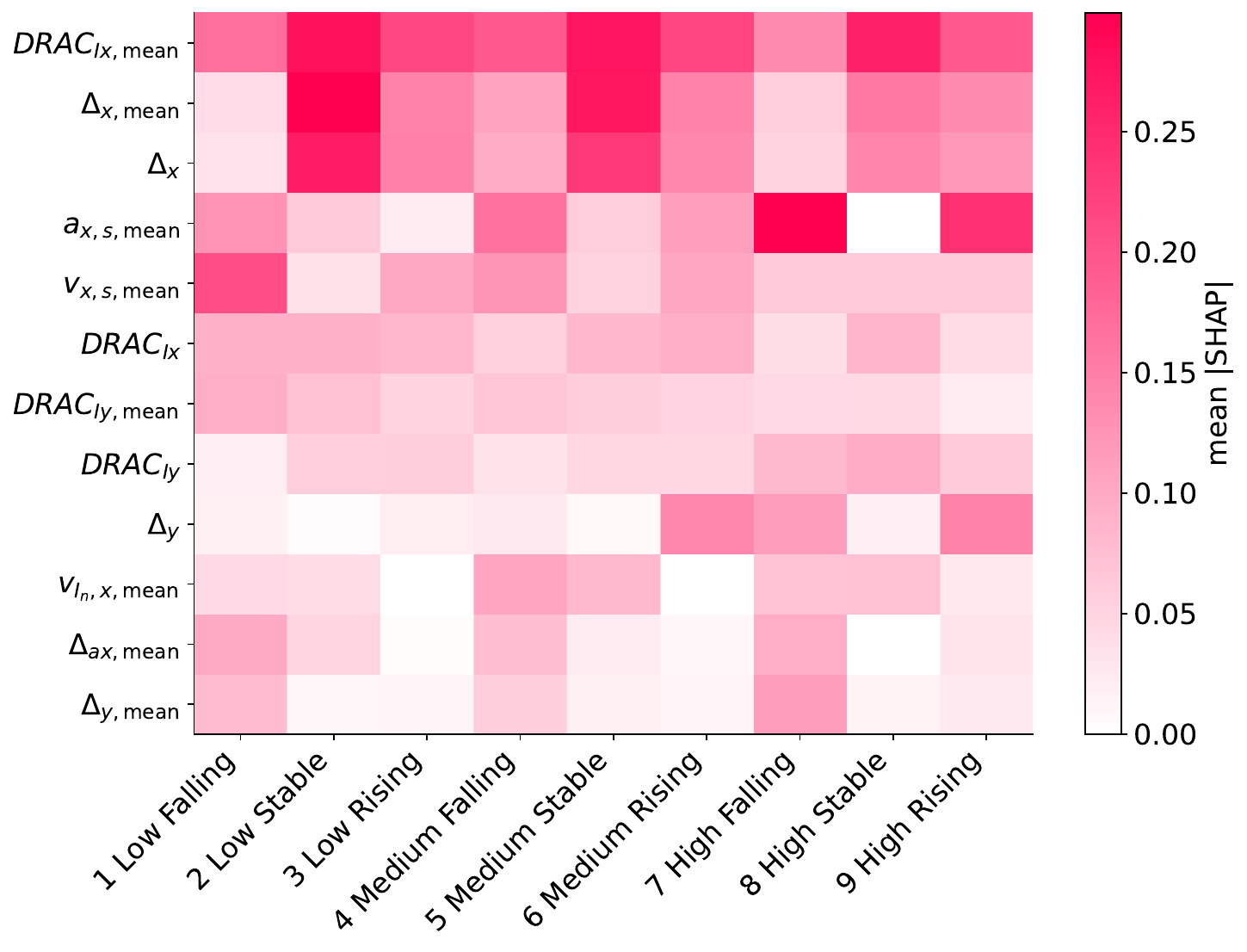}
        \caption{HB.}
        \label{fig:si:heatmap_hb}
    \end{subfigure}
    \hfill
    \begin{subfigure}[t]{0.49\textwidth}
        \centering
        \includegraphics[width=\linewidth]{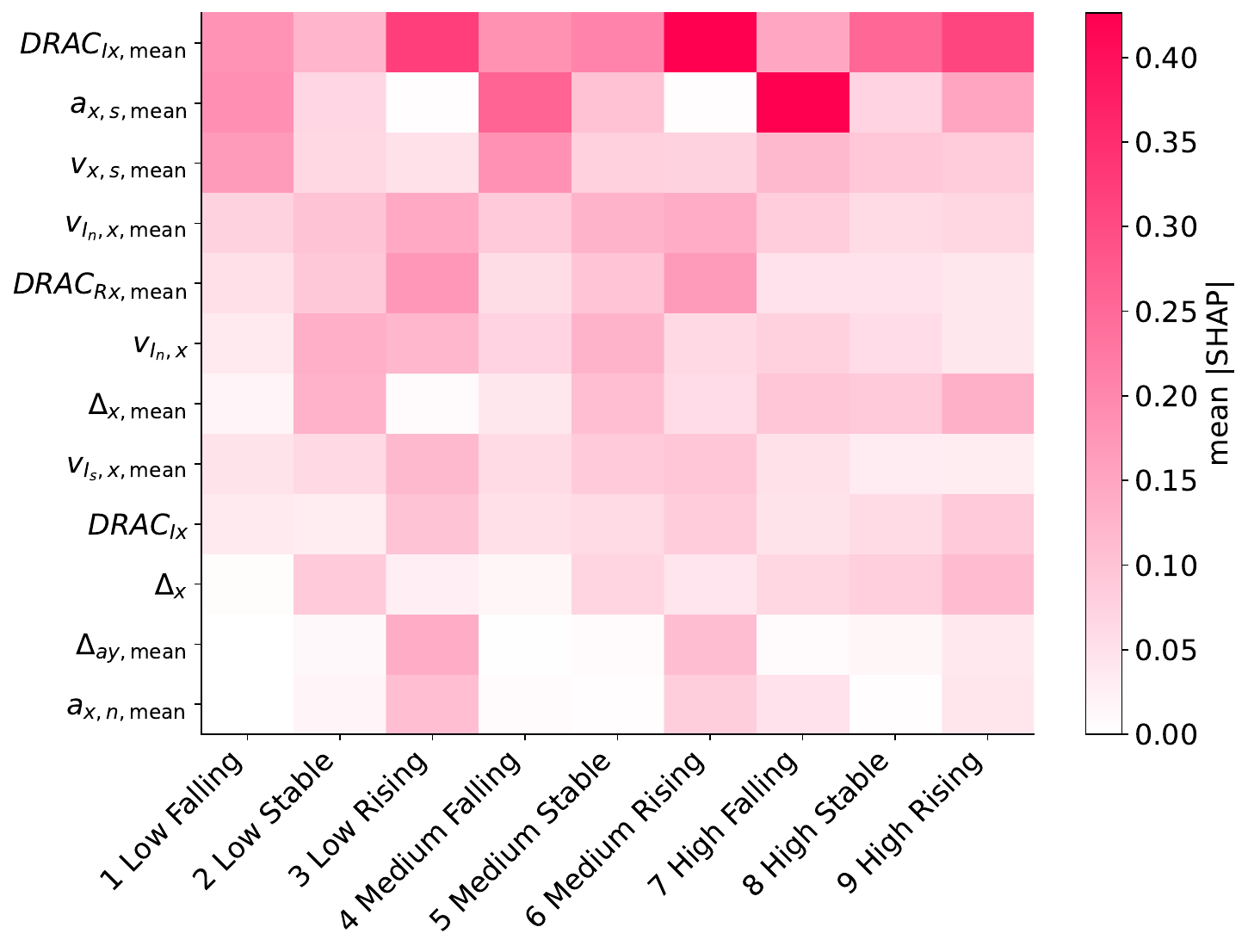}
        \caption{MB.}
        \label{fig:si:heatmap_mb}
    \end{subfigure}

    \vspace{0.6em}

    \begin{subfigure}[t]{0.49\textwidth}
        \centering
        \includegraphics[width=\linewidth]{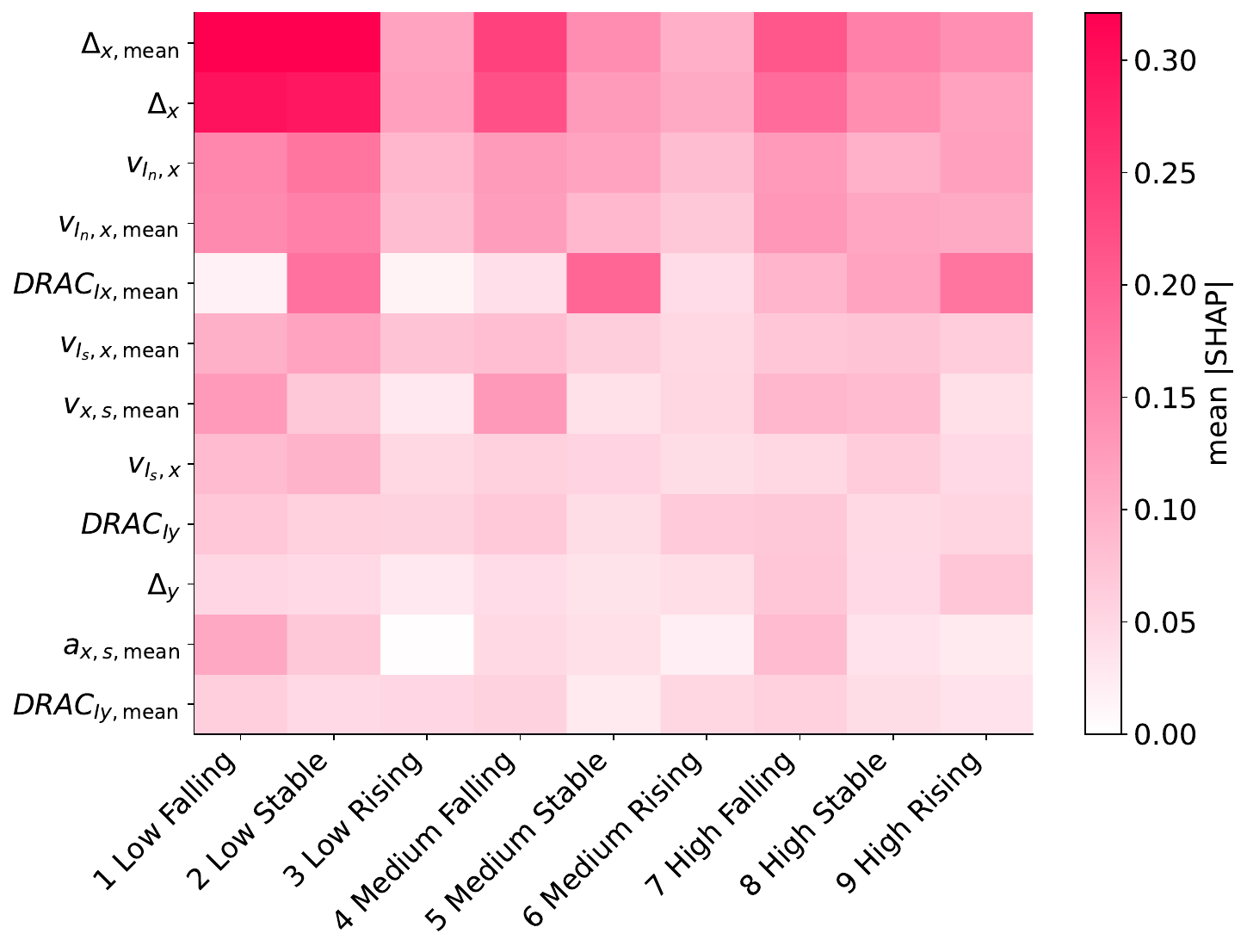}
        \caption{LC.}
        \label{fig:si:heatmap_lc}
    \end{subfigure}
    \hfill
    \begin{subfigure}[t]{0.49\textwidth}
        \centering
        \includegraphics[width=\linewidth]{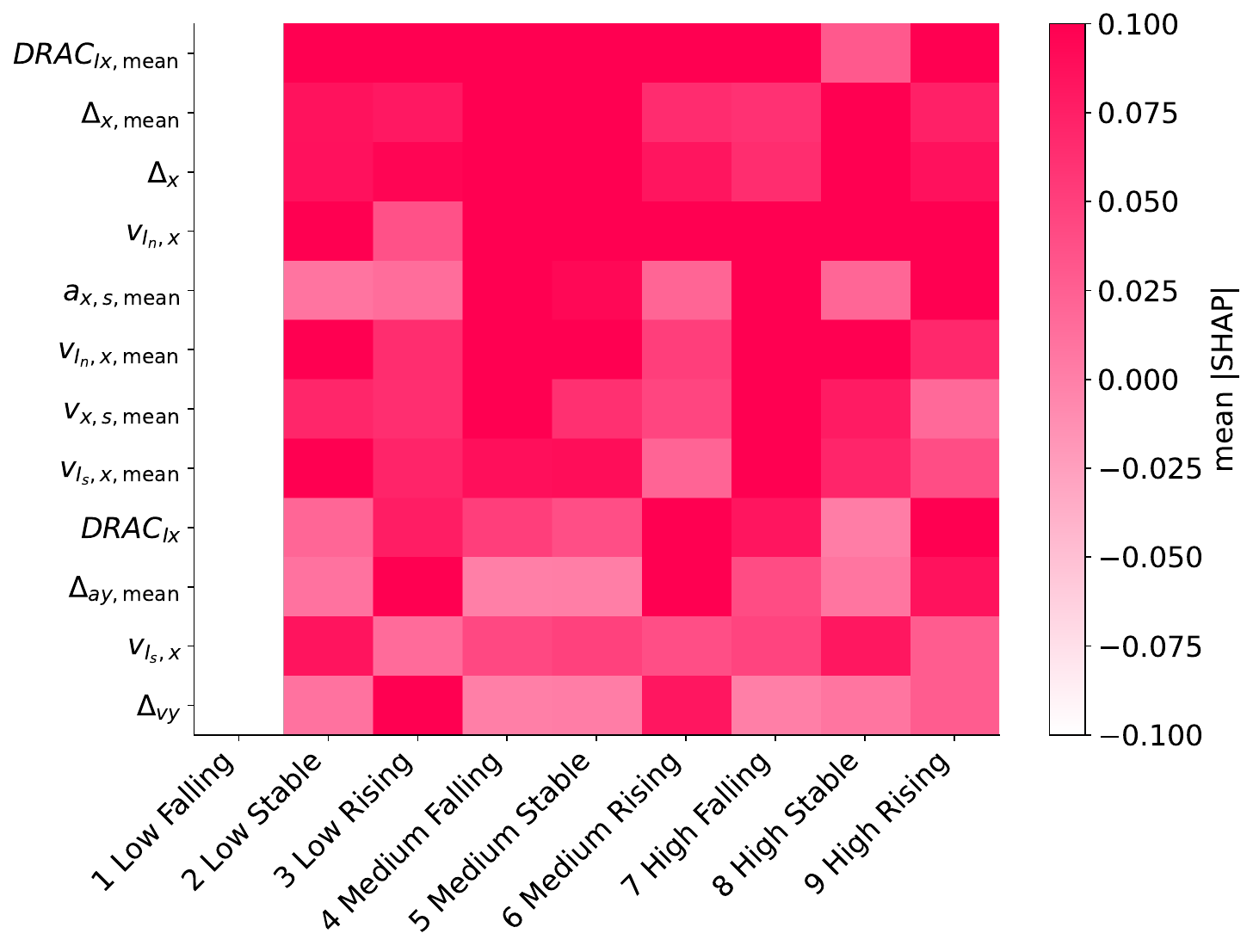}
        \caption{SVM.}
        \label{fig:si:heatmap_svm}
    \end{subfigure}

    \caption{Scenario specific heatmaps of mean absolute SHAP values for the top 12 features, summarised over nine inferred evolution states.}
    \label{fig:si:scenario_top12_shap_heatmaps}
\end{figure*}

\begin{figure}[h]  
    \centering

    \begin{subfigure}{0.49\textwidth}
        \centering
  
        \includegraphics[width=\linewidth]{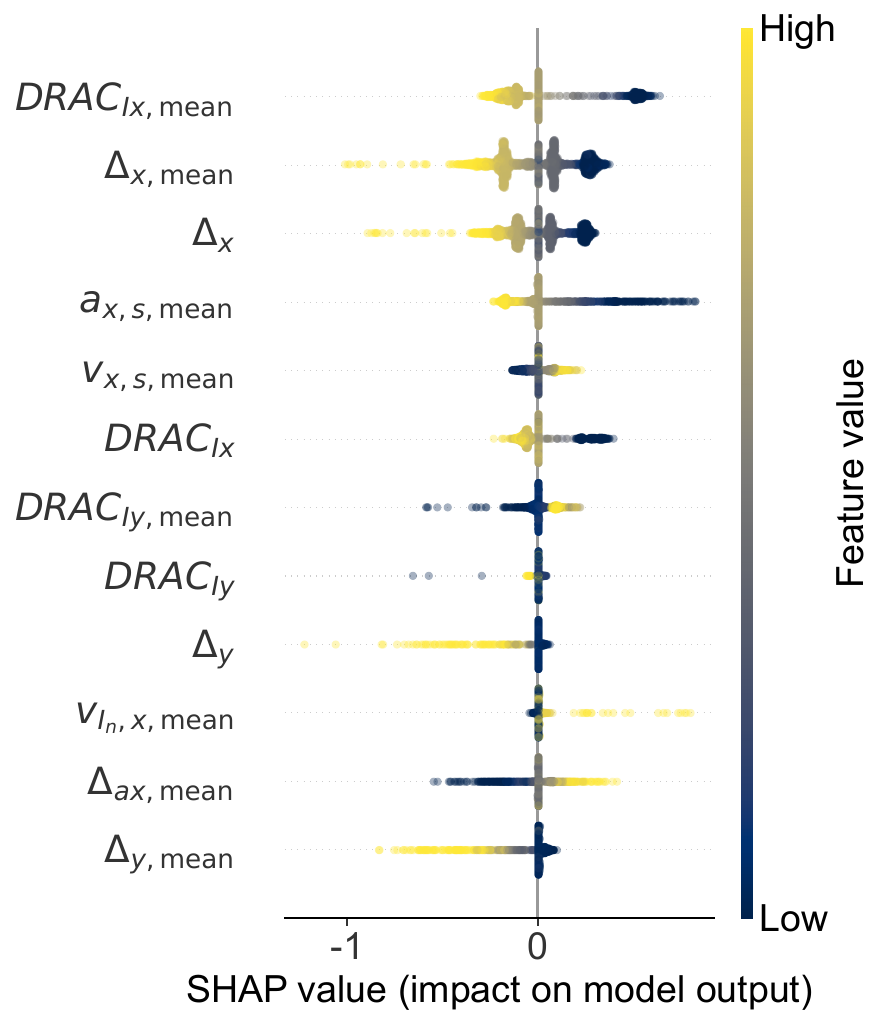}
        \caption{HB} 
        \label{fig:si:heatmap_hb}
    \end{subfigure}
    \hfill
    % ---  MB ---
    \begin{subfigure}{0.49\textwidth}
        \centering
        \includegraphics[width=\linewidth]{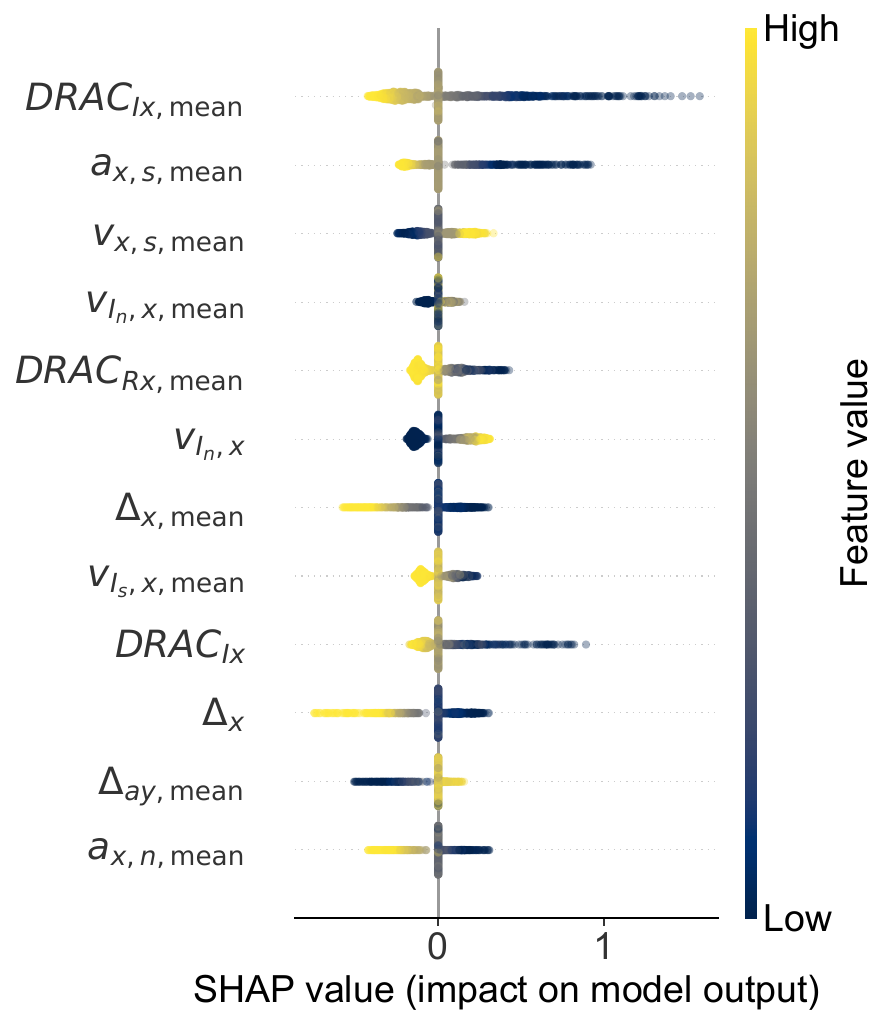}
        \caption{MB} 
        \label{fig:si:heatmap_mb}
    \end{subfigure}

    \vspace{1em} 

    % ---  LC ---
    \begin{subfigure}{0.49\textwidth}
        \centering
        \includegraphics[width=\linewidth]{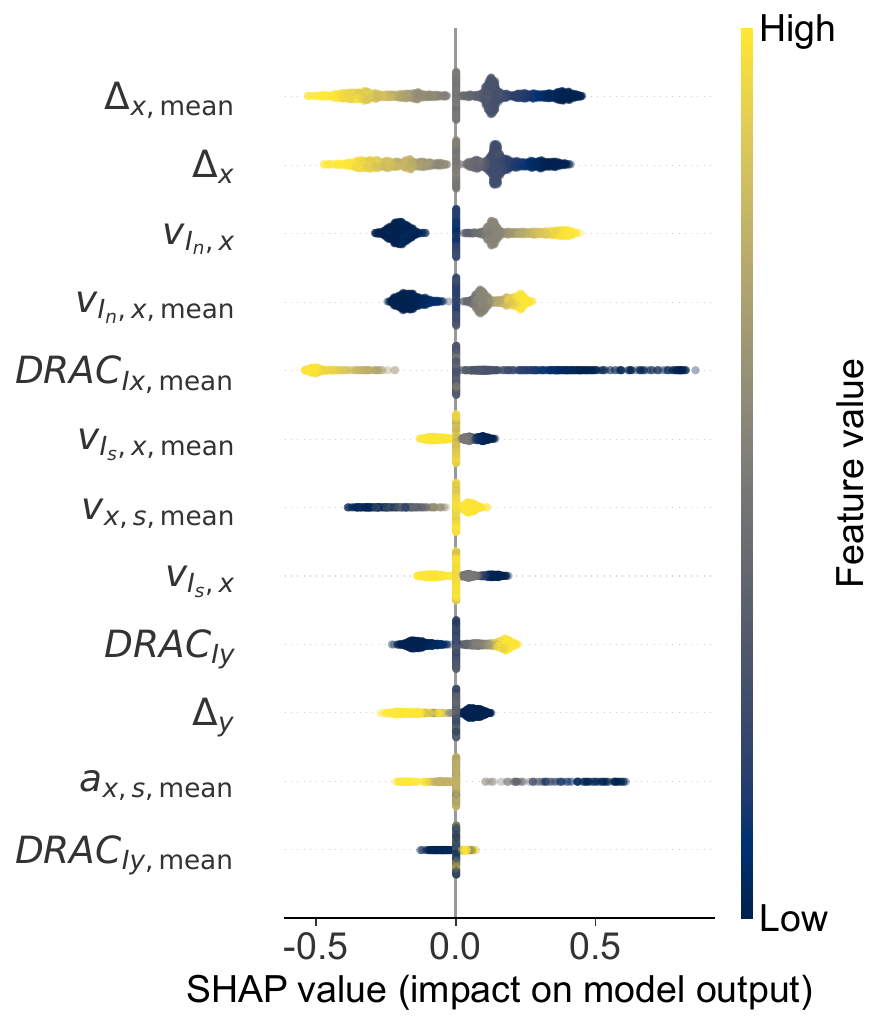}
        \caption{LC}
        \label{fig:si:heatmap_lc}
    \end{subfigure}
    \hfill
    % --- SVM ---
    \begin{subfigure}{0.49\textwidth}
        \centering
        \includegraphics[width=\linewidth]{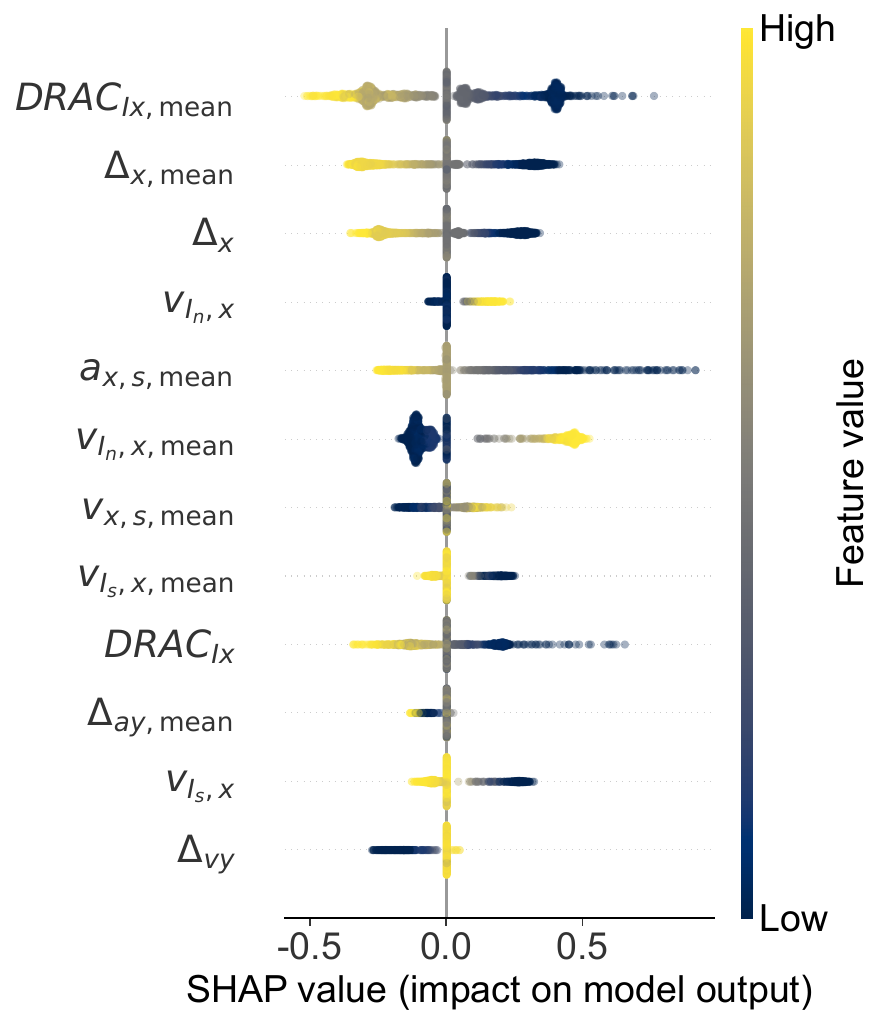}
        \caption{SVM}
        \label{fig:si:heatmap_svm}
    \end{subfigure}

    \caption{Scenario specific beeswarm summarised by mean values over events}
    \label{fig:si:scenario_shap_beeswarm}
\end{figure}

Scenario specific heatmaps of mean absolute Shapley values summarised over nine inferred evolution states are shown in Fig. \ref{fig:si:scenario_top12_shap_heatmaps}. Scenario specific beeswarm summarised by mean values over events are shown in Fig. \ref{fig:si:scenario_shap_beeswarm}. The exported Shapley value tables and event level summaries underlying this section, and time-based Shapley for all 105 events are included in the data repository \hyperlink{10.4121/242d9474-e522-4518-8917-8f284fc8a7a8}{10.4121/242d9474-e522-4518-8917-8f284fc8a7a8}.

\clearpage
% \bibliography{sn-bibliography.bib}
\makeatletter
\makeatother

% ==========================================

\end{document}